%% file: main.tex
\documentclass[aoas,preprint]{imsart}

\RequirePackage{amsthm,amsmath,amsfonts,amssymb}
\RequirePackage[authoryear]{natbib}
\RequirePackage{hyperref}
\RequirePackage{graphicx}

\usepackage[utf8]{inputenc} 
\usepackage[T1]{fontenc}    
\usepackage{url}            
\usepackage{booktabs}       
\usepackage{nicefrac}       
\usepackage{algorithm, algpseudocode, algorithmicx}
\usepackage{subcaption}
\usepackage{color,xcolor}
\usepackage{epsfig}
\usepackage{bm,bbm}
\usepackage{comment}
\usepackage{multirow}
\usepackage{multibib} 

\usepackage{tikz}
\usetikzlibrary{bayesnet}
\usetikzlibrary{arrows}
\usetikzlibrary{backgrounds}

\makeatletter
\usepackage{comment}
\let\wfs@comment@comment\comment
\let\comment\@undefined

\usepackage{easyReview}
\let\wfs@easyReview@comment\comment
\let\comment\@undefined

\newcommand\comment{%
    \ifthenelse{\equal{\@currenvir}{comment}}
    {\wfs@comment@comment}
    {\wfs@easyReview@comment}%
}
\makeatother

\startlocaldefs
\hypersetup{colorlinks,citecolor=blue,urlcolor=blue}
\newcites{supp}{REFERENCES}

\theoremstyle{remark}

\def\d{\mathrm d}

\def\E{\mathbb E}
\def\V{\mathbb V}

\def\I{\mathbbm 1}

\def\Normal{\operatorname N}

\def\argmax{\operatorname {argmax}}
\DeclareMathOperator*{\GP}{GP}
\DeclareMathOperator*{\MED}{Med}

\newcommand{\Av}[1]{\operatorname{Av}\left(#1\right)}

\newcommand{\pnorm}[2]{\left\|#2\right\|_{#1}}

\newcommand{\diag}[1]{\operatorname{diag}\left(#1\right)}

\def\Data{\mathcal D}
\def\Window{\mathcal W}
\def\Real{\mathbb R}

\def\Xspace{\mathbb{X}}
\def\Tspace{\mathbb{T}}

\def\xvec{\bm{x}}
\def\svec{\bm{s}}
\def\vel{\bm{\mathrm v}}
\def\bbBeta{\bm{B}}
\def\bbbeta{\bm{\beta}}
\def\bbxi{\bm{\xi}}
\def\bbXi{\bm{\Xi}}

\def\calM{\mathcal{M}}
\def\calL{\mathcal{L}}
\def\calS{\mathcal{S}}
\def\calX{\mathcal{X}}
\def\calT{\mathcal{T}}

\def\OHT{\mathrm{OHT}}
\def\TT{\mathrm{TT}_0}

\endlocaldefs

\begin{document}

\begin{frontmatter}
\title{Spatio-temporal Local Interpolation of Global Ocean Heat Transport using Argo Floats: A Debiased Latent Gaussian Process Approach}\runtitle{Spatio-Temporal Interpolation of Global Ocean Heat Transport}

\begin{aug}
\author[A]{\fnms{Beomjo} \snm{Park}\ead[label=e1,mark]{beomjo@stat.cmu.edu}},
\author[A]{\fnms{Mikael} \snm{Kuusela}\ead[label=e2,mark]{mkuusela@andrew.cmu.edu}},
\author[B]{\fnms{Donata} \snm{Giglio}\ead[label=e3]{donata.giglio@colorado.edu}}
\and
\author[C]{\fnms{Alison} \snm{Gray}\ead[label=e4]{argray@uw.edu}}
\address[A]{Dept. of Statistics \& Data Science, Carnegie Mellon University,
\printead{e1,e2}}

\address[B]{Dept. of Atmospheric and Oceanic Sciences,
University of Colorado Boulder,
\printead{e3}}

\address[C]{School of Oceanography,
University of Washington,
\printead{e4}}
\end{aug}

\begin{abstract}
The world ocean plays a key role in redistributing heat in the climate system and hence in regulating Earth's climate.
Yet statistical analysis of ocean heat transport suffers from partially incomplete large-scale data intertwined with complex spatio-temporal dynamics, as well as from potential model misspecification. We present a comprehensive spatio-temporal statistical framework tailored to interpolating the global ocean heat transport using in-situ Argo profiling float measurements. We formalize the statistical challenges using latent local Gaussian process regression accompanied by a two-stage fitting procedure. We introduce an approximate Expectation-Maximization algorithm to jointly estimate both the mean field and the covariance parameters, and refine the potentially under-specified mean field model with a debiasing procedure. This approach provides data-driven global ocean heat transport fields that vary in both space and time and can provide insights into crucial dynamical phenomena, such as El Ni{\~n}o \& La Ni{\~n}a, as well as the global climatological mean heat transport field, which by itself is of scientific interest. The proposed framework and the Argo-based estimates are thoroughly validated with state-of-the-art multimission satellite products and shown to yield realistic subsurface ocean heat transport estimates.
\end{abstract}


\end{frontmatter}

\input{sec1_intro.tex}
\input{sec2_data_background_ver2.tex}

\section{Statistical Methodology}\label{sec: Methods}
\input{sec3_1_method_overview.tex}

\input{sec3_1_LGPR.tex}

\input{sec3_1_LLGPR_twostage.tex}

\input{sec3_1_AM.tex}
\input{sec3_1_Detrend_ADA.tex}

\input{sec3_1_algorithm.tex}

\section{OHT Field Estimated from Argo Data}\label{sec:Results}
\input{sec5_1_estimate.tex}

\input{sec6_validation.tex}
\input{sec7_discussion.tex}

 \section*{Acknowledgements}
We would like to acknowledge high-performance computing support from Cheyenne (\href{https://www.doi.org/10.5065/D6RX99HX}{doi:10.5065/D6RX99HX}) provided by NCAR's Computational and Information Systems Laboratory, sponsored by the National Science Foundation. Donata Giglio acknowledges support from NASA (Award NNH20ZDA001N-PO). Donata Giglio and Mikael Kuusela acknowledge support from NOAA (Award NA21OAR4310261). Alison Gray acknowledges support from NASA (Award NNX80NSSC19K1252), the U.S. Argo Program through NOAA (Award NA15OAR4320063), and the Microsoft Faculty Fellowship program.
We are grateful to the Statistical Oceanography and STAMPS groups, and especially to Fred Bingham, Sarah Gille and Matt Mazloff for constructive discussions and suggestions related to Argo, Spray data and physical oceanography. We appreciate the constructive feedback by the Editor, the Associate Editor and the two anonymous reviewers which substantially improved the utility and readability of the paper.


\begin{supplement} 
\sname{Supplement A}
\stitle{Supplement to ``Spatio-temporal Local Interpolation of Global Ocean Heat Transport using Argo Floats: A Debiased Latent Gaussian Process Approach''}
\slink[doi]{COMPLETED BY THE TYPESETTER}
\sdatatype{.pdf}
\sdescription{We provide in the supplement additional details on quality control, analytic formulas regarding the covariance kernel, explicit derivations of the EM procedure, the predictive distribution for single depth OHT, and supporting figures, as well as extensive additional analyses where Spray glider profiles are jointly analyzed with Argo float profiles.}
\end{supplement}


\bibliographystyle{imsart-nameyear} 
\bibliography{references}       

\input{appendix}

\clearpage

\input{supplement.tex}

 \bibliographystylesupp{imsart-nameyear} 
 \bibliographysupp{references}

\end{document}

%% file: sec1_intro.tex
\section{Introduction}\label{sec: Intro}

The ocean plays a pivotal role in regulating Earth's climate on regional to global scales \citep[e.g.,][]{bryden_chapter_2001,macdonald_ocean_2013,stocker_chapter_2013}. Notably, it redistributes the excess heat taken up at the equator, transporting it to higher latitudes where it is released to the atmosphere \citep{trenberth_global_1994,ganachaud_improved_2000,trenberth_estimates_2001,forget_global_2019}.
Convergence and divergence of heat in the ocean also have impacts on regional sea level (via thermal expansion of sea water, e.g., \citet{forget_partition_2015}), with implications for local populations. Ocean heat transport can additionally regulate regional temperature extremes in the ocean, with implications for marine ecosystems. As an example of the latter, \citet{behrens_meridional_2019} describe a causal link between ocean heat content and the area and intensity of marine heatwaves in the Tasman Sea: ocean heat content fluctuations in the Tasman Sea are largely controlled by meridional transport of heat in the ocean; hence, better estimates of ocean heat transport can help improve forecasts of marine heatwaves, with potential implications for the management of ecosystems in Australasia \add{and beyond}. \add{Indeed, marine heatwaves are a global phenomenon and their relevance for marine ecosystems makes them the focus of several ongoing studies that investigate their generation, demise, and impacts on ocean biogeochemistry \citep[see, e.g.,][]{mogen_ocean_2022}}.

In this paper, we present a  statistical framework to characterize  ocean heat transport (OHT) over the global ice-free ocean during 2007--2018, based on direct observations of temperature and salinity in the upper 2000\,m of the ocean \add{from Argo profiling floats}. \add{While} obtaining an accurate picture of the heat transport within and across ocean basins is critical to understanding changes in the climate system and for data-driven policy and decision making in a changing climate, direct OHT estimates are typically made at only a few locations where suitable ship- or mooring-based observations are available and thus do not provide a global view. Historically, global OHT has been estimated indirectly by subtracting the atmospheric component from total heat transport estimates \citep{trenberth_global_1994,trenberth_estimates_2001}, leveraging top-of-the-atmosphere radiation measurements from satellites. The Argo array of profiling floats, in contrast, collects observations of temperature and salinity in the upper 2000\,m of the open ocean  with unprecedented spatio-temporal coverage \citep{jayne_argo_2017}. \add{In-situ Argo profiles} provide an extraordinary opportunity to quantify, on a global scale, the spatial and temporal variability of upper ocean heat transport\add{, both via observation-only based mapping methods and via data assimilation \citep[e.g.,][]{dong_importance_2011}.}

When Argo measurements are used in scientific analyses, a vast majority of literature relies on spatio-temporally interpolated temperature and salinity maps that convert the Argo measurements sampled irregularly in space and time to a regular spatio-temporal grid \citep[e.g.,][]{roemmich_20042008_2009,good_en4_2013}. \add{Gridded products are key to leveraging point observations to better understand the climate system. As one example, gridded products of ocean heat content based on ocean observations are the most efficient approach to estimate the 
energy imbalance at the top of the atmosphere.
This is the case as over $93\%$ of the excess energy that is gained by the Earth in response to the positive energy imbalance at the top of the atmosphere accumulates into the ocean in the form of heat \citep{meyssignac_measuring_2019}. Gridded OHT products are essential to understand how this heat energy is redistributed in the climate system \citep[see, e.g.,][]{colin_de_verdiere_meridional_2019,sun_global_2019,mcmonigal_reduction_2022}. Argo-based OHT estimates was used, in particular, to investigate the mechanisms behind the observed basin-scale warming in the Indian Ocean \citep{mcmonigal_reduction_2022}. Furthermore, gridded heat and salt transport estimates provide insight on how ocean circulation regulates regional water properties (including in marine protected areas) and sea level \citep[][]{giglio_understanding_2013,kawai_heat_2021}. These estimates are also useful to validate climate models that produce projections of how the Earth's climate may change in the future under different emission scenarios \citep[see, e.g.,][]{li_assessment_2020}.} 

However, unlike temperature and salinity, interpolating OHT faces a critical challenge from the fact that OHT---a vertical integral of essentially the product between temperature and velocity---is only partially observed by common oceanographic instruments, including Argo floats. Even though each float records temperature directly, the velocity, and thus OHT, is not directly measured (and cannot be derived from a single observation) but rather has to be inferred as the gradient of a variable computed from the in-situ observations. Such latent construction constitutes the crux of a statistical challenge distinct from archetypal spatio-temporal interpolation problems.

The latent nature of the problem is intertwined with the classical challenges in modern large-scale spatio-temporal statistics: spatio-temporal local dependency, global heterogeneity, and model misspecification, not to mention the large volume of in-situ Argo data \citep[see e.g.,][]{cressie_statistics_2011}. In particular, (1) both the latent velocity field and the OHT field are globally non-stationary spatio-temporal processes; (2) sharp ocean fronts are insufficiently identified when model misspecification is not properly addressed; (3) the massive number of irregularly-spaced, sparse spatio-temporal observations demands computationally efficient methods that are able to account for both the variability and the underlying spatio-temporal structure of the data.

To overcome these challenges, we propose a two-stage statistical framework based on debiased local Gaussian process regression (LGPR), extending the work of \citet{kuusela_locally_2018} on Argo temperature fields. The framework is a comprehensive suite of statistical techniques tailored to OHT interpolation, in that we formalize the statistical challenges into a latent LGPR model accompanied by a two-stage fitting procedure, introduce an approximate Expectation-Maximization (EM) algorithm \citep{dempster_maximum_1977} to jointly estimate both the mean field and the covariance parameters, and refine the potentially misspecified model with a debiasing procedure. The two-stage procedure solves the spatio-temporally correlated latent variable problem by predicting the latent velocity fields on the first stage using LGPR with a related oceanographic variable whose realizations are directly measured for each Argo profile. 
Our approach is unifying since the same LGPR framework succinctly represents both velocity and OHT fields.

We improve the LGPR approach of \citet{kuusela_locally_2018} by simultaneously estimating both the mean field and the covariance parameters with an iterative EM algorithm in a computationally efficient manner. \citet{kuusela_locally_2018} focus on estimating a local space-time covariance model from mean-centered temperature observations where the mean field was estimated with ordinary least squares (OLS). 
Joint estimation on both mean and covariance parameters is imperative in OHT estimation, as we need to estimate the actual mean field of the latent process not only the mean-centered field. Joint estimation of both mean and the covariance parameters is not uncommon in spatial statistics. For instance, the sub-optimality of OLS in a regression kriging context is typically resolved using generalized least squares (GLS), which accounts for the spatio-temporal correlation of the residuals \citep{cressie_statistics_1993}. A similar iterative GLS approach was also adopted to estimate velocities from Argo data in \citet{gray_method_2015}. Our approximate EM algorithm shares the same spirit but requires a separate treatment since the LGPR model localizes the spatio-temporal covariances seasonally along the temporal axis within the span of the spatio-temporal mean. 

Predicting the latent velocity field with an under-specified model may result in a concerning bias. It is vital to correct  the bias since a bias in the latent field would propagate to the second stage, degrading the final OHT interpolation. By formalizing an approach previously used by oceanographers \citep{roemmich_20042008_2009}, we provide an intuitive debiasing procedure by estimating the bias from the predicted field and then correcting the estimated bias in an iterative manner. This data-driven debiasing procedure is shown to capture sharper ocean fronts bearing crucial importance in ocean dynamics and to improve prediction and interpolation, as confirmed by a validation study based on satellite data. While our approach has close connections to iterative bias-correction in classical regression modeling \citep{kuk_asymptotically_1995, guerrier_asymptotically_2020} and to accounting for model discrepancy in Bayesian computer model calibration \citep{kennedy_bayesian_2001,bayarri_computer_2007,brynjarsdottir_learning_2014}, it has, to the best of our knowledge, not been previously embraced by the spatio-temporal statistics community.

Our work aligns with the oceanographic community's interests yet does not address all of the challenges in characterizing global OHT with in-situ Argo measurements. Currently, the Argo fleet does not fully resolve the narrow western boundary currents that are a key component of the global OHT, nor does it sample below 2000 m on a global scale (although expansions of the array to address both of these deficiencies are being planned).  In addition, the portion of the velocity field directly driven by the winds (i.e., the Ekman velocity) cannot be estimated from measurements of temperature and salinity, despite playing a non-negligible role in OHT. Thus, Argo observations must be integrated with other datasets for full-depth, cross-basin estimates of OHT. Even though our paper focuses only on Argo-based OHT estimates, in Section~\ref{sec: Supp_Spray} of the Supplementary Material \citep{park_supplement_2020}, we provide improved estimates in the western North Atlantic Ocean by applying our proposed framework to data from both Argo floats and Spray gliders \citep{rudnick_spray_2016}.  An alternative approach for estimating OHT from Argo observations \citep{colin_de_verdiere_meridional_2019} contends with these issues by combining float- and ship-based datasets.  That method, however, entails solving two Poisson equations over the entire domain, and thus the results depend heavily on the adhoc specification of accurate boundary conditions.

The rest of the paper is organized as follows. Section~\ref{sec:Background_Data} defines ocean heat transport and gives a brief overview of the related scientific context and the Argo dataset. In Section~\ref{sec: Methods}, we present the complete framework for quantifying global ocean heat transport fields based on Argo data. This includes the spatio-temporal model specification, estimation, and refinement procedures. Section~\ref{sec:Results} presents the estimated latent velocity and OHT fields and illustrates a scientific application of the resulting OHT estimates in the context of the El Ni\~{n}o--Southern Oscillation. Section~\ref{sec: Validation} validates both our proposed method and the resulting estimates using state-of-the-art satellite products. Section~\ref{sec: discussion} discusses the results and implications along with future research directions. Our code is publicly available online at {\url{https://github.com/beomjopark/OHT_analysis}} for reproducibility and re-use of the proposed framework.

%% file: sec2_data_background_ver2.tex
\section{Scientific Background and Data}\label{sec:Background_Data}

Before describing our statistical methodology, we provide a brief introduction to the computation of OHT, as well as relevant details of the Argo profiling float dataset. The reader is referred to \citet{macdonald_ocean_2013} for a detailed review of OHT and its impact in the climate system and to \citet{wong_argo_2020} for a thorough treatment of the Argo dataset.

Fundamentally, the ocean transports heat because it moves water with velocity $\vel$ while containing an amount of heat proportional to its potential temperature $\theta$ (where $\theta$ removes the thermodynamic effect of pressure from the in-situ temperature $T$).  This heat flux can be integrated vertically at any location in the ocean to produce an estimate of OHT.  We thus define OHT at a given spatial coordinate $\xvec=(x,y) \in \Xspace \subseteq \Real^2$ and temporal coordinate $t \in \Tspace \subset \Real_+$ as
\begin{align}\label{eq: OHT_def}
    \OHT(\xvec,t) &=  C_p \int \theta (\xvec, t, z) \vel (\xvec, t, z) \rho (\xvec, t, z) \,\d z 
            = C_p  \int \frac{ \theta (\xvec, t, p) \vel (\xvec, t, p)}{g (\xvec, p)} \d p, 
\end{align}
where $\rho$ is in-situ water density, $C_p$ is the specific heat capacity of seawater, and $g$ is gravitational acceleration.  As shown here, the vertical integral can be computed in depth $z$ or, using the hydrostatic relation, can be expressed as a function of pressure $p$.  Given the range in density of seawater, pressure $p$ (in units of decibars, dbar) and depth $z$ (in units of meters, m) are approximately equal and are often used interchangeably in the oceanographic literature.  Nonetheless, in-situ oceanographic observations, including those from the Argo profiling floats, predominately measure pressure and not depth, and thus here we mainly adopt the dependence on $p$.  

As we can see from Equation~\eqref{eq: OHT_def}, a direct calculation of OHT relies on the vertical structure of both temperature $T$ (from which $\theta$ is derived) and velocity $\vel$. While temperature measurements have been collected throughout the global ocean from ships and moorings as well as autonomous platforms, the direct observation of subsurface ocean velocity is much more challenging.  As a result, direct velocity measurements remain incredibly sparse in the global ocean below the surface \citep{scott_total_2010}. One common approach to address this issue exploits the fact that on large spatio-temporal scales, ocean flows are generally constrained to follow lines of constant pressure at any particular depth (i.e., the velocity is \emph{geostrophic}).  By combining this relationship with the assumption of hydrostatic balance, which is valid for large-scale geophysical flows, the vertical structure in the velocity field can be inferred from horizontal variations in the density field, given a known velocity field at just a single pressure level \citep{talley_descriptive_2011,gray_global_2014}. Because seawater density $\rho$ depends only on $T$, $p$, and salinity $S$, all relatively easy to measure beneath the ocean surface, this transformation provides a crucial way to estimate velocity, and correspondingly OHT, based on observations throughout the global ocean.

Mathematically, at any space-time point $(\xvec,t)$, the geostrophic velocity at pressure $p$ can be expressed as the sum of $\vel_{\mathrm{ref}}$, the velocity at a fixed reference pressure $p_0$, and $\vel_{\mathrm{rel}}$, the difference between the geostrophic velocities at $p$ and $p_0$ as follows:
\begin{align} \label{eq: vrel_def}
    \vel (p) = \vel_{\mathrm{ref}} (p_0) + \vel_{\mathrm{rel}} (p)  
        = \vel_{\mathrm{ref}} (p_0) + \frac{1}{f} R \cdot \nabla_{\xvec} \Psi (p),
\end{align}
where $R = [0, -1; 1, 0]$, $f = 2\Omega \sin(y)$ is the Coriolis parameter which depends on the rotational rate of the Earth ($\Omega= 7.2921 \times 10^{-5}$ rad $s^{-1}$) and latitude $y$, and the horizontal derivative operator $\nabla_{\xvec} (\cdot) = \left[ \frac{\partial}{\partial x}, \frac{\partial}{\partial y} \right]^\top$.  The dynamic height anomaly $\Psi$ at any one space-time location is computed from the vertical integral of the inverse of density $\rho$,
\begin{align} \label{eq: dynamicHeight}
    \Psi(p) := \Psi(S, T, p) = - \int_{p_0}^{p} \left( \frac{1}{\rho(S(p^*), T(p^*), p^*)} - \frac{1}{\rho(S_O, 0^\circ {\rm C}, p^*)} \right) \d p^*,
\end{align}
where $S_O = 35.16504$ g kg$^{-1}$, and the limits of the integration are the reference pressure $p_0$ and the level of interest $p$.

Bringing together Equations~\eqref{eq: OHT_def} -- \eqref{eq: dynamicHeight}, concurrent measurements of $T(p)$ and $S(p)$, together with an estimate of $\vel_{\mathrm{ref}}$ at $p_0$, can be used to compute an observation-based estimate of OHT. While historically such observations have been sparse and unevenly sampled in space and time, over the past two decades the international oceanographic community has built a global array of autonomous instruments that provides exactly these measurements with unprecedented spatio-temporal coverage. The Argo array \citep{roemmich_design_1998,riser_fifteen_2016} consists of nearly 4000 autonomous profiling floats that collect subsurface measurements of $T$, $S$, and $p$ in the upper 2000\,m of the ocean globally, with near-uniform sampling every $3^\circ \times 3^\circ \times 10$ days in space and time. The number of floats has continuously increased since initial deployments began in the early 2000s, reaching the designed spatial coverage in 2007. The strength of Argo comes from its high sampling density and global, nearly uniform spatio-temporal coverage, along with its high-precision in-situ measurements \citep{riser_fifteen_2016}. Each float follows a pre-determined cycle in which it starts by descending to a parking depth of 1000 dbar, then drifts for 9 days with the predominant currents at that depth, and subsequently sinks to a profiling depth of 2000 dbar before slowly ascending to the surface while measuring ocean variables with vertical resolution of up to 2 dbar for modern floats \citep{roemmich_design_1998}. The set of measurements during the ascent, along with the spatial location and time stamp for each cycle (determined from satellite positioning systems while at the surface), is called a \emph{profile}. These data are transmitted to shore-based computing systems via satellite communications and made freely available to the public in near real time.

The dataset used in this study is based on Argo $T(p)$ and $S(p)$ profiles from throughout the global ocean, from January 2007 to December 2018, obtained from a January 2019 snapshot of the Argo Global Data Assembly Center \citep[GDAC,][]{argo_argo_2020}; see Figure~\ref{fig: location_temp}. Quality control criteria detailed in Section~\ref{sec: QC} of the Supplementary Material \citep{park_supplement_2020} along with those of \citet{kuusela_supplementary_2018} are applied to filter out problematic profiles.  At each profile location, $\Psi$ is computed from the measured $T$ and $S$ profiles according to Equation~\eqref{eq: dynamicHeight} with the TEOS-10 software library \citep{mcdougall_getting_2011}, which is also used to calculate $\theta$ from the measured variables.  The reference pressure used to compute $\Psi$ was set to 900\,dbar to align with our choice of $\vel_{\mathrm{ref}}$ (described below).  The final dataset $\Data$ consists of 1,140,693 $\theta$-$\Psi$ profiles that passed the quality control, with pressure levels ranging from 10\,dbar to 900\,dbar. This corresponds to $89.6\%$ of all available profiles which passed the initial quality control of \citet{kuusela_supplementary_2018}. We avoided estimating any variables at pressure levels too close to the surface, i.e., less than 10\,dbar, since only $67\%$ of profiles could be retained in that case, due to a lack of very shallow observations in many profiles.

\begin{figure}[!htb]
    \begin{subfigure}{.49\textwidth}
        \centering
        \includegraphics[trim={240 280 150 200},clip,width=0.9\textwidth]{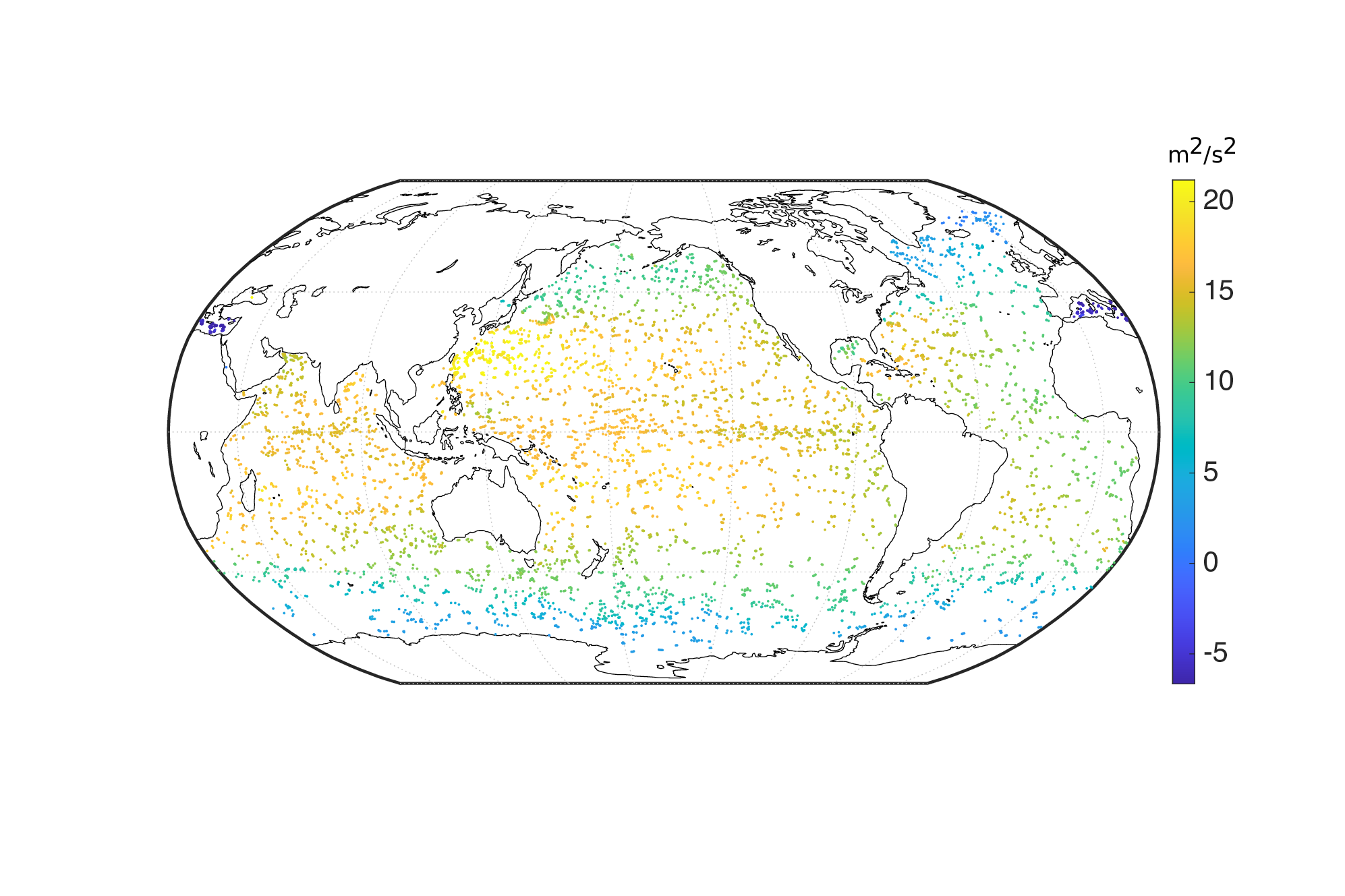}
        \caption{Near-surface $\Psi$}\label{fig: location_temp}
    \end{subfigure}
    \begin{subfigure}{.49\textwidth}
        \centering
        \includegraphics[width=0.48\textwidth]{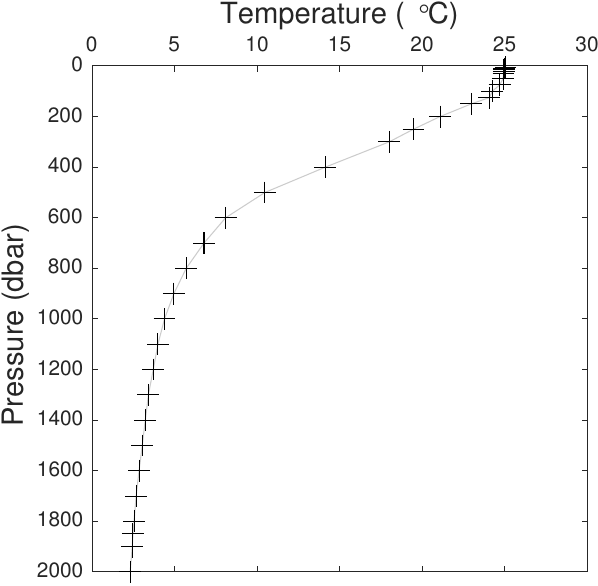}
        \includegraphics[trim={0 0 25.5 0},clip,width=0.48\textwidth]{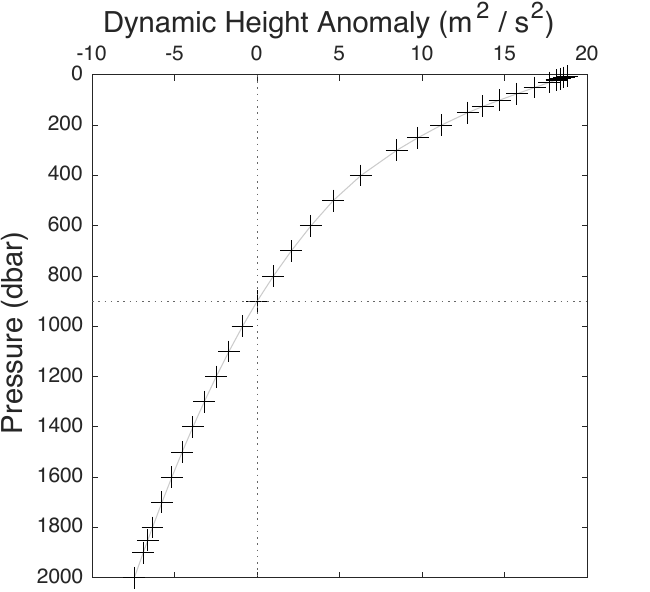}
        \caption{$\theta$-$\Psi$ profile }\label{fig: profile} 
    \end{subfigure}        
    \caption{Visual illustration of the Argo dataset. (a) Locations of profiles collected in February 2017, colored by their dynamic height anomalies $\Psi$ at 10 dbar referenced to $p_0 = 900$ dbar. (b) A $\theta$-$\Psi$ profile for float ID 5900208 observed at $19.4^\circ$S, $154^\circ$E, and 4 am UTC on Sep. 26, 2017. The dotted line on the $\Psi$ profile indicates the reference pressure $p_0 = 900$ dbar.}
\end{figure}

While the Argo dataset can be used to determine $\vel_{\mathrm{rel}}$ according to Equation~\eqref{eq: vrel_def}, a complete estimate of the absolute velocity $\vel$, and consequently OHT, also requires an estimate of the reference velocity $\vel_{\mathrm{ref}}$. \add{However, estimating $\vel_{\mathrm{ref}}$ requires a separate treatment since $\theta$-$\Psi$ profiles does not contain direct information on $\vel_{\mathrm{ref}}$.} In this study, we assume that the reference velocity is given as there are existing well-studied products for the absolute geostrophic velocity at the sea surface or at the Argo floats' parking depth  \citep[see, e.g.,][]{lebedev_yomaha07_2007,willis_combining_2008,ollitrault_andro_2013,gray_global_2014}. For the empirical analyses in Section~\ref{sec:Results}, we adopt the reference geostrophic velocity \add{estimates and mapping error estimates} derived from Argo float trajectories at $p_0 = 900$ dbar \citep{gray_global_2014} at all profile spatio-temporal coordinates based on their nearest-neighbor grid point in the data product. These estimates are solely based on direct observations of the Argo float trajectories, which aligns well with our goal to quantify the geostrophic velocity and OHT based on autonomous in-situ observations. We note that the quality of the reference velocity estimate directly impacts the accuracy \add{and uncertainty} of the resulting estimate of absolute velocity and hence heat transport; improving the reference velocity field is, however, beyond the scope of the present work.

%% file: sec3_1_method_overview.tex
\subsection{Overview}\label{sec: Methods_Overveiw}

We first overview each component of the statistical methodology and explain how they bind together in a unified framework.
The main procedural challenge can be understood as a combination of two classical statistical problems: spatio-temporal interpolation and latent variable modeling. Given $\Psi$ profiles at some spatio-temporal coordinates, the velocity $\vel$ can be understood as a spatio-temporally dependent latent function in which the dependency structure is heterogeneous across the ocean and the time span. The OHT field, the final quantity of interest, presents similar spatio-temporal challenges as well. Neglecting these unique characteristics of the spatio-temporal (latent) variables could result in suboptimal OHT predictions.

To overcome these challenges, a two-stage procedure based on local Gaussian process regression (LGPR) is introduced. LGPR applied particularly to the Argo dataset \citep{kuusela_locally_2018} has shown outstanding interpolation performance compared to that of previous state-of-the-art methods. We extend the work of \citet{kuusela_locally_2018} by considering \emph{latent LGPR}, which is specifically tailored to solving the statistical complications in estimating the OHT field. 
Based on the scientific framework in the previous section, the first stage of procedure estimates the dynamic height anomaly $\Psi$ field at a series of fixed pressure levels, of which the spatial gradients provide the latent relative velocity $\vel_{\mathrm{rel}}$ field according to Equation~\eqref{eq: vrel_def}.  Next, the results of this step are combined with an independent estimate of $\vel_{\mathrm{ref}}$ to compute spot OHT values at the space-time locations of the Argo profiles using Equation~\eqref{eq: OHT_def}.  This integral can be calculated across any range of pressure levels, providing the capability to examine the contribution of different water layers to the total OHT. Conditional on the predicted spot OHT, these estimated OHT values are then interpolated to a regular spatio-temporal grid in the second stage of the LGPR procedure.
We detail the LGPR framework in Section~\ref{sec: Methods_Model} and the latent LGPR with the two-stage procedure in Section~\ref{sec: Methods_Two-step}.

We further improve the LGPR approach of \citet{kuusela_locally_2018}, which focuses on estimating a local space-time covariance model from detrended temperature observation whose mean field was estimated using OLS, by simultaneously estimating both the mean and the covariance parameters with an \emph{approximate EM} algorithm. The procedure shares similarities with GLS. However, our EM procedure is able to account for the overlapping local moving windows of the LGPR covariance structure in a computationally efficient fashion when estimating the mean field. We detail the procedure in Section~\ref{sec: Methods_AM}.

Predicting the gradient field from incomplete observations with a potentially under-specified mean field model may result in a concerning bias. By formalizing a procedure previously used by \citet{roemmich_20042008_2009}, we provide in Section~\ref{sec: Debias} an intuitive \emph{debiasing} procedure that effectively mitigates the bias in the predicted gradient and, if needed, the target field. The procedure captures the asymptotically valid bias field by correcting which improves the calibration of both gradient and target field.

%% file: sec3_1_LGPR.tex
\subsection{Spatio-temporal LGPR model}\label{sec: Methods_Model}

We briefly review the LGPR model originally proposed for Argo mapping in \citet{kuusela_locally_2018} motivated by \citet{haas_kriging_1990,haas_local_1995}, and illustrate the similarities and differences when adopting LGPR specifically for OHT interpolation.
Consider a real-valued spatio-temporal random field of a quantity of interest $\{\Upsilon (\xvec,t,p)\}_{\xvec\in \Xspace, t \in \Tspace}$ observed at a spatial location $\xvec = (x,y)$ in the open ocean $\Xspace \subseteq \Real^2$ with longitude $x$ and latitude $y$ in degrees; time $t \in \Tspace \subseteq [0, 365]$ in yeardays; and at some fixed pressure $p$. Hereafter, we will use $\svec = (\xvec, t)$ to denote a spatio-temporal coordinate. \add{The response field $\Upsilon$ can be either the dynamic height anomaly $\Psi$ or the Ocean Heat Transport $\OHT$, depending on the context, with the same model structure.} We express the field as:
\begin{align}    \label{eq: PsiModel}
    \Upsilon (\xvec,t,p) &= m(\xvec,t,p) + a(\xvec,t,p) + \epsilon (\xvec,t,p),
\end{align}
where $m(\xvec,t,p)$ denotes a large-scale climatological \emph{mean field} with a seasonal cycle; $a (\xvec,t,p)$ denotes an \emph{anomaly field}, i.e., a transient deviation from the climatological mean, and $\epsilon$ is a fine-scale \emph{nugget} effect. The term \emph{mean}, denoted by $m$, is adopted to specify $\E [\Upsilon (\xvec, t,p)]$, the deterministic mean of the process $\Upsilon$, whereas the term \emph{anomaly}, and the notation $a$, refers to a residual process centered at zero. We drop $p$ hereafter for brevity whenever the argument does not depend on the choice of $p$.

In this paper, we consider a \emph{locally semiparametric} model in the sense that the mean field is assumed to be locally parametric whereas the anomaly field is locally nonparametric---specifically, a locally stationary Gaussian process. Nevertheless, both the mean and the anomaly field are actually nonparametric models since the semiparametric distinction happens only at local neighborhoods. Local polynomial regression \citep{fan_local_1997}, which we employ for the mean field, is already in itself a nonparametric method. The locally semiparametric model not only improves estimation efficiency by confining the parameter space but also matches our intent that the mean field explains the systematic large-scale patterns whereas the anomaly field captures the transient patterns.

The nugget effect $\epsilon$ is assumed to locally be a Gaussian white noise process with mean zero and variance $\sigma^2_\epsilon$ and independent of the anomaly field $a$. This distributional assumption leads to a closed-form predictive distribution, enabling convenient uncertainty quantification. Even though the Gaussian nugget is widely adopted in the literature, \citet{kuusela_locally_2018} pointed out that the Gaussian nugget may be insufficient to account for the heavy-tailed nugget distribution of subsurface temperature data in certain parts of the ocean. An extension to a heavy-tailed Student nugget \citep{kuusela_locally_2018} is possible. However, we only focus on the Gaussian nugget in this paper for simplicity.

We let the pilot model of the large-scale mean field $m(\xvec,t)$ to be a \emph{local polynomial regression} \citep{fan_local_1997} with uniform weights \citep{stone_optimal_1980}. In particular, within a small circular spatial window $\Window_{\lambda_G}(\xvec^*) = \{\xvec : \| \xvec - \xvec^* \|_G \le \lambda_G \}$, where $\|\cdot\|_G$ denotes the distance in WGS84 coordinates and $\lambda_G$ is a positive bandwidth that controls the size of the spatial neighborhoods in estimating the coefficients, we let
\begin{align}\label{eq: mean_lq}
\begin{split}
  m(\xvec,t) = \beta_0 &+ \beta_x x_c + \beta_y y_c + \beta_{xy} x_c y_c + \beta_{x^2} x_c^2 + \beta_{y^2} y_c^2 \\ &+ \sum_{k=1}^{K} \left[ \beta_{c_k} \cos \left( \frac{2 \pi k t}{365} \right) + \beta_{s_k} \sin \left( \frac{2 \pi k t}{365} \right) \right], 
\end{split}
\end{align}
where $x_c := x - x^*$ and $y_c := y - y^*$ are spatial coordinates centered around $x^*$ and $y^*$, and $K$ is a predefined maximum number of harmonics. The first line in Equation~\eqref{eq: mean_lq} captures the local spatial structure of the mean field, while the second line models the seasonal cycle within the window.
This regression model with $K=6$ has been successfully adopted in the oceanographic literature to model the mean field of Argo observations \citep{ridgway_ocean_2002,roemmich_20042008_2009}, albeit with slight different estimation method.

 The anomaly field is modeled using a zero-mean locally stationary Gaussian process which is \add{i.i.d.} over the years and whose distance metric is defined as the Mahalanobis distance both in terms of space and time \citep{kuusela_locally_2018}. Let $\svec^* = (\xvec^*,t^*)$ be a space-time (intra-annual) grid point for which a prediction is desired. Within a small spatio-temporal window $\widetilde{\Window}_{\bm{\lambda}}(\svec^*) = \Window_{\lambda_G} (\xvec^*) \times [t^* - \lambda_t, t^* + \lambda_t]$ around $\svec^*$, we let
\begin{align} \label{eq:spatioTemporalModel}
  a_i 
    \overset{\mathrm{i.i.d.}}{\sim} \GP (0, k(\svec_1, \svec_2; \bm{\xi})),
    \qquad i=1,\ldots,I,
\end{align}
where the index $i$ refers to years, $k(\svec_1, \svec_2; \bm{\xi}) = k(x_1 - x_2, y_1 - y_2, t_1 - t_2;\bm{\xi})$ is a stationary space-time covariance function depending on non-negative \add{hyper}parameters $\bm{\xi} = (\phi, \xi_x, \xi_y, \xi_t)^\top$ detailed below and $\bm{\lambda} = (\lambda_G, \lambda_t)$ are positive bandwidth parameters with an additional parameter $\lambda_t$ to control the size of the temporal neighborhood.
\begin{figure}[!htb]
      \begin{subfigure}{.49\textwidth}
        \centering
        \includegraphics[trim={0 0 0 0},clip,width=\textwidth]{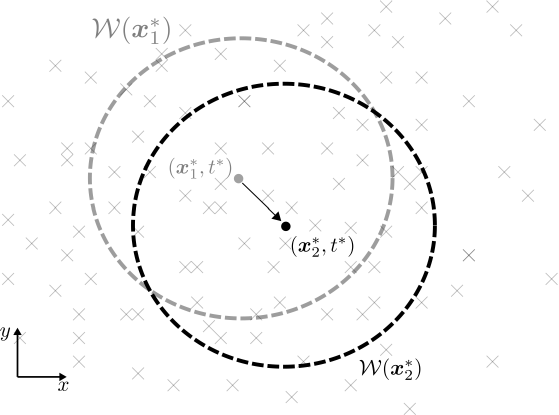}
        \caption{Spatial window $\Window$}\label{fig: window_x}
    \end{subfigure}
    \centering
    \begin{subfigure}{.49\textwidth}
        \centering
        \includegraphics[trim={0 0 0 0},clip,width=\textwidth]{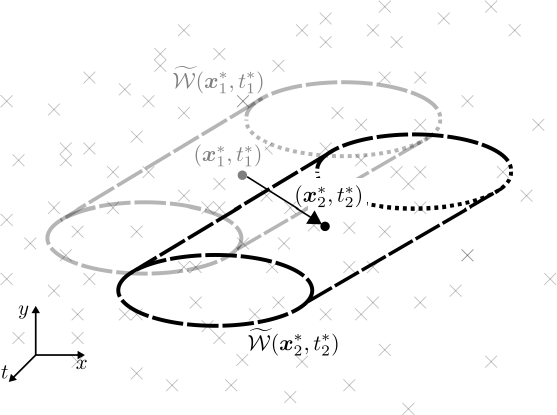}
        \caption{Spatio-temporal window $\widetilde{\Window}$}\label{fig: window_xt}
    \end{subfigure}
    \caption{Local windows adopted in the spatio-temporal LGPR model.}\label{fig: Windows}
\end{figure}

Figure~\ref{fig: Windows} illustrates the circular local windows $\Window$ and cylindrical windows $\widetilde{\Window}$ adopted for the mean field and the anomaly field, respectively. In $\Real^2 \times [0,365]$ the spatio-temporal window $\widetilde{\Window}_{\bm{\lambda}}(\svec^*)$ is a subset of $\Window_{\bm{\lambda}} (\xvec^*)$ for any spatial location $\xvec^*$ so the anomaly field flexibly captures the local interannual temporal signal on top of the parametric climatological seasonal cycle in Equation~\eqref{eq: mean_lq}. This nested construction is the reason we use the iterative EM instead of GLS for jointly estimating the relevant parameters, as we will see in Section~\ref{sec: Methods_AM}. We also note that the circular window used here is more natural than the square window used in \citet{kuusela_locally_2018}.

\add{Bandwidth choice entails a trade-off in both computational and statistical aspects. The larger the size of the window, the larger the computational cost. Given that the computational complexity grows in a cubic order to the number of observations within $\widehat{\Window}$, the bandwidths affect $O(\lambda_G^{6} \lambda_t^{3})$ computational cost. While access to high performance computing makes estimation and prediction feasible for larger bandwidth parameters, larger window sizes do not equate to better prediction due to a bias-variance trade-off. Considering that ocean dynamics are globally non-stationary, excessively large windows are more likely to violate the assumption that the Gaussian process is stationary within the window, resulting in a concerning bias. On the contrary, too small window size suffers from a higher estimation variance or even fail to make a prediction, e.g., near the coastal boundary, due to scarce data within the window. Therefore, it is recommended to choose window sizes with which the computation and the locally stationary assumption are both feasible without losing essential boundary dynamics.}

Care has to be taken in specifying the local windows $\Window_{\bm{\lambda}}$ and $\widetilde{\Window}_{\bm{\lambda}}$ for the $\OHT$ field ($\Upsilon = \OHT$) near the equator since geostrophic balance, and thus Equation~\eqref{eq: vrel_def}, does not hold as the Coriolis parameter $f$ approaches zero. We threshold the windows to ameliorate this issue by masking out the tropical latitude band $[-\zeta, \zeta]$ for some positive parameter $\zeta$. 
More refined methods might be possible, such as using a $\beta$-plane approximation \citep{lagerloef_tropical_1999};
these are, however, beyond the scope of the present study. 

Unlike \citet{kuusela_locally_2018}, in which an exponential covariance function was used, we choose the Mat{\'e}rn covariance function \citep{stein_interpolation_1999} to ensure that the process is differentiable which is required for estimating the velocities. Since a Gaussian process with Mat{\'e}rn covariance with smoothness parameter $\nu$ is $\lceil \nu \rceil - 1$ times differentiable, we set $\nu$ to be $3/2$ to ensure first-order differentiability. Specifically,
\begin{align}\label{eq: Matern_def}
    k \left( \svec_1, \svec_2; \bm{\xi} \right) &=  \phi \left( 1 + \sqrt{3} \pnorm{\bm{A}^{-1}}{\Delta \svec } \right) \exp \left( - \sqrt{3}\pnorm{\bm{A}^{-1}}{\Delta \svec } \right),
\end{align}
where $\phi$ is the GP variance, $\pnorm{\bm{A}^{-1}}{\Delta \svec } = \sqrt{\Delta \svec^\top \bm{A}^{-1} \Delta \svec}$ is the Mahalanobis norm with $\Delta \svec = \svec_1 - \svec_2$ and $\bm{A} = \bm{A}(\bm{\xi})$ is a positive definite matrix parameterized by $\bm{\xi}$. Non-diagonal elements of $\bm{A}$ represent rotation of the spatio-temporal space although at the expense of three additional parameters. Given that we estimate the Gaussian process locally, the number of parameters increases in the order of the number of local windows. A diagonal covariance parameter matrix $\bm{A} = \diag{\xi_x^2, \xi_y^2, \xi_t^2}$ is therefore chosen to efficaciously restrict the parameter space since we did not see empirical improvements in our application from adding extra off-diagonal parameters, agreeing with \citet{kuusela_locally_2018}.

A convenient feature of a Gaussian process is that the first-order derivative is still a Gaussian process \citep{banerjee_directional_2003,rasmussen_gaussian_2006}. That is, the joint process $[a_i, \nabla_{\xvec} a_i]$ is a multivariate Gaussian process: For any $\svec_1, \svec_2 \in \widetilde{\Window}_{\bm{\lambda}} (\svec^*)$,
\begin{align}\label{eq: joint_GP}
  \begin{bmatrix}
    a_i \\ \nabla_{\xvec} a_i
  \end{bmatrix}
  \overset{\mathrm{i.i.d.}}{\sim} \GP \left( \bm{0}, \begin{bmatrix}
    k (\svec_1, \svec_2) & \nabla_{\xvec_2} k (\svec_1, \svec_2)^\top \\
    \nabla_{\xvec_1} k (\svec_1, \svec_2) & \nabla_{\xvec_1} \nabla_{\xvec_2} k(\svec_1, \svec_2) \\
  \end{bmatrix} \right),
\end{align}
where the analytic forms for the gradient and the Hessian of the covariance function are provided in Section~\ref{sec: Matern_derivative_detail} of Supplementary Material \citep{park_supplement_2020}. This feature leads to an important consequence: by Equation~\eqref{eq: vrel_def}, the geostrophic velocity field $\vel$ retains the same LGPR structure~\eqref{eq: PsiModel}, where the mean field is given by the local spatial coefficients $\beta_x$ and $\beta_y$ of the local polynomial model~\eqref{eq: mean_lq}, and the anomaly field is a locally stationary Gaussian process given in~\eqref{eq: joint_GP}. \add{Thus, the predictive distribution of $\vel (\svec^*)$, where the year of $t^*$ is $i$, is
\begin{align*}
  \vel (\svec^*) | \Data 
    \sim \Normal \bigg(
     & \mu_{\vel_{\rm ref}} (\svec^*) + f^{-1} R \left[ \nabla_{\xvec} m_\Psi (\svec^*; \bbbeta_\Psi) + \nabla_{\xvec} k_{\Psi,i}^\top (\svec^*) \bm{K}_{\Psi,i}^{-1} (\Psi (\svec_{i\cdot}) - m_\Psi (\svec_{i\cdot})) \right],\\
     &\sigma^2_{\vel_{\rm ref}} (\svec^*) + f^{-2} R \left[ \nabla_{\xvec}\nabla_{\xvec} k (\svec^*, \svec^*) - \nabla_{\xvec} k_{\Psi,i}^\top (\svec^*) \bm{K}_{\Psi,i}^{-1} \nabla_{\xvec} k_{\Psi,i} (\svec^*) \right] R^\top
      \bigg),
\end{align*}
where $\mu_{\vel_{\rm ref}}$ and $\sigma^2_{\vel_{\rm ref}}$ are the reference velocity estimate and its mapping uncertainty, $\svec_{i\cdot}$ are the spatio-temporal coordinates of $\svec_\Data$ within $\widetilde{\Window} (\svec^*)$ for the $i$-th year, $\svec_\Data$ is a collection of all observed spatio-temporal coordinates in $\Data$, $k_{\Psi,i} (\svec^*) = [k (\svec, \svec^*; \bbxi_\Psi)]_{\svec \in \svec_{i\cdot}}$, $\bm{K}_{\Psi, i} = [k(\svec_j, \svec_k; \bbxi_\Psi)]_{\svec_j, \svec_k \in \svec_{i\cdot}} + \sigma^2_{\Psi,\epsilon} I_{\svec_{i\cdot}}$ is the associated $|\svec_{i\cdot}| \times |\svec_{i\cdot}|$ Gram matrix of $k$ plus the nugget variance. This notation will be repeatedly used hereafter for both $\Psi$ and $\OHT$ depending on the context.
}

%% file: sec3_1_LLGPR_twostage.tex
\subsection{Latent LGPR and two-stage estimation procedure}\label{sec: Methods_Two-step}

\add{Our overarching inferential goal is to compute the predictive mean $\E [\OHT \;|\; \Data]$ for point prediction, and ultimately the predictive distribution of $\OHT \;|\; \Data$.}
The main complication in estimating the $\OHT$ field using definition~\eqref{eq: OHT_def} is that $\vel$ is a latent spatio-temporal field whose realizations are not observable by the floats. Only potential temperature $\theta$ and dynamic height anomaly $\Psi$ profiles are observed, as in Section~\ref{sec:Background_Data}. In this section, we link the final field of interest $\OHT$ with the $\Psi$ field via two-stage estimation approach.

Within a small spatio-temporal window $\widetilde{\Window}_{\bm{\lambda}}(\svec^*)$ around $\svec^*$, OHT can be expressed as a \emph{latent LGPR} model as follows.
 \begin{align}\label{eq: LLGPR}
 \begin{split}
  \Psi_i (p) 
    &\overset{\mathrm{i.i.d.}}{\sim} \GP \left( m_{\Psi}, k(\svec_1, \svec_2 ; \bbxi_{\Psi}) \right), \qquad  p= p_1, \dots, p_P,\\
  \OHT_i \;|\; \bm{\vel}_{1}, \dots, \bm{\vel}_{I}
    &\overset{\mathrm{i.i.d.}}{\sim} \GP \left( m_{\OHT}, k(\svec_1, \svec_2 ; \bbxi_{\OHT}) \right)   
 \end{split}
\end{align}
where $m_{\Upsilon}$ is a mean field~\eqref{eq: mean_lq} of $\Upsilon=\Psi$ or $\Upsilon=\OHT$ within the spatial window $\Window_{\lambda_G}(\xvec^*)$, and $\bm{\vel}_i$ is a $\svec_{i\cdot} \times P$-dimensional matrix with respect to $P$ pressure levels.

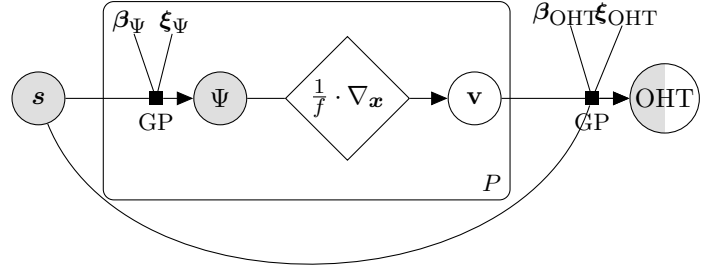
\begin{figure}[!htb]
\centering
\begin{tikzpicture}

  \node[obs]          (s)   {$\svec$}; %
  \node[obs, right=1.7 of s] (psi)   {$\Psi$}; %

  \node[const, above=0.5 of psi, xshift=-1.2cm] (mpsi) {$\bm{\beta}_{\Psi}$} ; %
  \node[const, above=0.5 of psi, xshift=-0.6cm]  (vpsi) {$\bbxi_{\Psi}$} ; %
  \factor[left=of psi] {s-f} {below:$\GP$} {mpsi, vpsi} {} ; %

  \node[det, right=0.5 of psi]   (dot)   {$\frac{1}{f} \cdot \nabla_{\xvec}$} ; %
  \node[latent, right=0.5 of dot]  (v)   {$\vel$}; %

  \node[latent, right=1.7 of v, path picture={\fill[gray!25] (path picture bounding box.south) rectangle (path picture bounding box.north west);}]          (oht)   {$\OHT$}; %

  \node[const, above=0.5 of oht, xshift=-1.3cm] (moht) {$\bm{\beta}_{\OHT}$} ; %
  \node[const, above=0.5 of oht, xshift=-0.5cm]  (voht) {$\bbxi_{\OHT}$} ; %
  \factor[left=of oht] {v-f} {below:$\GP$} {moht, voht} {} ; %

  \factoredge {s} {s-f} {psi} ; %
  \factoredge {v} {v-f} {oht} ; %
  \factoredge[bend right=70] {s} {v-f} {oht} ; %

  \edge[-] {psi} {dot} ;
  \edge[->] {dot} {v} ;

  \plate {} {%
    (s-f)(s-f-caption) %
    (mpsi)(vpsi)
    (psi)(dot)(v) %
  } {$P$} ;

\end{tikzpicture}

  \caption{Graphical representation of latent LGPR for OHT quantification. A grey node indicates an observed variable, while a white node indicates a latent variable. Each variable represents a $|\svec_\Data|$-dimensional vector. A diamond-shaped node denotes a deterministic operation.}
  \label{fig: LLGPR_graph}
\end{figure}

Figure~\ref{fig: LLGPR_graph} illustrates the latent structure of \eqref{eq: LLGPR}. Notice that the OHT variable, $\OHT \propto \int \theta(p) \cdot \vel(p) \d p$, is only half colored since the temperature $\theta$ is observed, whereas the velocity $\vel$ is not. The spatio-temporal dependence of both $\Psi$ and $\OHT$ is encoded with edges stemming from $\svec$. By Equation~\eqref{eq: joint_GP}, $\vel$ retains the spatio-temporal LGPR structure, and we can leverage the prediction of $\vel$ under the LGPR model to obtain an estimate of the unobserved velocity. This model effectively incorporates the key spatio-temporal properties, i.e., the complex spatio-temporal dependence structure and the global non-stationarity, of both the latent field $\vel$ and the final quantity of interest $\OHT$ in a data-driven manner.

\add{Consider the simplest case where $P = 1$, i.e., the heat transport at a fixed pressure $p^*$ which is $\OHT(\svec) = C_p / g (\xvec, p^*) \cdot \theta (\svec, p^*) \cdot \vel (\svec, p^*)$. Since $\OHT (\svec) \;|\; \Data \propto (\OHT (\svec) \;|\; \vel (\svec_{i\cdot}), \Data) \cdot (\vel (\svec_{i\cdot}) \;|\; \Data)$ and the model~\eqref{eq: LLGPR} implies that the predictive distributions of both $\OHT (\svec) \;|\; \vel (\svec_{i\cdot})$ and $\vel (\svec_{i\cdot})$ are Gaussian, the predictive distribution of $\OHT$ is therefore also Gaussian with a closed-form mean and variance.} 
 Details are provided in Section~\ref{sec: predictive_simple} of the Supplementary Material \citep{park_supplement_2020}.

For multiple pressure levels, the predictive mean $\E [\OHT \;|\; \Data]$, used as a point predictor, can be similarly derived using the law of iterated expectations. For any $\svec^{\star}$ within $\widetilde{\Window} (\svec^*)$,
\begin{align} \label{eq: predictive_mean}
  \E [\OHT_i (\svec^{\star}) \;|\; \Data] 
    &= \E \left[ \E [\OHT_i (\svec^{\star}) \;|\; \widehat{\vel} (\svec_{i\cdot}, p_1), \dots, \widehat{\vel} (\svec_{i\cdot}, p_P), \Data] \;|\; \Data \right],\\
    &=  m_{\OHT} (\svec^{\star}) +  k_{\OHT} (\svec^{\star}, \svec_{i\cdot}) {\bm{K}}_{\OHT,i}^{-1} \left[\widetilde{\OHT} (\svec_{i\cdot}) - m_{\OHT} (\svec_{i\cdot}) \right], \nonumber\\
  \widetilde{\OHT} (\svec) &=  C_p \int_{p_1}^{p_P} \frac{\theta(\svec, p) \cdot \E [\vel(\svec, p) | \Data]}{g(\xvec, p)} \d p,
   \qquad \forall \svec \in \svec_{\Data}. \label{eq: OHT_tilde}
\end{align}

Procedurally, this can be viewed as a \emph{two-stage} method where we first construct the predicted 
OHT data set $\widetilde{\Data} = \{ (\svec, \widetilde{\OHT} (\svec)) : \svec \in \svec_{\Data}\}$ in the first stage. We then compute the conditional mean $\E (\OHT \;|\; \widetilde{\Data})$ using the generated dataset in the second stage; see Algorithm~\ref{alg: OHT_full}. In practice, $\Psi$ can only be evaluated at a finite set of $P$ pressure levels, which leads us to approximate the vertical integral when computing $\widetilde{\OHT}$. 
We employ piecewise cubic Hermite interpolation (PCHIP, \citet{fritsch_monotone_1980}) followed by numerical integration. PCHIP is well-suited for this task since it constructs a piecewise cubic interpolant that respects the monotonicity of the data, thereby avoiding spurious bumps typical of alternative interpolation methods \citep{barker_two_2020}.

\add{
The predictive variance of $\OHT$ can be expressed using the law of total variance. For any $\svec^{\star}$ within $\widetilde{\Window} (\svec^*)$,
\begin{align} \label{eq: predictive_var}
\begin{split}
  \V [\OHT_i (\svec^{\star}) \;|\; \Data]  
    =& \left[ \phi_{\OHT} + \sigma_{\epsilon,\OHT}^2 -  k_{\OHT} (\svec^{\star}, \svec_{i\cdot}) {\bm{K}}_{\OHT,i}^{-1} k_{\OHT} (\svec_{i\cdot}, \svec^{\star}) \right]\\    
    &+ k_{\OHT} (\svec^{\star}, \svec_{i\cdot}) {\bm{K}}_{\OHT,i}^{-1} \V \left[ {\OHT} (\svec_{i\cdot}) \;|\; \Data \right]{\bm{K}}_{\OHT,i}^{-1}   k_{\OHT} (\svec_{i\cdot}, \svec^{\star}) .
\end{split}
\end{align}
This decomposition shows that the predictive variance of $\OHT$ is a combination of (i) variation solely from the second stage (the first line), and (ii) the uncertainty that propagates from the first stage to the second stage (the second line). Even though the point predictor $\E [\OHT \;|\; \Data]$ can be obtained without approximations using Equation~\eqref{eq: predictive_mean}, the predictive variance would require the knowledge of the vertical correlation to compute $\V \left[ {\OHT} (\svec_{i\cdot}) \;|\; \Data \right]$.} This ultimately necessitates an approximation or a conservative upper bound to the predictive variance. Incorporating the vertical correlation in addition to spatio-temporal correlation is still an active area of research \citep[see e.g.][]{yarger_functional-data_2020}. 

%% file: sec3_1_AM.tex
\subsection{Approximate Expectation-Maximization algorithm}\label{sec: Methods_AM}

Given the Argo data\footnote{Even though $\Data$ is originally defined as a collection of the triplets $(\theta, \Psi, \svec)$ as in Algorithm~\ref{alg: OHT_full}, we redefine $\Data$ as duplets with a slight abuse of notation for this section to better focus on the procedure.} $\Data := \big\{ (\Upsilon_{ij}, \svec_{ij}) : i=1,\dots,I; j=1,\dots, n_i \big\}$ (either $\Psi$ or $\OHT | \bm{\Psi}$), we seek to estimate a collection of $\bbbeta$ denoted by $\bbBeta = \{ \bbbeta (\xvec^*) : \xvec^* \in \calX \}$ and a collection of $\bbxi$ denoted by $\bbXi = \{\bbxi (\svec^*) : \svec^* \in \calS \}$, where $\calS = \calX \times \calT$ is a set of target spatio-temporal coordinates $\svec^* = (\xvec^*, t^*)$, since the LGPR model specifies the covariance structure on the spatio-temporal window $\widetilde{\Window} (\svec^*)$ nested within the spatial window $\Window (\xvec^*)$ on which the mean field structure is defined. 
 We wish to find the parameters that maximize the likelihood function $\calL (\bbBeta, \bbXi)$; however, a closed-form solution is not available for our LGPR model. We therefore employ an approximate EM algorithm \citep{dempster_maximum_1977}, resulting in a \emph{block coordinate ascent} algorithm, to jointly estimate all of the parameters.

We update the parameters at iteration $l = 0,1,\dots$ as follows:
\begin{align}
    \bbBeta^{(l+1)} &= {\arg\!\max}_{\bbBeta} \widetilde{\calL} (\bbBeta \;|\; \bbXi^{(l)}) \tag{\text{E-Step}} \label{eq: beta_update} \\
    \bbXi^{(l+1)} &= {\arg\!\max}_{\bbXi} \calL (\bbXi \;|\; \bbBeta^{(l+1)}), \tag{\text{M-Step}} \label{eq: xi_update}
\end{align}
where the initial guess $\bbXi^{(0)}$ corresponds to a set of identity covariance matrices, and therefore assuming that the process is spatio-temporally uncorrelated within each spatial window $\Window$. $\widetilde{\calL}$ is an approximated $\calL$ which we will detail subsequently. At first glance, the above steps look like an alternating maximization (AM) algorithm \citep{csiszar_information_1984}, which indeed can be viewed as a special case of the EM algorithm as first suggested by \citet{neal_view_1998}. See Section~\ref{sec: AMview_EM} of the Supplementary Material \citep{park_supplement_2020} for how they are connected under our setup. This EM algorithm is a generalization of \citet{kuusela_locally_2018} since the MLE of the covariance parameters in \citet{kuusela_locally_2018} corresponds to the EM algorithm with $l = 0$, which ignores the spatio-temporal correlation when estimating the mean field. \add{Empirical improvement over \citet{kuusela_locally_2018} in predictive performance is demonstrated in Section~\ref{sec: EM_iteration} of the Supplementary Material.}

The \ref{eq: xi_update} is essentially obtaining the maximum likelihood estimator (MLE) of $\bbXi$ from the residuals $\widetilde{\Upsilon}^{(l+1)}_{ij} := \Upsilon_{ij} - \widehat{m}^{(l+1)} (\svec_{ij}), \forall i, j$, where the estimated mean field $\widehat{m}^{(l+1)}$ is constructed based on the parameters $\bbbeta^{(l+1)} (\xvec^*)$ updated in the previous \ref{eq: beta_update}. For every $\svec^* \in \calS$, 
\begin{align*}
  \mathcal{L} \left( \bbxi (\svec^*) \;|\; \bbbeta^{(l+1)} (\xvec^*) \right) = \prod_{i=1}^I p \left( \widetilde{\bm{\Upsilon}}^{(l+1)}_i ; \bbxi (\svec^*) \right),
\end{align*}
where $\widetilde{\bm{\Upsilon}}^{(l+1)}_i$ is a vector of $\widetilde{\Upsilon}^{(l+1)}_{ij}$'s within the window $\widetilde{\Window} (\svec^*)$ in a specific year $i$, and $p(\widetilde{\bm{\Upsilon}}^{(l+1)}_i ; \bbxi (\svec^*))$ is the pdf of the multivariate Gaussian distribution with zero mean and covariance matrix $\bm{K}_i (\bbxi (\svec^*)) + \sigma^2_\epsilon (\svec^*) \bm{I}_{n_i}$.
 To solve the \ref{eq: xi_update}, we adopted the BFGS quasi-Newton algorithm \citep{nocedal_updating_1980} in the empirical studies in Sections~\ref{sec:Results} and \ref{sec: Validation}.

\ref{eq: beta_update} updates the deterministic mean field accounting for the spatio-temporal correlation of the residuals $\widetilde{\bm{\Upsilon}}^{(l)}$ learned in the previous \ref{eq: xi_update}. This step is analogous to the GLS estimator in regression kriging literature to resolve the sub-optimality of OLS \citep{cressie_statistics_1993} and shares a similarity with iterative GLS \citep{gray_method_2015} adopted previously for Argo data. However, in our LGPR model, using the GLS estimator is not straightforward due to the nested temporal window $\widetilde{\Window}$ within the spatial window $\Window$, which limits the availability of the correlations at large temporal lags. To aggregate the local spatio-temporal covariance structures of $\widetilde{\Window}$ into the spatial window $\Window$, we employ the Vecchia approximation \citep{vecchia_estimation_1988} which confines the aggregated covariance structure by thresholding the temporal lag outside of each~$\widetilde{\Window}$ in the conditional distribution. We note that this approach is different from block covariance tapering, and the Vecchia approximation is known to have advantages over covariance tapering \citep{stein_statistical_2013}.

The Vecchia approximation \citep{vecchia_estimation_1988} is a natural choice both from the perspective of LGPR modeling and computational efficiency. Choosing uniform weights on each spatio-temporal window $\widetilde{\Window}$, we hard-threshold the conditional spatio-temporal dependency along the temporal axis in the anomaly field within $2 \lambda_t$ temporal lag of the target time point $t^*$. This implies that the LGPR model assumes observations within $\widetilde{\Window}$ to be uncorrelated beyond the temporal window. Such a structure is reflected in the approximate likelihood function via Vecchia approximation. Additionally, this choice yields a closed-form \ref{eq: beta_update} resembling a GLS-like estimator, for which the details are given in Section~\ref{sec: AM_detail} of Supplementary Material.

The overall EM procedure (\ref{eq: beta_update} and \ref{eq: xi_update}) leads to a computationally efficient algorithm since these steps can be viewed as a \emph{gather-and-broadcast} algorithm. The \ref{eq: xi_update} can be performed fully in parallel across each $\widetilde{\Window}$ once the residuals have been broadcast to each computing node. The \ref{eq: beta_update} then gathers the estimated covariance structures from each computing node and updates the aggregated mean parameter within $\Window$. This parallelization leads to major computational benefits since the main computational bottlenecks of the procedure are the numerical optimizations required in the \ref{eq: xi_update} as opposed to the \ref{eq: beta_update} where the closed form solution is fast to compute.

%% file: sec3_1_Detrend_ADA.tex
\subsection{Debiasing the mean field}\label{sec: Debias}

In this section, we describe a simple debiasing procedure to account for potential mean field model misspecification. We have noticed that climate scientists oftentimes compute the empirical mean of the estimated anomaly fields across years and add that back to the mean field to make the resulting estimate of the anomaly fields temporally centered at zero  \citep[e.g.,][]{roemmich_20042008_2009}. We formalize this procedure and demonstrate that it is a legitimate approach to (partially) identifying model misspecifications and correcting them. For the LGPR model~\eqref{eq: PsiModel}, model misspecification may arise in both the mean and the covariance structure. Specifically, we focus on a potentially misspecified mean field, since inferring the climatological mean is of key interest in this application, and a bias arising from mean field misspecification propagates to the localized anomaly fields which leads to biased inference and prediction of the anomalies. The LGPR model~\eqref{eq: PsiModel} utilizes a mean field model $m$ given in Equation~\eqref{eq: mean_lq} which is inspired by previous work in the oceanographic literature \citep{ridgway_ocean_2002,roemmich_20042008_2009}. Even though this model is known to work well for simple oceanographic variables, such as temperature and salinity, the model may have trouble representing the mean of the $\OHT$ or $\Psi$ fields with sharp fronts and other localized patterns, which further motivates us to use a bias-correction procedure in this application.

The predictive mean $\widehat{\Upsilon} (\svec^{**}) := \E (\Upsilon (\svec^{**}) \;|\; \Data)$, for any $\svec^{**} \in \widetilde{\Window} (\svec^*)$, is an unbiased estimator of the true mean field $\E (\Upsilon (\svec^{**})) = m(\svec^{**})$ if the assumed mean field model for $\Upsilon$ is well-specified following the construction~\eqref{eq: PsiModel}. That is, when the year of $\svec^{**}$ is~$i$,
\begin{align}\label{eq: marginal_predictive}
    \E \left[ \widehat{\Upsilon}_i (\svec^{**}) \right]
     = \E \left[ m(\svec^{**}) + k_i^\top (\svec^{**}) \bm{K}_i^{-1} (\Upsilon_i (\svec_{i\cdot}) - m (\svec_{i\cdot}) ) \right]
     \overset{(\star)}{=} m(\svec^{**}).
\end{align}
Suppose the analyst was oblivious to the true mean field $m$ and misspecified the mean field model as $\E_{\mathsf{A}} (\Upsilon (\svec^{**})) := m (\svec^{**}) + B (\svec^{**})$ by introducing a non-zero bias field $B$. Here $\E_{\mathsf{A}}$ denotes the assumed expectation under the analyst's model. The predictive mean $\widehat{\Upsilon} (\svec^{**})$ under the misspecified model becomes
\begin{align*}
    \widehat{\Upsilon}_i (\svec^{**}) = \E_{\mathsf{A}} (\Upsilon (\svec^{**})) + k_i^\top (\svec^{**}) \bm{K}_i^{-1} \left( \Upsilon_i (\svec_{i\cdot}) - \E_{\mathsf{A}} (\Upsilon_i (\svec_{i\cdot})) \right). 
\end{align*}
Then, $(\star)$ in Equation~\eqref{eq: marginal_predictive} no longer holds but instead
\begin{align}\label{eq: kriged_bias_error}
    \E \left[ \widehat{\Upsilon}_i (\svec^{**}) \right]  - m(\svec^{**}) 
    = B (\svec^{**}) - k_i^\top (\svec^{**}) \bm{K}_i^{-1} B (\svec_{i\cdot}).
\end{align}
As the observations $\svec_{i\cdot}$ get denser within $\widetilde{\Window} (\svec^*)$, Equation~\eqref{eq: kriged_bias_error} essentially converges to zero, and thus $\widehat{\Upsilon}_i \rightarrow \Upsilon_i$ for every year $i$ under infill asymptotics, {despite the misspecification of the mean field $m$}. See \citet[Chapter~4, Theorem~8]{stein_interpolation_1999} for a rigorous statement.

Given $I$ years of observations, we estimate $B(\svec^{**})$ using the negative average anomaly
\begin{align}
\widehat{B}(\svec^{**}) 
    &= - \frac{1}{I} \sum_{i=1}^{I}  k_i^\top (\svec^{**}) \bm{K}_i^{-1} \left( \Upsilon_i (\svec_{i\cdot}) -  \E_{\mathsf{A}} (\Upsilon (\svec^{**})) \right)  \label{eq:biasEst}\\
    &= \E_{\mathsf{A}} (\Upsilon (\svec^{**})) - \frac{1}{I} \sum_{i=1}^{I} \widehat{\Upsilon}_i (\svec^{**})
    \xrightarrow[I \to \infty]{\text{infill}}    
     \E_{\mathsf{A}} (\Upsilon (\svec^{**})) - \E (\Upsilon (\svec^{**})) = B(\svec^{**}). \nonumber
\end{align}
This leads to a bias-corrected mean field $\E_{\mathsf{A}}^\text{new}(\Upsilon) = m + B - \widehat{B}$, which asymptotically converges to the true mean field $m$ assuming that the $\Upsilon$ fields are observed densely enough for every year and that we have observations from a large enough number of years.

The mean field misspecification affects the estimation of both the mean and the anomaly fields since we assumed the true mean field is $m + B$ when initially computing $\widehat{m}$, and consequently assumed $\E_{\mathsf{A}} (a_i (\svec_{i\cdot})) = 0$, when in reality $\E (a_i (\svec_{i\cdot})) = - B(\svec_{i\cdot})$, when estimating the covariance parameters $\bbXi$ before correcting the bias. After the bias is identified, we re-estimate the covariance parameters $\bbXi$ based on the corrected residuals $\Upsilon_i (\svec_{i\cdot}) - \widehat{m} (\svec_{i\cdot}) + \widehat{B} (\svec_{i\cdot})$, for all $i=1,\dots,I$, utilizing the \ref{eq: xi_update} of the EM procedure. The re-estimation step ensures that the covariances are computed from the correct model structure~\eqref{eq: PsiModel} under which the anomaly field is truly centered at zero asymptotically. We then recompute the interpolated fields $\widehat{\Upsilon}$ based on the updated covariance parameters.

The proposed debiasing method is directly applicable not only to $\Psi$ or $\OHT | \bm{\Psi}$ in the latent LGPR model~\eqref{eq: LLGPR} but also to the latent velocity field $\vel$ without additional computational burden.
Recall that applying the deterministic operation $\frac{1}{f} \cdot \nabla_{\xvec}$ on the $\Psi$ field yields the $\vel$ field. Since the operation only consists of linear operators, our bias estimate for the $\vel$ field is
\begin{align} \label{eq: bias_derivative}
    \frac{1}{f} \cdot \widehat{\nabla_{\xvec^{**}} B} (\svec^{**}) = -\frac{1}{f} \cdot \frac{1}{I} \sum_{i=1}^I \nabla_{\xvec^{**}} k_i^\top (\svec^{**}; \widehat{\bbxi}_{\Psi}) \bm{K}_i^{-1} (\widehat{\bbxi}_{\Psi}) (\Psi (\svec_{i\cdot}) - \widehat{m}_{\Psi}(\svec_{i\cdot})).
\end{align}
As the analytic form of the gradient of the Mat{\'e}rn covariance function is available (see Section~\ref{sec: Matern_derivative_detail} of the Supplementary Material \citep{park_supplement_2020}), the additional computational burden to calculate the bias of $\vel$ is marginal in the process of computing the bias of $\Psi$.

In Sections~\ref{sec:Results} and \ref{sec: debias_effect}, we empirically show the importance of accounting for the possible bias in both $\Psi$ and $\OHT$. Debiasing the $\Psi$ field especially yields a considerable improvement on the latent $\vel$ field when we do not have direct observations to fit a model for $\vel$.

%% file: sec3_1_algorithm.tex
\subsection{Complete OHT interpolation framework}\label{sec: Methods_algorithm}

Algorithm~\ref{alg: OHT_full} summarizes the full two-stage procedure we have described throughout this section. Procedures \textsc{LGPR} and \textsc{Debias} summarize the proposed approximate EM algorithm and bias-correction as described in Sections~\ref{sec: Methods_AM} and~\ref{sec: Debias}, respectively. The computational complexity of our framework is dominated by the procedure \textsc{LGPR}, and thus is analogous to that of \citet{kuusela_locally_2018}. Given $n_i$ observations for each year $i=1,\dots I$, the computational complexity of global Gaussian process regression is $O (\sum_{i=1}^I n_i^3)$ due to computing the inverse Gram matrices $\bm{K}_i^{-1}$. The computation of \textsc{LGPR} is localized to $|\calS|$ target grid points at which the windows are centered, with each window containing $w \cdot n_i$ observations for each year $i$ ($w$ is the fraction of data contained in the window). With the computations parallelized to $C$ threads, the computational complexity of Algorithm~\ref{alg: OHT_full} is $O (P \cdot |\calS| \cdot C^{-1} \cdot w^3 \cdot \sum_{i=1}^I n_i^3)$.

\begin{algorithm}[!ht]
\linespread{0.90}\selectfont
    \caption{Two-stage OHT interpolation framework} \label{alg: OHT_full}
\begin{flushleft}
     \hspace*{\algorithmicindent} \textbf{Input:} Data $\Data = \{\Data_1, \dots, \Data_P\}$ where $\Data_p = \{(\theta_{ij} (p), \Psi_{ij}(p), \svec_{ij}): i=1,\dots,I; j=1,\dots,n_i\}$ (Denote $\svec_{\Data} := \{\svec_{ij}:\forall i, j\}$); Spatio-temporal target $\svec^*$ \\
\end{flushleft}

\begin{algorithmic}[1]
\Function{LGPR}{$\Upsilon$, $\Data$, $\bm{S}$} \Comment{Target response $\Upsilon$ can be $\Psi$ or $\OHT$}
    \Repeat
        \State Estimate mean field coefficients $\widehat{\bbBeta}_\Upsilon$ in the \eqref{eq: beta_update}
        \State Estimate covariance coefficients $\widehat{\bbXi}_\Upsilon$ in the \eqref{eq: xi_update}
    \Until{Converge}
    \State \Return $\widehat{\Upsilon} (\bm{S})$ field
\EndFunction
\Statex
\Procedure{Debias}{$\Upsilon$, $\Data$, $\widehat{\bbXi}_{\Upsilon}$}
    \State Compute the bias $\widehat{B}$ with \eqref{eq:biasEst}.
    \State Debias $\widehat{\Upsilon}$ and $\widehat{\nabla_{\xvec} \Upsilon}$, respectively, using $\widehat{B}$ and \eqref{eq: bias_derivative}.
    \State Re-estimate $\widehat{\bbXi}_{\Upsilon}$ with \eqref{eq: xi_update} using the bias-corrected residuals.
    \State Update $\widehat{\Upsilon}$ field based on re-estimated $\widehat{\bbXi}_{\Upsilon}$.
\EndProcedure
\Statex
\State Estimate pilot predictions $\widehat{\Psi} (\svec_{\Data}, p)$ $\gets$ \Call{LGPR}{$\Psi$, $\Data_p$, $\svec_{\Data}$} for fine grid of $p$.
\State \Call{Debias}{$[\widehat{\Psi}  - \widehat{m}_{\Psi}] (\svec_{\Data}, p)$, $\Data_p$, $\widehat{\bbXi}_{\Psi}$} for fine grid of $p$.
\State Construct a dataset $\widetilde{\Data}$ whose response variable is predicted OHT, $\widetilde{\OHT} (\svec_{\Data})$, using \eqref{eq: OHT_tilde}.
\State Map $\widehat{\OHT}(\svec^*)$ $\gets$ \Call{LGPR}{$\widetilde{\OHT}$, $\widetilde{\Data}$, $\svec^*$}.
\State \Call{Debias}{$[\widehat{\OHT} - \widehat{m}_{\OHT}] (\svec^*)$, $\widetilde{\Data}$, $\widehat{\bbXi}_{\OHT}$}.
\end{algorithmic}
\end{algorithm}

%% file: sec5_1_estimate.tex
In this section, we present empirical results applying the two-step estimation procedure described in Section~\ref{sec: Methods} to the Argo dataset to produce output fields on a spatio-temporal grid $\calX \times \calT$ where $\calX$ is a $1^\circ \times 1^\circ$ spatial grid and $\calT$ is a regularly spaced monthly temporal grid centered on the 15th day of each month. Each quantity of interest is compared before and after applying the debiasing procedure described in Section~\ref{sec: Debias}. The bandwidth parameter $\lambda_G$ for the spatial window is set to $442$ km (approximately $4^\circ$), and $\lambda_t$ for the temporal window is set to $1.5$ months. \add{All computations in the subsequent sections are performed on Cheyenne, a high performance computing cluster at NCAR with 36 CPU nodes with 109 GB of RAM. It takes on average 25 min each to execute a single EM iteration and to make predictions on $\calS$ for each field.}

In the subsequent Sections~\ref{sec: Vel_result} and \ref{sec: OHT_result}, we present the time-averaged quantities:
\begin{align}\label{eq: time_averaged_OHT}
    \Av{\Upsilon} (\xvec) = \frac{1}{|\calT|} \int_{\calT} \E [\Upsilon (\xvec, \tau)] \d \tau ,
\end{align}
where $\Upsilon$ can be $\vel$ or $\OHT$ depending on the context. Our product actually generates a monthly varying spatial map, however, we present time-averaged quantities which succinctly summarize spatial mean variability without loss of generality.

\subsection{Geostrophic Velocity \texorpdfstring{$\hat{\vel}$}{\textbf{v}}}\label{sec: Vel_result}

\begin{figure}[htb]
    \begin{subfigure}{.49\textwidth}
        \centering
        \includegraphics[trim={20 0 0 38},clip,width=\textwidth]{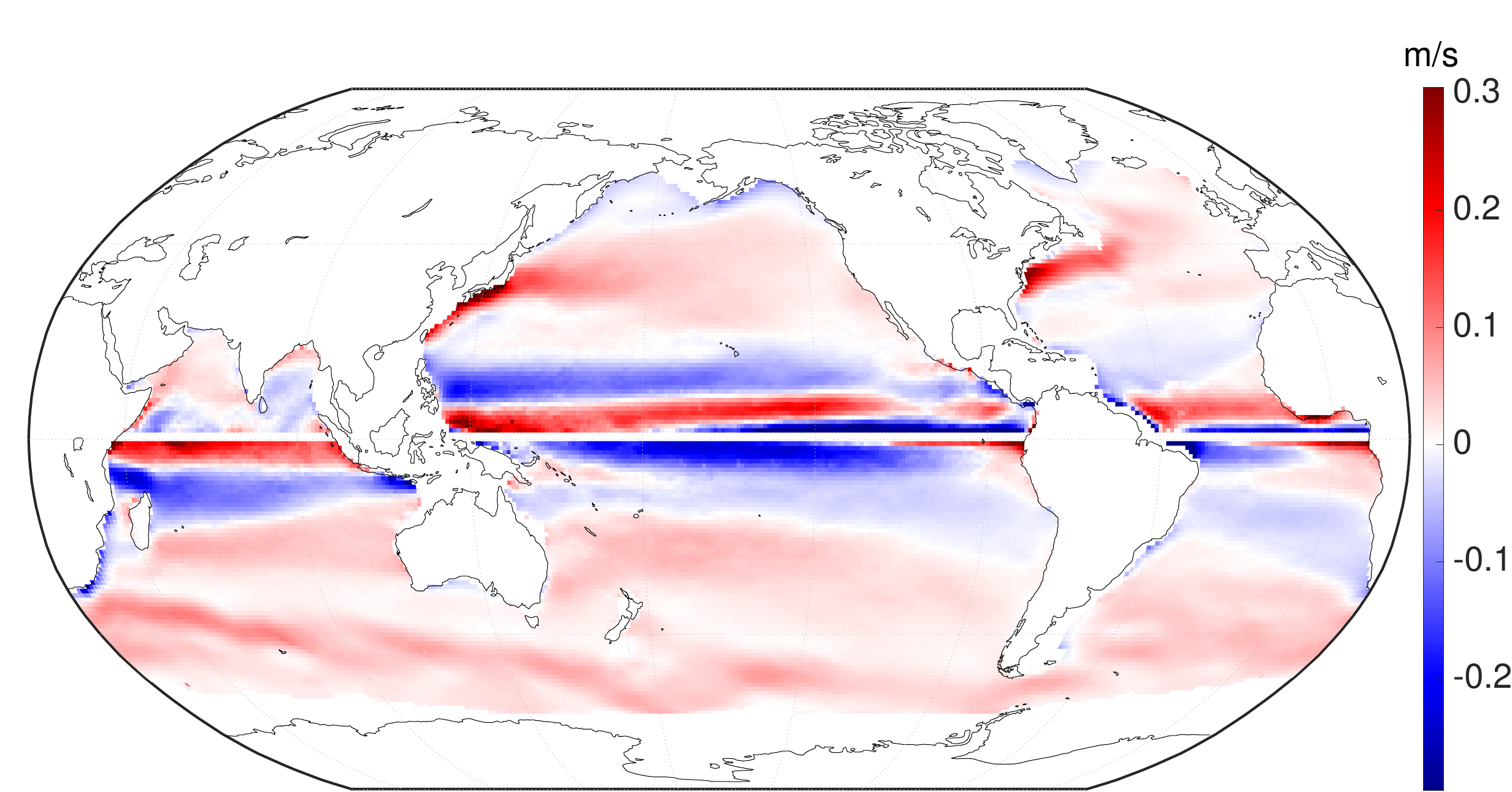}
        \caption{Zonal (initial)}\label{fig: pre_mean_zonal10_Argo}
    \end{subfigure}
    \centering
    \begin{subfigure}{.49\textwidth}
        \centering
        \includegraphics[trim={20 0 0 38},clip,width=\textwidth]{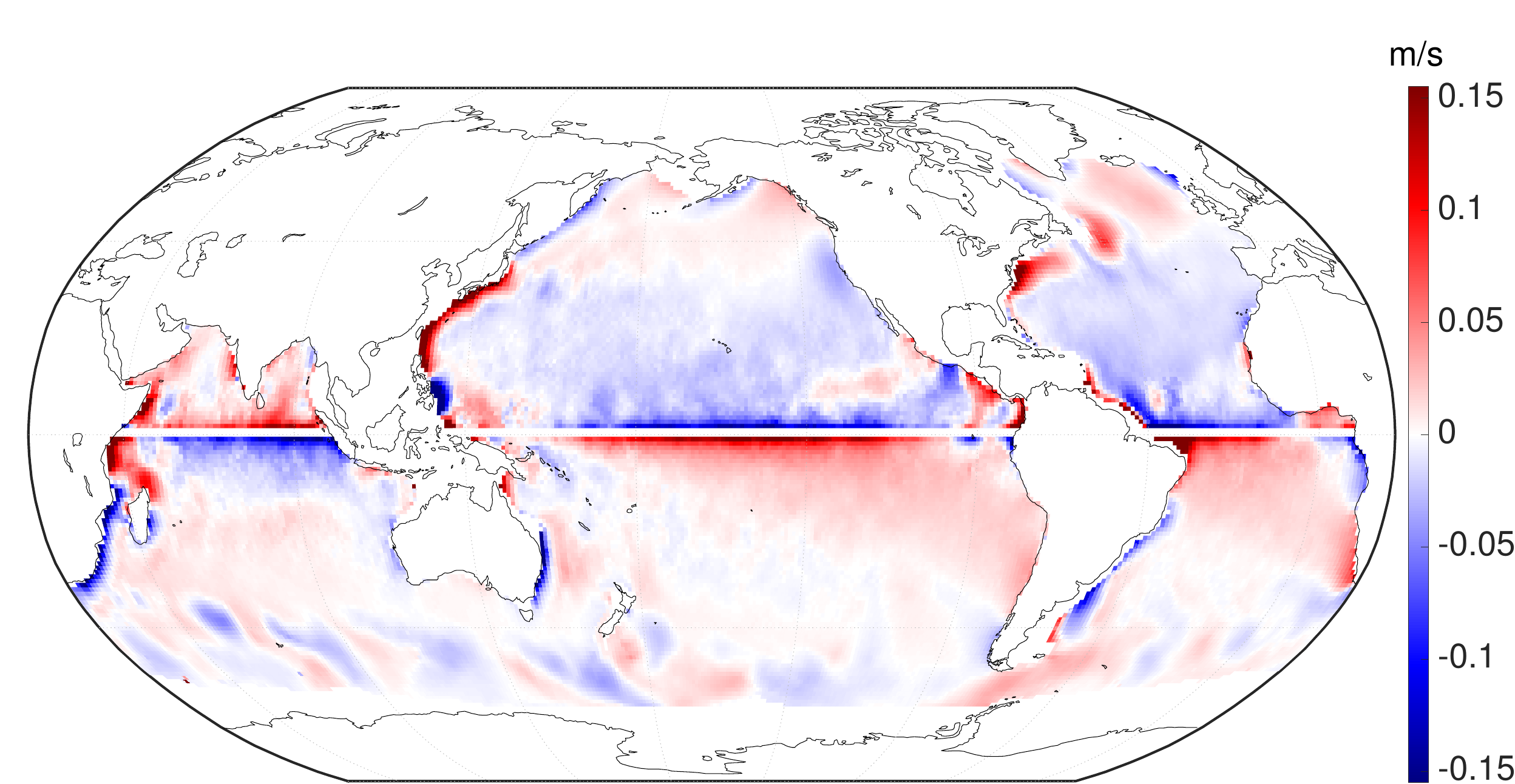}
        \caption{Meridional (initial)}\label{fig: pre_mean_meridional10_Argo}
    \end{subfigure}

    \begin{subfigure}{.49\textwidth}
        \centering
        \includegraphics[trim={20 0 0 38},clip,width=\textwidth]{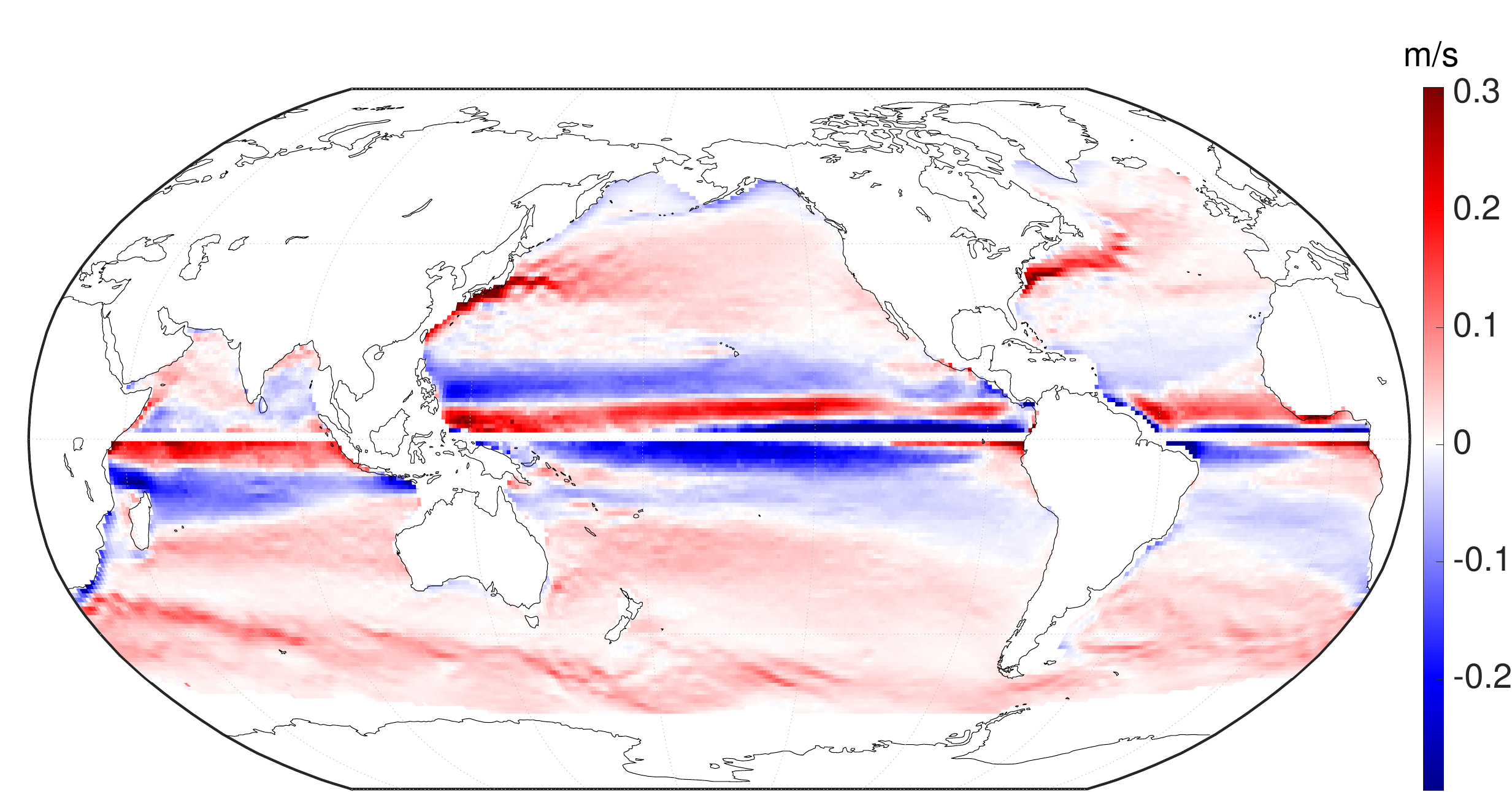}
        \caption{Zonal (debiased)}\label{fig: mean_zonal10_Argo}
    \end{subfigure}
    \centering
    \begin{subfigure}{.49\textwidth}
        \centering
        \includegraphics[trim={20 0 0 38},clip,width=\textwidth]{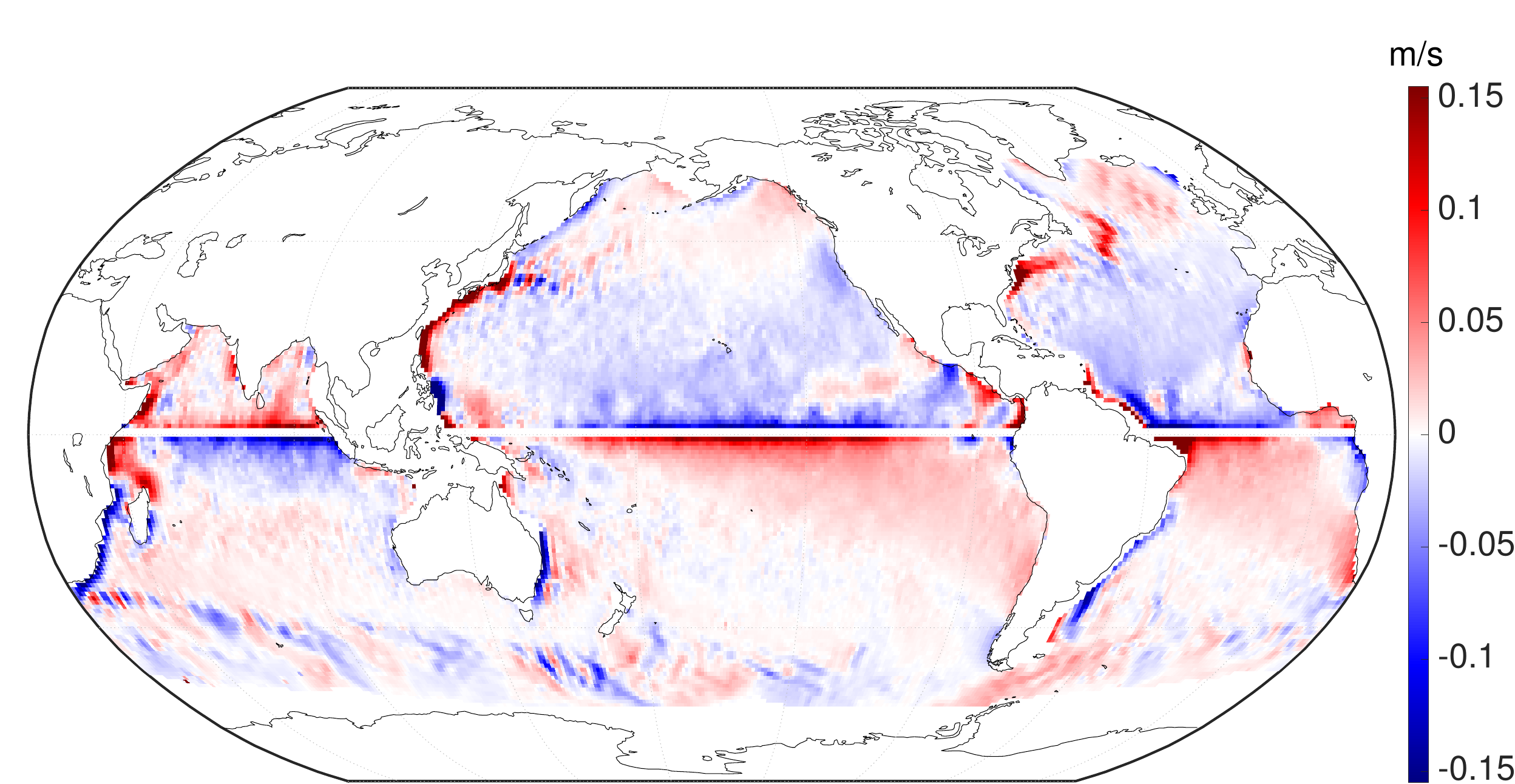}
        \caption{Meridional (debiased)}\label{fig: mean_meridional10_Argo}
    \end{subfigure}
    \caption{Estimated mean geostrophic velocity $\Av{\hat{\vel}_{\mathrm{rel}}}$ at 10 dbar relative to 900 dbar. Red/blue corresponds to the east/west direction in the zonal quantities and north/south direction in the meridional quantities for every map figure here and below. }\label{fig: mean_geovel10_Argo}
\end{figure}

Recall from Section~\ref{sec: Methods_Two-step} that we only need spot-predicted velocities $\vel$ from the first step; however, it is worthwhile to visualize the interpolated latent $\vel$ field to see if the latent field is well-represented. The estimated mean field for the relative geostrophic velocity $\Av{\hat{\vel}_{\mathrm{rel}}}$ from the first step can be found in Figure~\ref{fig: mean_geovel10_Argo}. Figures~\ref{fig: pre_mean_zonal10_Argo} and \ref{fig: pre_mean_meridional10_Argo} show the non-bias-corrected initial zonal and meridional velocity estimates, respectively. Figures~\ref{fig: mean_zonal10_Argo} and \ref{fig: mean_meridional10_Argo} show the estimated mean field after the debiasing procedure. In all figures, we mask out $\pm 2^\circ$ equatorial bands, where geostrophic balance is invalid. The estimates depict the major ocean currents in each basin, including Equatorial Currents, the Antarctic Circumpolar Current, and (at least partially) the western boundary currents and their extensions. The debiasing procedure captures higher-order local features that are not described by the local second degree polynomial, without introducing spurious noise. This is highlighted in the Kuroshio Current (off the coast of Japan) and the Agulhas Return Current (near the southern tip of Africa), where meanders are clearly visible in Panel (d) that are not present in Panel (b) before the bias is corrected. These meanders are known to be quasi-stationary and are also observable in satellite products (see Section \ref{sec: Appendix_more_fig} of Supplementary Material), which indicates that these local features are in fact part of the real signal. 

\begin{figure}[!htb]
    \centering
    \includegraphics[width=0.8\textwidth]{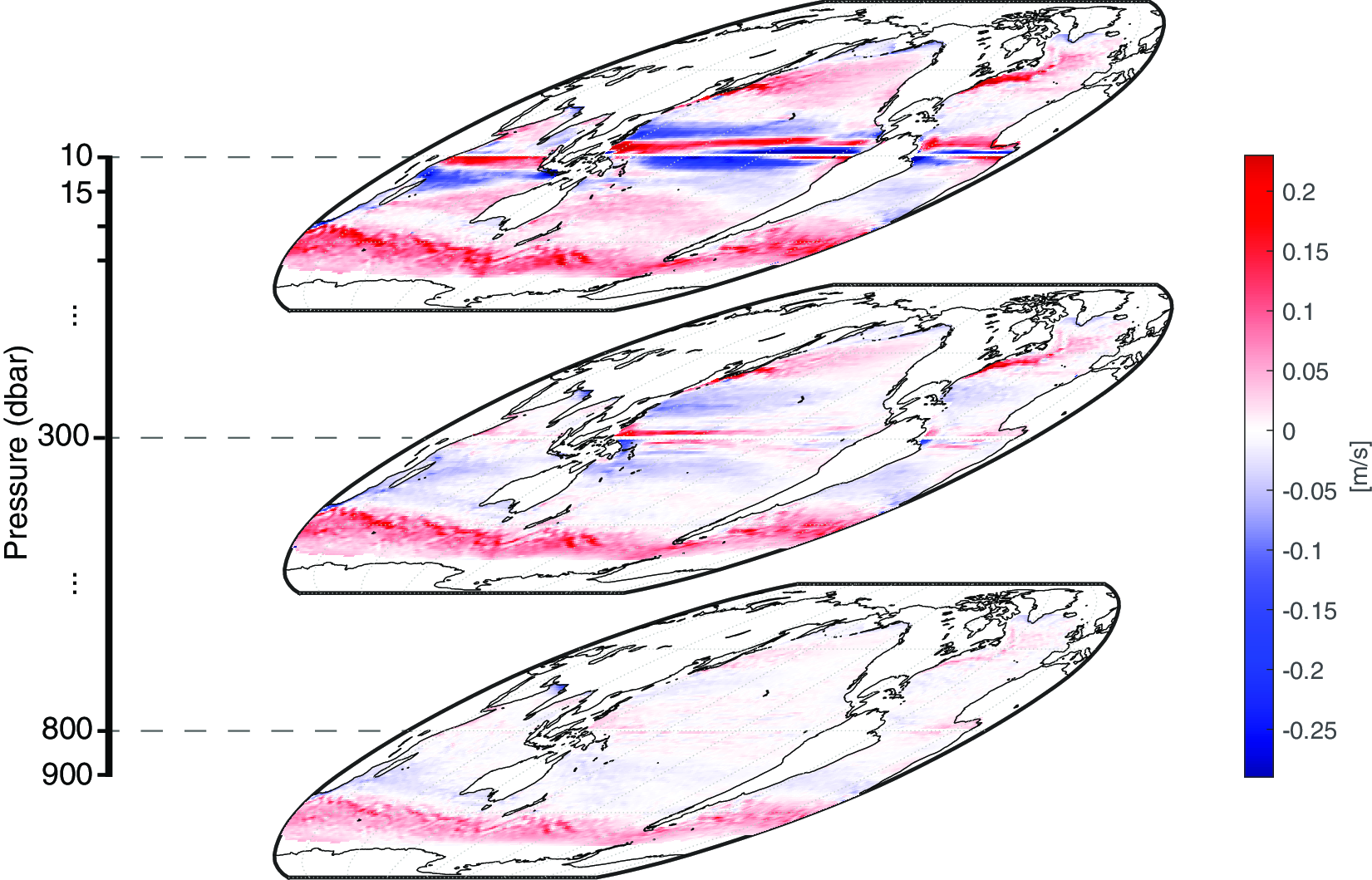}
    \caption{Estimated mean zonal velocities at multiple pressure levels (10, 300 and 800 dbar) 
    .}\label{fig: mean_zonvel10_900_Argo}
\end{figure}

Even though we have only presented the velocity field estimated at $10$ dbar in Figure~\ref{fig: mean_geovel10_Argo}, we emphasize that the relative velocity field is estimated at 17 different pressure levels.  The vertical structure of the resulting velocity estimate is illustrated in Figure~\ref{fig: mean_zonvel10_900_Argo}. Note that the relative velocity field in the continent-free Southern Ocean retains much of its strength even at 800~dbar, as opposed to the other basins, where the relative geostrophic velocities generally decay more quickly with depth.

\subsection{Heat Transport \texorpdfstring{$\widehat{\OHT}$}{OHT}}\label{sec: OHT_result}

\begin{figure}[!htb]
    \begin{subfigure}{.49\textwidth}
        \centering
        \includegraphics[trim={30 0 0 80},clip,width=\textwidth]{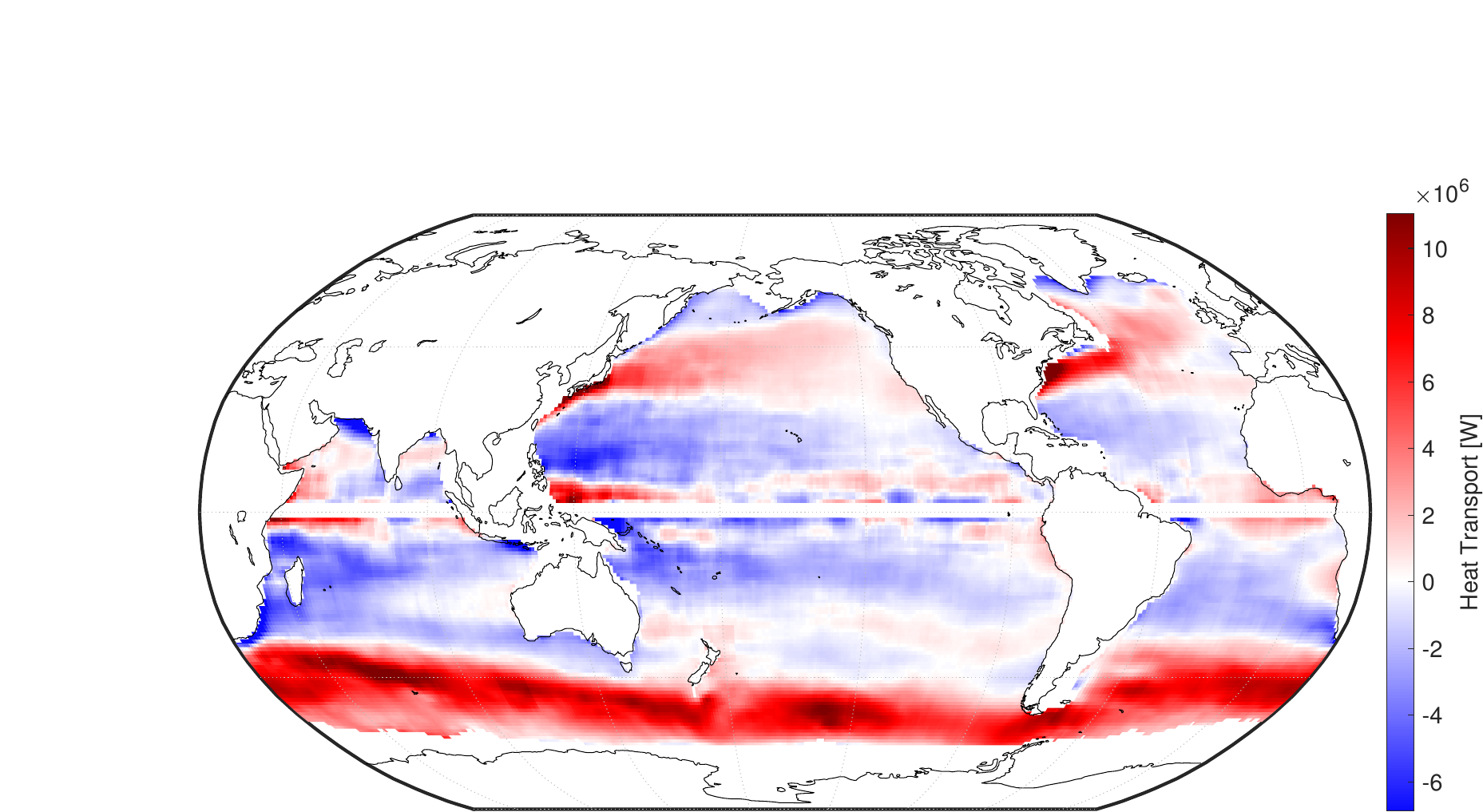}
        \caption{Zonal (initial)}\label{fig: pre_mean_zonht10_900_Argo}
    \end{subfigure}
    \centering
    \begin{subfigure}{.49\textwidth}
        \centering
        \includegraphics[trim={30 0 0 80},clip,width=\textwidth]{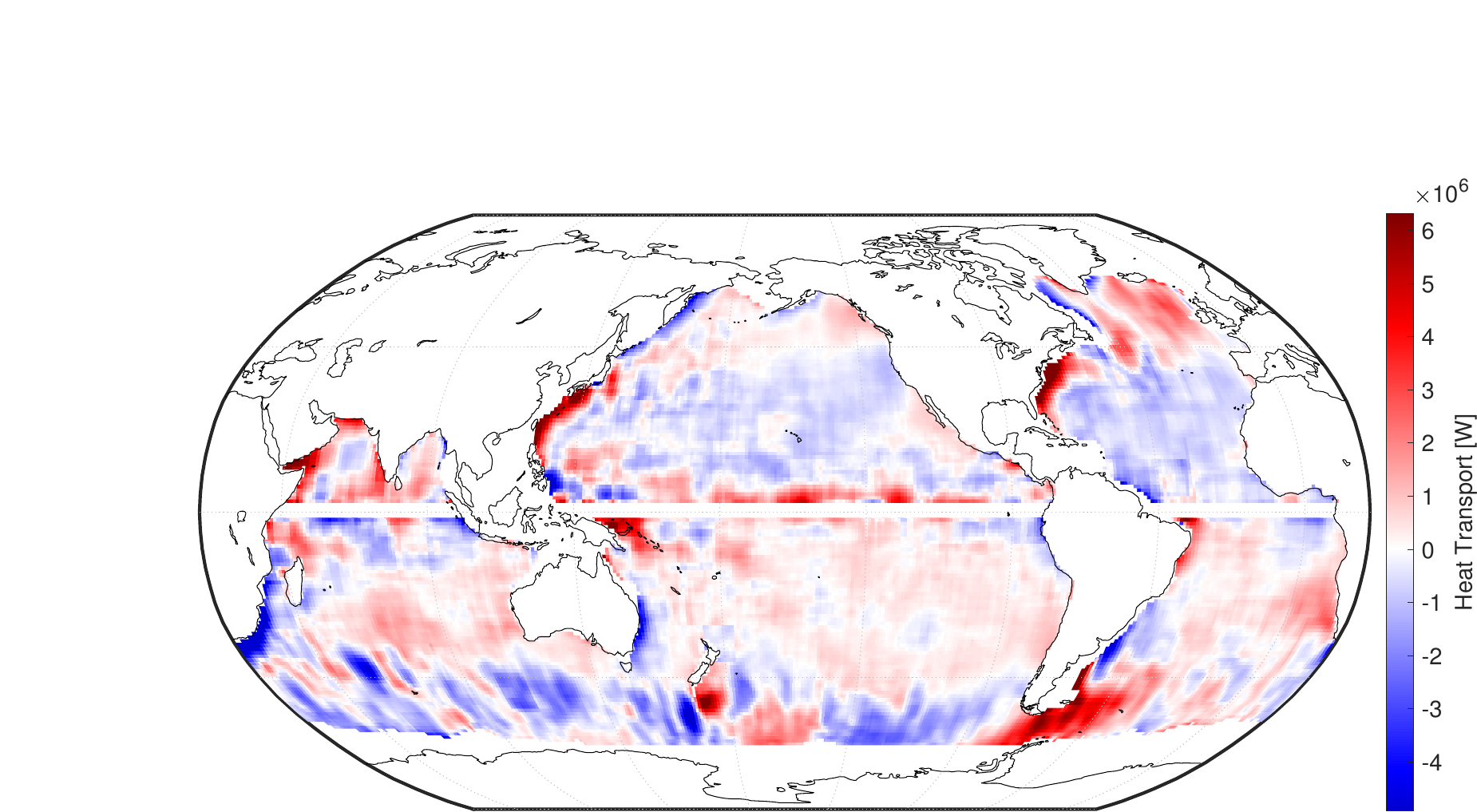}
        \caption{Meridional (initial)}\label{fig: pre_mean_merht10_900_Argo}
    \end{subfigure}

    \begin{subfigure}{.49\textwidth}
        \centering
        \includegraphics[trim={30 0 0 80},clip,width=\textwidth]{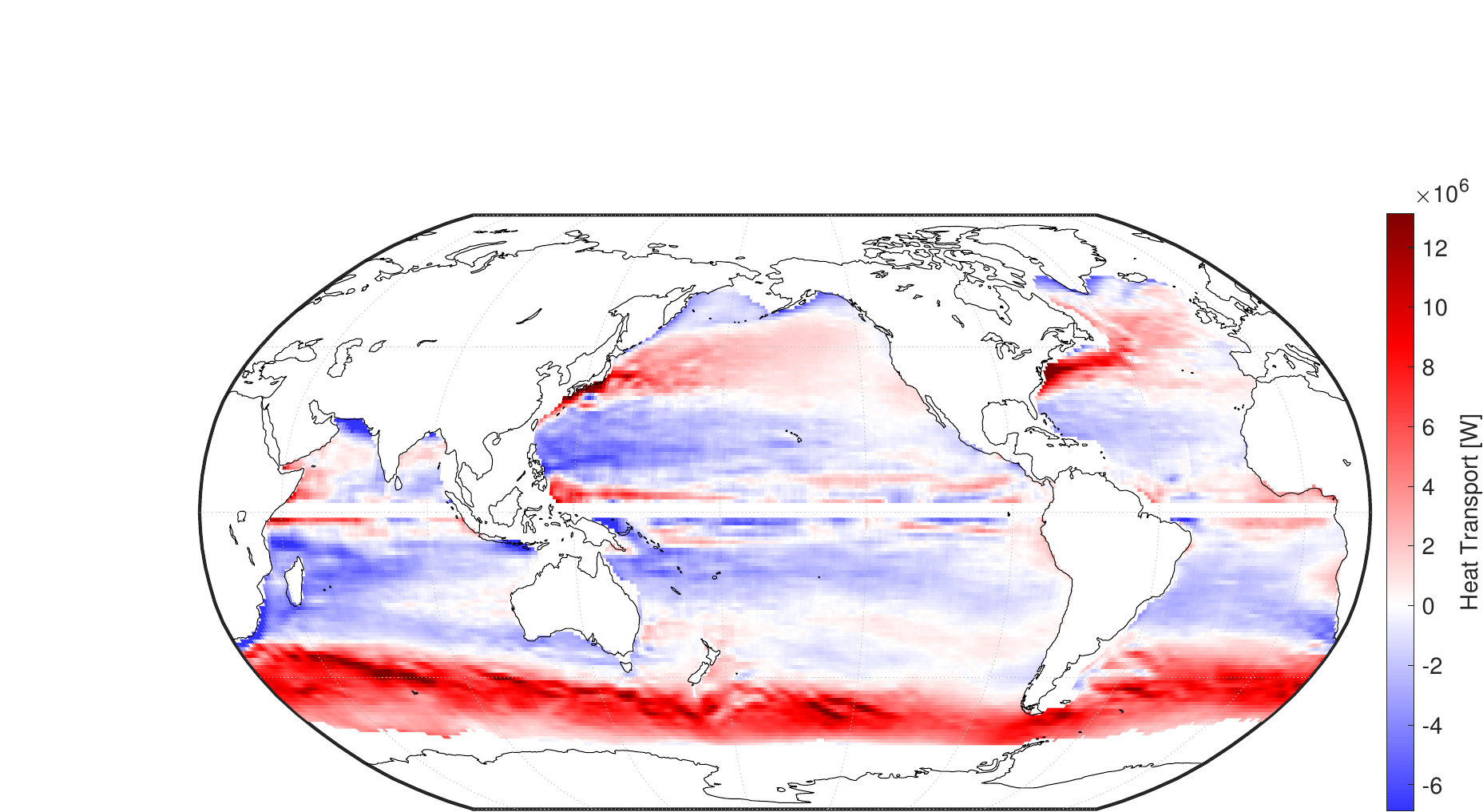}
        \caption{Zonal (debiased)}\label{fig: mean_zonht10_900_Argo}
    \end{subfigure}
    \centering
    \begin{subfigure}{.49\textwidth}
        \centering
        \includegraphics[trim={30 0 0 80},clip,width=\textwidth]{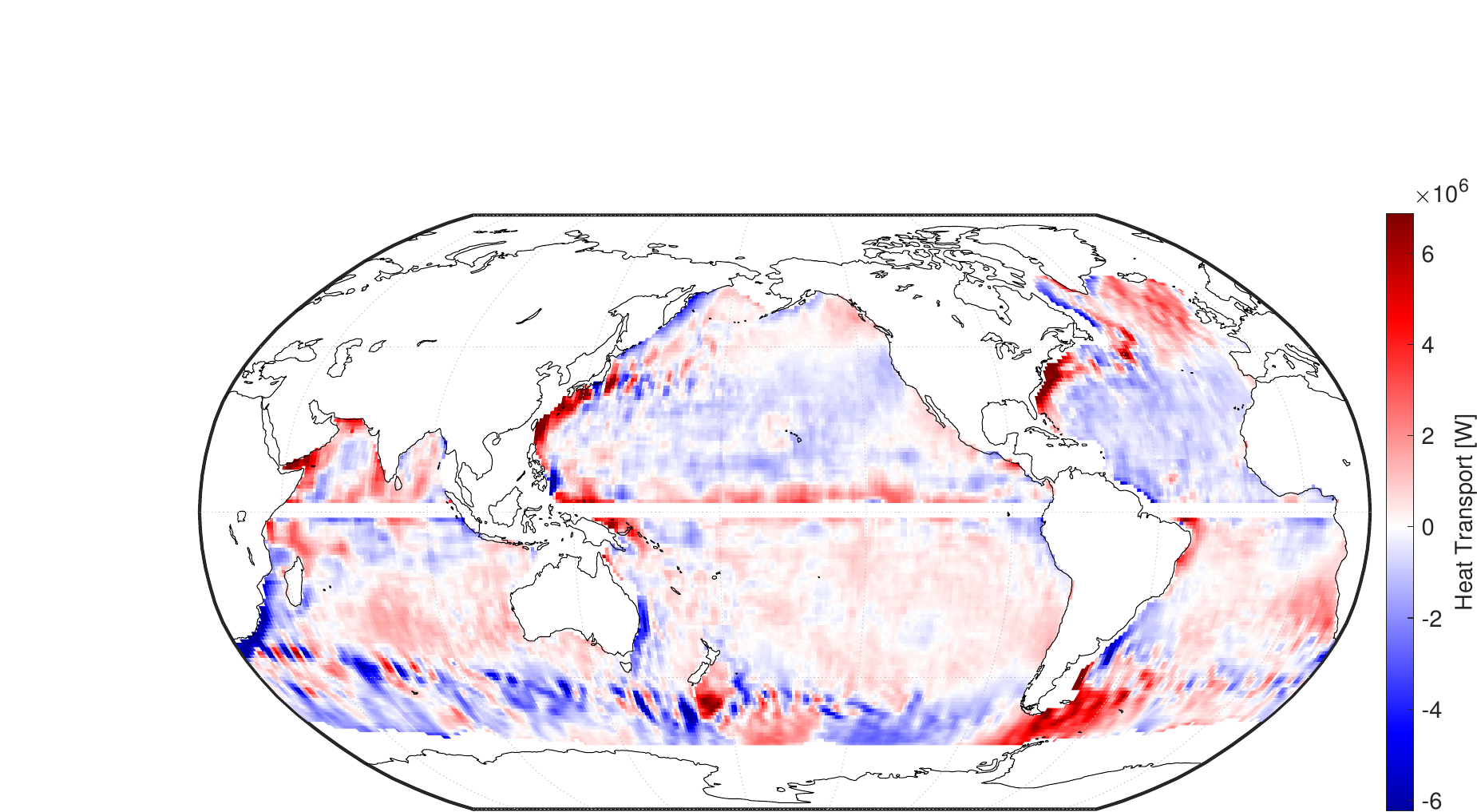}
        \caption{Meridional (debiased)}\label{fig: mean_merht10_900_Argo}
    \end{subfigure}
    \caption{Estimated mean heat transport $\Av{\widehat{\OHT}}$ between $10$ dbar to $900$ dbar.}\label{fig: mean_ht10_900_Argo}
\end{figure}

Figure~\ref{fig: mean_ht10_900_Argo} shows the estimated mean field of zonal and meridional heat transport $\Av{\widehat{\OHT}}$ between $10$ dbar to $900$ dbar computed using the two-step procedure in Section~\ref{sec: Methods_Two-step}. The heat transport fields largely resemble the geostrophic velocity fields, although temperature-driven features are noticeable in Figure~\ref{fig: mean_ht10_900_Argo}. In addition, the vertical structure of the currents has a significant impact on these vertically integrated estimates.  For instance, the Antarctic Circumpolar Current in the Southern Ocean becomes much more evident in the OHT estimate than it is in the velocity estimates, consistent with its role as the strongest, most depth-independent current in the global ocean. On the contrary, heat transport in the tropical Pacific does not stand out in Figure~\ref{fig: mean_ht10_900_Argo}, even as the velocities in this region dominate the near-surface flow shown in Figure~\ref{fig: mean_geovel10_Argo}. The varying contributions to the total OHT from transport in different depth layers can only be seen by resolving the vertical structure of the flow, as done in this work.

\subsection{Heat Transport Anomalies and El Ni{\~n}o-Southern Oscillation}

The previous two sections illustrate the time-averaged mean fields, which by themselves are important for spatially resolving the global heat transport. In this section, we show the utility of quantifying the spatio-temporal OHT anomalies in the context of the El Ni{\~n}o-Southern Oscillation (ENSO), an important recurring phenomenon in the climate system.

\begin{figure}[!htb]
  \centering
    \includegraphics[trim={0 0 0 0},clip, width=0.72\textwidth]{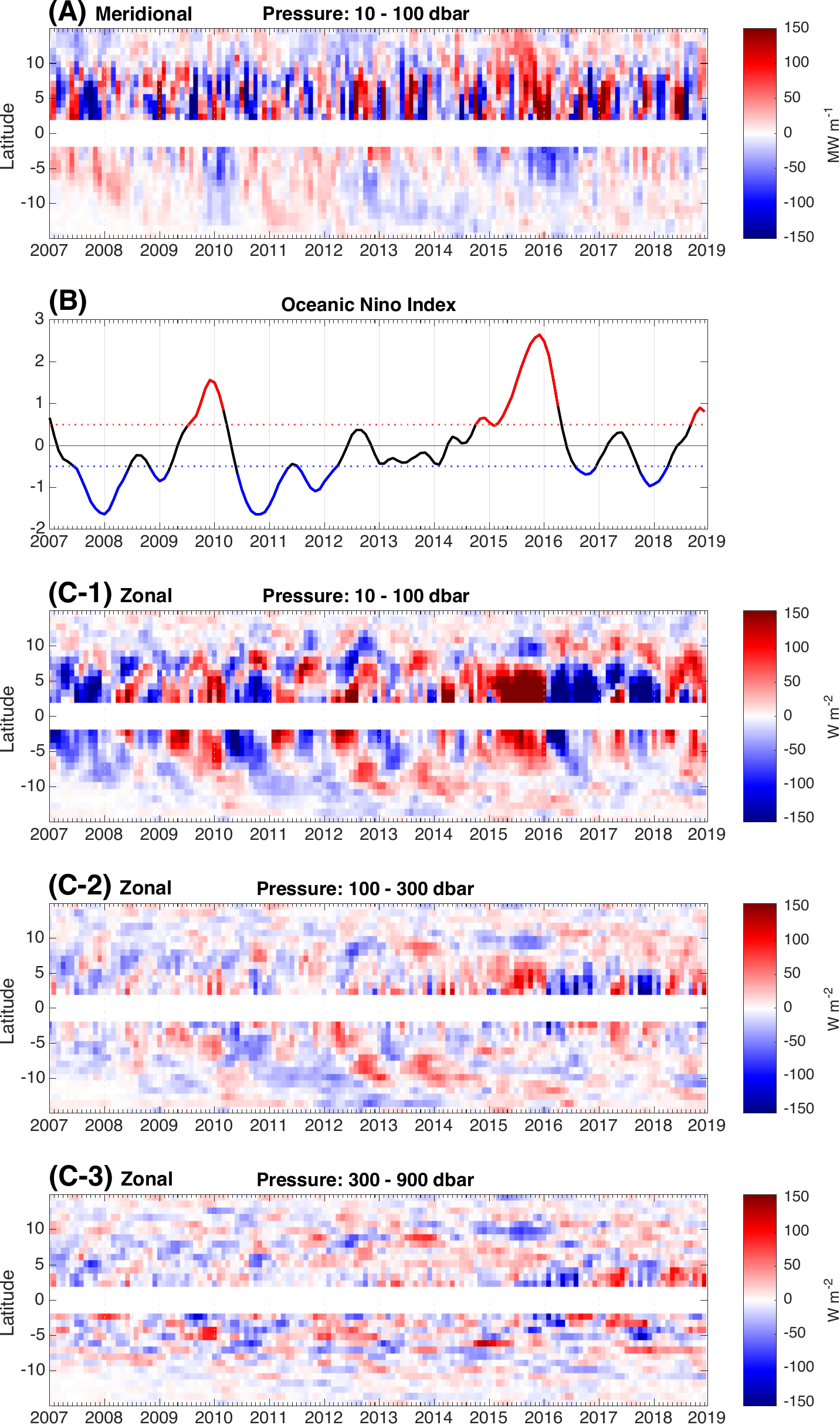}
\caption{Ocean heat transport anomaly over the longitude band of El Ni{\~n}o 3.4 region has a close connection to the Oceanic Ni{\~n}o Index. Panel (A): Total meridional heat transport anomaly at the 10--100 dbar pressure range. Red/blue means north/south direction. Panel (B): Historical ONI. Panels (C): Averaged zonal heat transport anomaly for pressure layers 10--100 dbar, 100--300 dbar, and 300--900 dbar (from top to bottom). Red/blue means east/west direction.}\label{fig: oht_anomaly_ONI}
\end{figure}

ENSO is a natural mode of climate variability that influences Earth's hydrological cycle and global weather patterns through teleconnections. ENSO alternates between a warm phase (El Ni{\~n}o) and a cold phase (La Ni{\~n}a), which are associated with changes in atmospheric circulation and ocean temperature. The state and intensity of ENSO can be described using NOAA's Oceanic Ni{\~n}o Index \citep[ONI,][]{glantz_reviewing_2020}, which is a 3-month running mean of ERSST.v5 Sea Surface Temperature (SST) anomalies in the east-central tropical Pacific between $5^\circ$N--$5^\circ$S, $120^\circ$--$170^\circ$W (El Ni{\~n}o 3.4 region). An event is classified as an El Ni{\~n}o (La Ni{\~n}a) when ONI is above (below) the threshold of $\pm 0.5^\circ$C for a minimum of 5 consecutive overlapping seasons. See Figure~\ref{fig: oht_anomaly_ONI}, Panel (B) for a time series of the ONI.

Figure~\ref{fig: oht_anomaly_ONI} shows Hovm{\"o}ller diagrams of the OHT anomaly across latitude and time. We present the total heat anomaly transported across the longitudes of the Ni{\~n}o 3.4 region for meridional OHT (Panel A), whereas the anomaly averaged over all longitudes of the Ni{\~n}o 3.4 region is presented for zonal OHT (Panels C1--C3). Panel A and Panel C1 provide a kinematic view of anomalous heat transport in the upper ocean between 10--100 dbar. The El Ni{\~n}o phase is associated with anomalous meridional transport of heat away from the Equator (red / blue in Northern / Southern hemisphere in Panel A) and dominant eastward anomalous heat transport (red in Panel C1), and vice versa for the La Ni{\~n}a phase. These observations are consistent with our scientific understanding of ENSO.  During El Ni{\~n}o conditions, the normal upwelling of cold waters along the equator is reduced, yielding upper ocean temperatures that are warmer than average, while increased upwelling in the La Ni{\~n}a phase results in cooler waters in the surface layer of the tropical Pacific \citep{mcphaden_nino_2020}. Furthermore, variations in anomalous upper-ocean currents have been observed during the development of ENSO.  \citet{ren_upper-ocean_2017} found that, at the equator, eastward (positive) zonal current anomalies strengthened in early 2015 before the anomalous currents turned to the west (negative) by 2016, in general agreement with the estimate presented in Panel C1.

From Panels C1--C3, we can observe that the anomalous heat transport associated with ENSO occurs predominantly in the upper layer of the ocean and that the patterns are subdued in the deeper parts of the ocean, which matches with earlier studies of ocean heat content variability \citep[see, e.g.,][]{trenberth_insights_2016}. During the 2015--16 super El Ni{\~n}o episode, the strongest El Ni{\~n}o in history, the anomalous zonal heat transport exhibits a coherent pattern that extends to the deeper 100--300 dbar layer. 
Compared to conventional indices or the rate of change in ocean heat content \citep{trenberth_insights_2016}, Figure~\ref{fig: oht_anomaly_ONI} reveals intriguing, complex spatial variability (the study of which remains outside the scope of this paper). For example, the meridional component of the anomalous heat transport has a much larger inter-hemispheric asymmetry than does the zonal component.

%% file: sec6_validation.tex
\section{Validation with Satellite Observations}\label{sec: Validation}

In this section, we provide empirical validation of our method and the resulting estimates by comparing with estimates based on satellite observations. The ultimate goal of the validation is to show our estimates align well with the existing products widely used by the oceanographic community, and ascertain the strength of the proposed method, i.e, the two-stage procedure together with the debiasing procedure.

Satellite data offer an excellent tool for validating our gridded near-surface OHT estimates, as satellites capture high-resolution snapshots of SST and sea surface height (SSH, which can be used to estimate geostrophic velocity at the surface). Higher resolution is a clear advantage of satellite observations compared to sparse in-situ data collected from research vessels, Argo floats, and moorings; in-situ subsurface measurements are, however, crucial for characterizing OHT over the depth of the water column, as in Definition~\eqref{eq: OHT_def}. In this section, we use the \emph{surface temperature transport}  $\TT(\svec) = \theta(\svec, z_0) \vel(\svec, z_0)$ instead of $\OHT$, with $z_0$ equal to 0 dbar for satellite based products (which are available only at the surface) and 10 dbar for our Argo-based in-situ product (the shallowest depth we considered), ignoring terms that do not impact the comparison. 

\add{While it would be desirable to further validate the subsurface estimates, it is not possible to perform the comparable analysis with satellites since they cannot observe subsurface temperatures and heights (and thus subsurface velocities and OHT). Alternative global subsurface observing systems whose spatial and temporal resolution is comparable to that of Argo are non-existent. The validation study at the surface, meanwhile, serves as a relevant benchmark, since the surface---the farthest from the reference pressure $p_0$---is the hardest depth to estimate using our method. This is because $\Psi(p)$ is computed as a vertical integral from $p_0$ to $p$ as in Equation~\eqref{eq: dynamicHeight}. If our estimate can resolve the true field at the surface, it should conceivably perform at least as well at other pressures between surface and $p_0$.}

We adopted two separate satellite gridded products for SST and SSH distributed by the EU Copernicus Marine Environment Monitoring Service (CMEMS). For SST, the European Space Agency (ESA) SST Climate Change Initiative (CCI) and Copernicus Climate Change Service (C3S) reprocessed Level-4 product \citep{merchant_satellite-based_2019} at daily $0.05^\circ$ degree spatial resolution is considered. For SSH and its derived geostrophic velocity, sea level TAC-DUACS Level-4 Delayed-Time product \citep{taburet_duacs_2019} is adopted. This product has a quarter-degree spatial resolution, along with daily temporal resolution. The DUACS product contains state-of-the-art surface geostrophic velocity estimates mainly based on multimission satellite altimetry over the global ocean, although in-situ Argo profiles and surface drifters are also used in part to estimate the Mean Dynamic Topography \citep{rio_new_2018}. However, the impact of in-situ observations on the DUACS product is negligible in validating the proposed framework and the estimates from Argo data.


\subsection{Comparing the OHT pipelines}

The primary reason we propose a two-stage method is that Argo floats do not directly measure velocity. Such a limitation requires us to first estimate the velocity and then combine the resulting estimates with in-situ temperature observations before interpolating in any space and time coordinates. The performance of the proposed procedure therefore depends on both the velocity estimation error and the OHT mapping error. Using the satellite-based SST and SSH products, we separately analyze the errors associated with each of these components.

The first step is to establish the \emph{ground truth}, defined here as the best possible gridded surface temperature transport field at $1^\circ \times 1^\circ$ spatial resolution. For this purpose, we compute the product of the gridded SST and velocity fields at $0.25^\circ \times 0.25^\circ \times 1$ day resolution and then upscale the result to the target resolution using natural-neighbor interpolation \citep{sibson_brief_1981}. This upscaled \emph{ground truth} is not influenced by any of our proposed interpolation methods.  See Section~\ref{sec: Appendix_more_fig} of the Supplementary Material \citep{park_supplement_2020} for the resulting ground truth time-averaged $\TT$ field.

A key advantage of utilizing high-resolution satellite products is that we can obtain SST $(\theta)$, SSH\footnote{Although SSH and dynamic height anomaly are not the same, in this section we also use $\Psi$ to denote SSH, with abuse of notation, since they fulfill the same purpose here.} $(\Psi)$, and velocity $(\vel)$ in any spatio-temporal location up to the resolution each product can resolve. We generated \emph{pseudo}-observations $\Data_{\rm{Pseudo}} = \{(\theta (\svec_{ij}), \Psi(\svec_{ij}), \vel (\svec_{ij})) : \svec_{ij} \in \svec_{\Data}, \; \forall i,j \}$ at the same spatio-temporal locations as the Argo array $\svec_\Data$ by taking the nearest high-resolution spatio-temporal grid point of SST, SSH, and velocity, respectively. Since the nearest high-resolution grid-point from any observed locations in $\svec_\Data$ is at most $0.177^\circ \times 0.5$ days away, the approximation error is marginal to the sampling resolution of the Argo array. By construction, these pseudo-observations match the sampling resolution; hence, surface temperature transport estimates derived from in-situ Argo profiles and from pseudo-observations are commensurable, allowing us to assess our method in comparison to the \emph{ground truth}.

We consider three candidate methods $\calM_j, j\in\{1, 2, 3\}$ to estimate the surface temperature transport field $\widehat{\TT}(\svec^*)$ as follows:
\begin{align*}
    \calM_1: \widehat{\theta \cdot \vel} (\svec^*), \qquad
    \calM_2: \widehat{\theta \cdot \widehat{\vel}} (\svec^*), \qquad
    \calM_3: \widehat{\theta} (\svec^*) \cdot \widehat{\vel} (\svec^*)
\end{align*}
where $\widehat{(\cdot)} (\svec^*)$ is used to denote the estimate of $(\cdot)$ at any spatio-temporal point $\svec^*$ given the data $\Data_{\rm{Pseudo}}$. Our proposed procedure from Section~\ref{sec: Methods_Two-step} corresponds to $\calM_2$, where we estimate $\vel$ from $\Psi$ and interpolate the $\TT$ field based on the in-situ $\theta \cdot \widehat{\vel}$. All results hereafter are based on estimates after debiasing on all stages.

 $\calM_1$ is a hypothetical procedure where we assume that $\vel$ can be obtained without estimation which is not feasible in practice (except at the surface where we have access to satellite-based $\vel$ fields). Given $\Data_{\rm{Pseudo}}$, $\calM_1$ only requires a second stage procedure that reduces to the local Gaussian process method from \citet{kuusela_locally_2018}. Thus, the performance of $\calM_1$ signifies the idealized interpolation capability of the local Gaussian process method when sparse spatio-temporal measurements are fully observed.
 Meanwhile, $\calM_3$ is an alternative approach detailed in Appendix~\ref{sec: Appendix_alternative_decomp}, where the two gridded products $\widehat{\theta}$ and $\widehat{v}$ can only be accessed separately. Such a situation frequently arises in oceanographic data analysis, in which case this approach is deemed a conventional norm. In this scenario, two separate interpolations---one for $\theta$ and the other for $\vel$---are needed; OHT is computed as the product of the two gridded fields.

\begin{figure}[!htbp]
    \centering
    \begin{subfigure}[b]{.49\textwidth}
        \centering
        \caption{Zonal}\label{fig: DUACS_zon}%
        \includegraphics[trim={0 0 0 0},clip,width=\textwidth, origin=c]{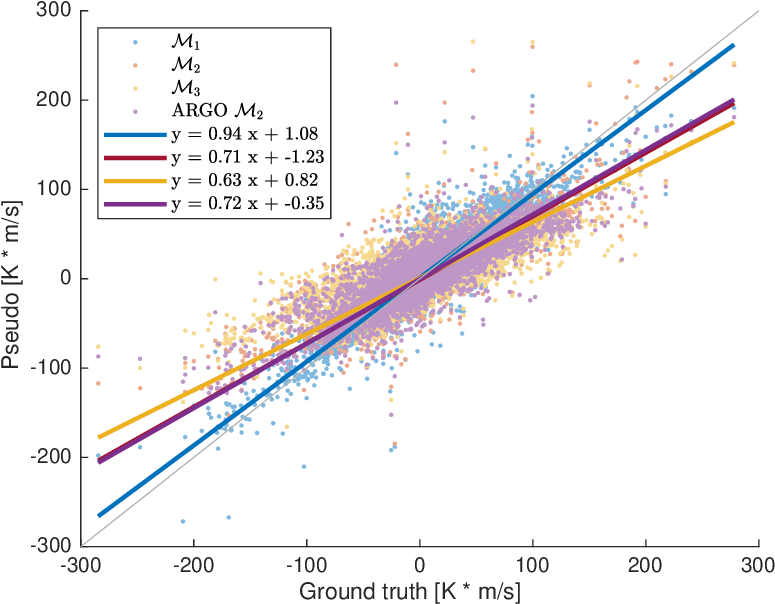}%
    \end{subfigure}
    \begin{subfigure}[b]{.49\textwidth}
        \centering
        \caption{Meridional}\label{fig: DUACS_mer}%
        \includegraphics[trim={0 0 0 0},clip,width=\textwidth, origin=c]{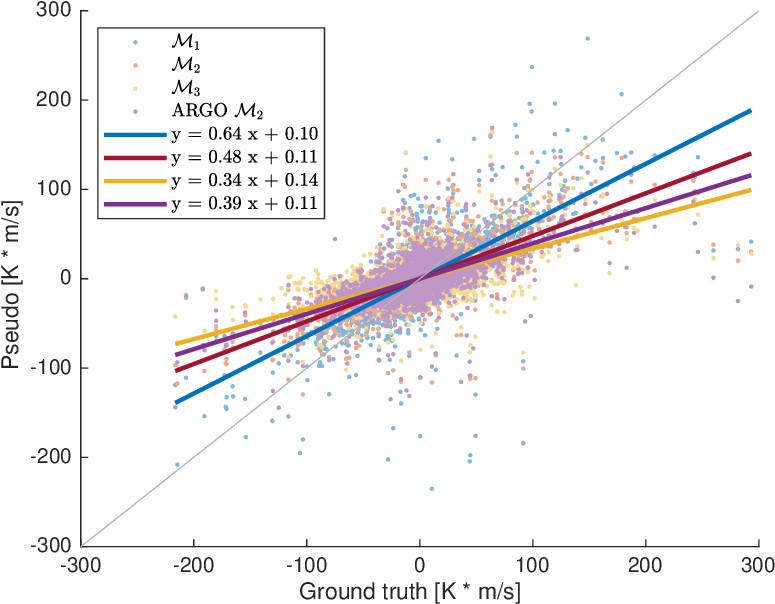}%
    \end{subfigure}
    \caption{Calibration of time-averaged zonal and meridional surface temperature transport $\text{Av} (\TT)$ on $1^\circ \times 1^\circ$ spatial grid between ground truth and competing models. For each case, the estimated regression line is superimposed with the same color.}\label{fig: model_comparison}
\end{figure}

Figure~\ref{fig: model_comparison} shows the calibration between the time-averaged surface temperature transport field $\text{Av} (\widehat{\TT})$ on a $1^\circ \times 1^\circ$ spatial grid computed with the three methods $\calM_1, \calM_2, \calM_3$ and the ground truth. $\calM_1$ (in blue) clearly performs the best of the three competing models, as the velocity $\vel$ is fully observed in this case (i.e., the first stage estimation achieves zero error). Note that estimating the meridional OHT is an intrinsically harder problem than estimating the zonal OHT. This asymmetry most likely stems from the fact that across most of the open ocean, the meridional signal is substantially smaller than the zonal signal \citep{zheng_ocean_2009,forget_global_2019}, leading to a decrease in the signal-to-noise ratio.

\begin{table}[htbp] \centering 
  \caption{Prediction performance of ${\TT}$ estimation methods.}
  \label{tab: model_comparison} 
\begin{tabular}{@{\extracolsep{5pt}} llccc|c} 
\toprule
  &  & $\calM_1$ & $\calM_2$ & $\calM_3$ & Argo $\calM_2$ \\ 
\midrule
\multirow{ 4}{*}{Zonal} & RMSE           & $38.3$ & $40.8$ & $53.5$ & $41.3$ \\ 
                        & MAD            & $4.12$ & $4.23$ & $4.25$ & $4.28$\\ 
                        & MIGN          & $4.63$ & $4.66$ & $4.79^\dagger$ & $4.71$ \\  
                        & MCRPS          & $12.5$ & $12.5$ & $14.9^\dagger$ & $12.9$ \\  
\midrule
\multirow{ 4}{*}{Meridional} & RMSE & $37.1$ & $38.4$ & $42.6$ & {$39.1$} \\ 
                             & MAD  & $4.01$ & $4.09$ & $4.11$ & $4.12$ \\ 
                             & MIGN   & $4.58$ & $4.61$ & $4.66^\dagger$ & $4.69$ \\  
                             & MCRPS   & $11.9$ & $12.0$ & $13.1^\dagger$ & $12.0$ \\  
\bottomrule
\end{tabular}
\\ \hspace{17em} $\dagger$ Based on an approximation.
\end{table}

Under a more realistic regime where $\vel$ cannot be directly observed, the proposed two-stage procedure $\calM_2$ outperforms $\calM_3$ by leveraging the high-frequency signals maintained from the first stage, in terms of both better calibration to the ground truth (Figure~\ref{fig: model_comparison}) and four quantitative performance measures---root mean squared error (RMSE), median absolute deviation (MAD), 
median ignorance score \citep[MIGN,][]{roulston_evaluating_2002}, and median continuous ranked probability score \citep[MCRPS,][]{gneiting_probabilistic_2007}---as seen in Table~\ref{tab: model_comparison}. The performance metrics for $\Upsilon$ are calculated as follows:
\begin{align*}
    \operatorname{RMSE}  &= \sqrt{\frac{1}{|\calS|} \sum_{\svec \in \calS} \left( \Upsilon_{\text{GT}} (\svec) - \widehat{\Upsilon} (\svec) \right)^2}, \qquad
    \operatorname{MAD} = \MED_{\svec \in \calS} \left| \Upsilon_{\text{GT}} (\svec) - \widehat{\Upsilon} (\svec) \right|,\\    
    \operatorname{MIGN}  &= \MED_{\svec \in \calS} \left[ - \log p( \Upsilon_{\text{GT}} (\svec) \;|\; \Data \, ; \widehat{\bbBeta}, \widehat{\bbXi}) \right],\\
    \operatorname{MCRPS} &= \MED_{\svec \in \calS} \int \left[  F (\Upsilon \;|\; \Data \, ; \widehat{\bbBeta}, \widehat{\bbXi}) - \I (\Upsilon \ge \Upsilon_{\text{GT}} (\svec)) \right]^2 \d \Upsilon, 
\end{align*}
where $\Upsilon_{\text{GT}}$ is the ground truth field, $p(\cdot \;|\; \Data)$ and $F(\cdot\;|\; \Data)$ are the predictive probability density function and predictive cumulative distribution function, respectively. While RMSE and MAD primarily assess the deterministic accuracy of the predictions, MIGN and MCRPS measure the probabilistic accuracy by taking both the deterministic accuracy and precision into consideration. MIGN measures the goodness of fit of the predictive distribution to the ground truth, and MCRPS can be viewed as a generalized version of MAD \citep{gneiting_probabilistic_2007}. The predictive distribution of $\calM_1$ and $\calM_2$ is Gaussian with analytically tractable mean and variance; however, that of $\calM_3$ is no longer Gaussian and entails further approximations for computing the MIGN and MCRPS since the estimate is the product of two correlated Gaussian random variables. Here MIGN and MCRPS for $\calM_3$ are computed using an approximate predictive distribution assuming that $\widehat{\theta}$ and $\widehat{\vel}$ are jointly Gaussian. 

In addition to validating the two-stage procedure, comparing the surface temperature transport estimates derived from in-situ Argo profiles $\Data$ to those constructed from the pseudo observations $\Data_{\text{Pseudo}}$ further confirms that the results presented in Section~\ref{sec:Results} are optimal in terms of the proposed framework. In Figure~\ref{fig: model_comparison}, Table~\ref{tab: model_comparison}, we provide the calibration and point prediction performance of the Argo-based estimate (denoted as Argo $\calM_2$ to emphasize that $\calM_2$ is comparable) with reference to the ground truth. The estimates derived from actual Argo profiles are surprisingly close to those based on pseudo observations ($\calM_2$) and outperform $\calM_3$. A performance degradation between Argo $\calM_2$ and $\calM_2$ could arise mainly for two possible reasons: (A) Argo-based estimates are computed at a near surface pressure (10 dbar) unlike the ground truth and $\calM_2$ estimates, and (B) reference velocities at 900 dbar adopted from \citet{gray_global_2014} might be underestimated. The good agreement found here demonstrates that these effects are minor in comparison to the other factors contributing to the overall performance of our method.

Recall that $\calM_1$ is the idealized estimate when $\vel$ is perfectly known. In other words, if we could improve our $\vel$ estimate in the first stage, we might be able to achieve performance close to $\calM_1$ even with in-situ Argo profiles alone. One can consider a natural extension of this work that synthesizes the high-resolution satellite product with the in-situ Argo profile dataset \citep{rio_improved_2018}, which is an interesting direction for future improvement but well beyond the scope of the present study.  Our results do emphasize, however, that the coarse sampling resolution is the main cause of miscalibration in comparison to the satellite-based {ground truth}. The miscalibration arises mostly near western boundaries and in regions with sharp fronts, where abrupt changes occur and strong currents are present. These issues could be explained by (A) the innate bias incurred from the nonparametric regression approach and/or (B) the inability of the Argo fleet to sample the coastal shelves, where the floats cannot reach 2000 dbar. There have been numerous methods proposed in the statistics literature to reduce the innate bias of (A). We tried the non-local means \citep{arias-castro_oracle_2011} and concluded that the method brings a marginal improvement in identifying the fronts. On the other hand, in Section~\ref{sec: Supp_Spray} of the Supplementary Material \citep{park_supplement_2020}, we confirmed that including Spray glider observations \citep{sherman_autonomous_2001,rudnick_ocean_2016} mitigates the underestimation occurring in the narrow boundary current in the western North Atlantic.
These results indicate that the miscalibration stems more from the Argo array's sampling deficiencies in the coastal regions and highly variable boundary currents and less from a boundary bias due to the nonparametric regression approach.

\subsection{Effect of Debiasing Procedure}\label{sec: debias_effect}

\begin{figure}[!htbp]
    \centering
    \begin{subfigure}[b]{.325\textwidth}
        \centering
        \caption{SSH}\label{fig: DUACS_SSH_debias}%
        \includegraphics[trim={0 0 0 0},clip,width=\textwidth, origin=c]{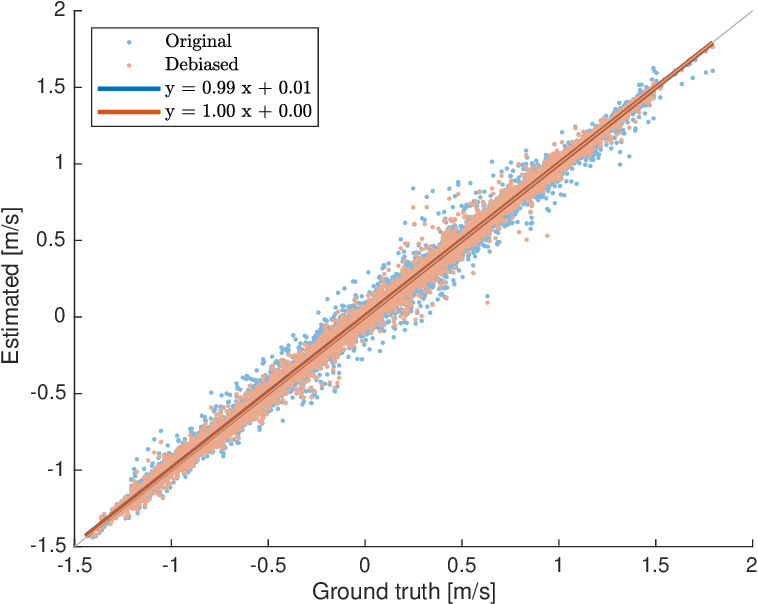}%
    \end{subfigure}
    \begin{subfigure}[b]{.325\textwidth}
        \centering
        \caption{Zonal $\vel$}\label{fig: DUACS_zon_vel_debias}%
        \includegraphics[trim={0 0 0 0},clip,width=\textwidth, origin=c]{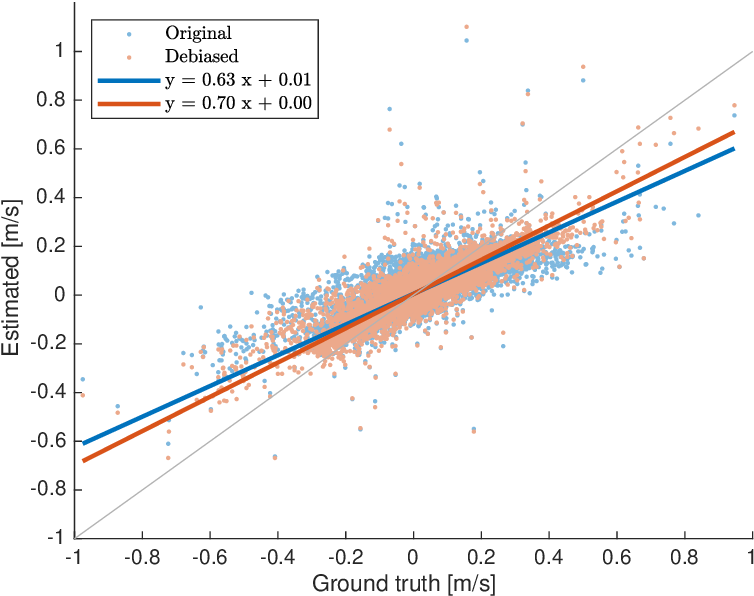}%
    \end{subfigure}
    \begin{subfigure}[b]{.325\textwidth}
        \centering
        \caption{Meridional $\vel$}\label{fig: DUACS_mer_vel_debias}%
        \includegraphics[trim={0 0 0 0},clip,width=\textwidth, origin=c]{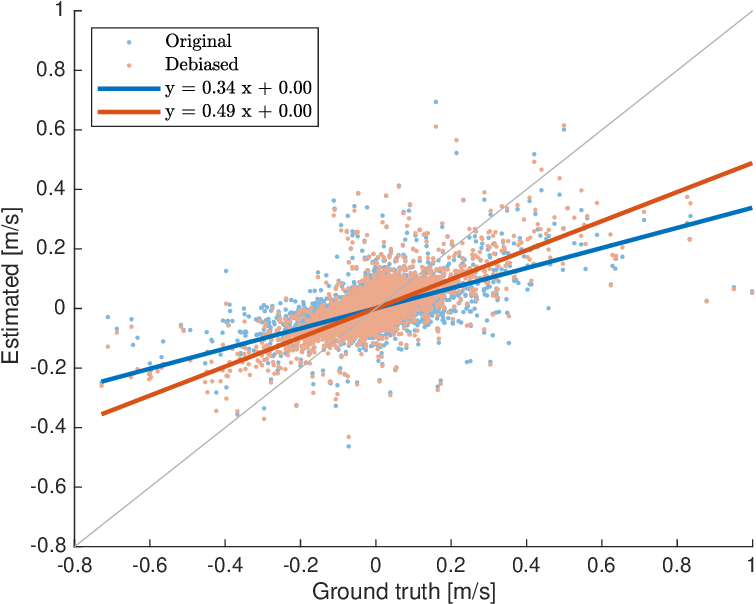}%
    \end{subfigure}
    \caption{Calibration of time-averaged $\Psi$ and $\vel$ between ground truth and estimates, respectively, with and without debiasing.}\label{fig: Psi_debias_comparison}
\end{figure}

In this section, we discuss the efficacy of the debiasing procedure proposed in Section~\ref{sec: Debias}. Figure~\ref{fig: Psi_debias_comparison} and Table~\ref{tab: Psi_debias_comparison} give the point prediction performance of the debiasing procedure on the first stage. Both show that the debiasing procedure improves the mapping of both zonal and meridional velocities. Notice that the local interpolation procedure yields a globally calibrated SSH ($\Psi$) field (Figure~\ref{fig: DUACS_SSH_debias}) as compared to the $\vel$ field (Figure~\ref{fig: DUACS_zon_vel_debias},\ref{fig: DUACS_mer_vel_debias}). While not surprising, this does demonstrate that predicting the gradient field from unobserved $\vel$ is a harder problem than estimating the underlying $\Psi$ field itself. This result also implies that the procedure may smooth more than the actual curvature in regions where sharp magnitude changes occur in the $\Psi$ field. Choosing a smaller bandwidth parameter may alleviate this concern but leads to unstable estimates because fewer observations are then available within each local spatio-temporal window. A data-driven bandwidth selector would be an appealing refinement \citep{fan_data-driven_1995,ruppert_multivariate_1994,de_brabanter_derivative_2013}. However, the associated improvements are known to be inconsistent depending on the data, not to mention that our current concern lies mainly in the derivative estimation, rather than estimating the observable response $\Psi$.

\begin{table}[!htbp] \centering 
  \caption{Point prediction performance of $\text{Av} (\widehat{\vel})$ by the debiasing procedure regarding RMSE, MAD, and correlation $\rho$. The best results are highlighted in bold.}
  \label{tab: Psi_debias_comparison} 
\begin{tabular}{@{\extracolsep{5pt}} lcc|cc} 
\toprule
    & \multicolumn{2}{c|}{Zonal}      & \multicolumn{2}{c}{Meridional} \\ \cmidrule{2-5}
    & Without Debias & Debias $\Psi$ & Without Debias & Debias $\Psi$  \\ 
\midrule
 RMSE & $0.059$ & \bm{$0.047$} & $0.046$ & \bm{$0.039$}  \\ 
 MAD  & $0.018$ & \bm{$0.015$}  & $0.012$ & \bm{$0.010$} \\ 
 $\rho$ & $0.781$ & \bm{$0.873$} & $0.536$ & \bm{$0.708$}\\
\bottomrule
\end{tabular} 
\end{table}

\begin{figure}[!htb]
    \centering
    \begin{subfigure}[b]{.49\textwidth}
        \centering
        \caption{Zonal}\label{fig: DUACS_zon_TT_debias}%
        \includegraphics[trim={0 0 0 0},clip,width=\textwidth, origin=c]{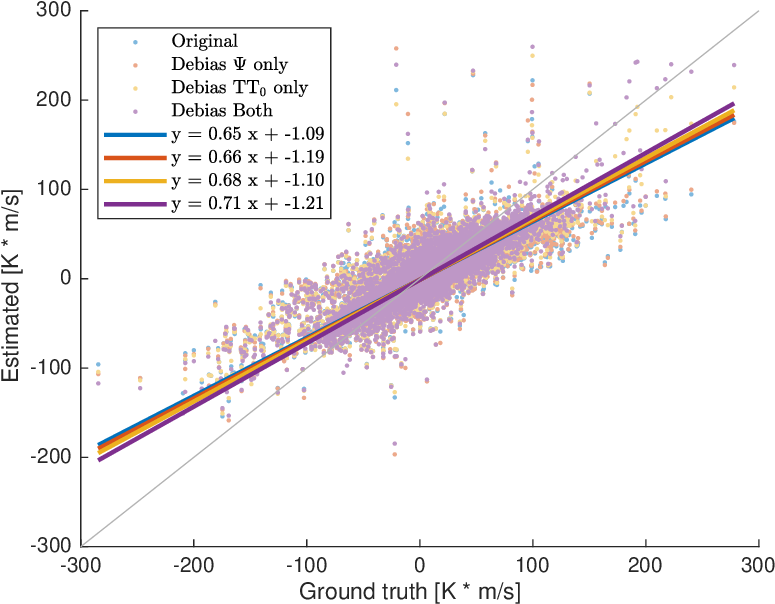}%
    \end{subfigure}
    \begin{subfigure}[b]{.49\textwidth}
        \centering
        \caption{Meridional}\label{fig: DUACS_mer_TT_debias}%
        \includegraphics[trim={0 0 0 0},clip,width=\textwidth, origin=c]{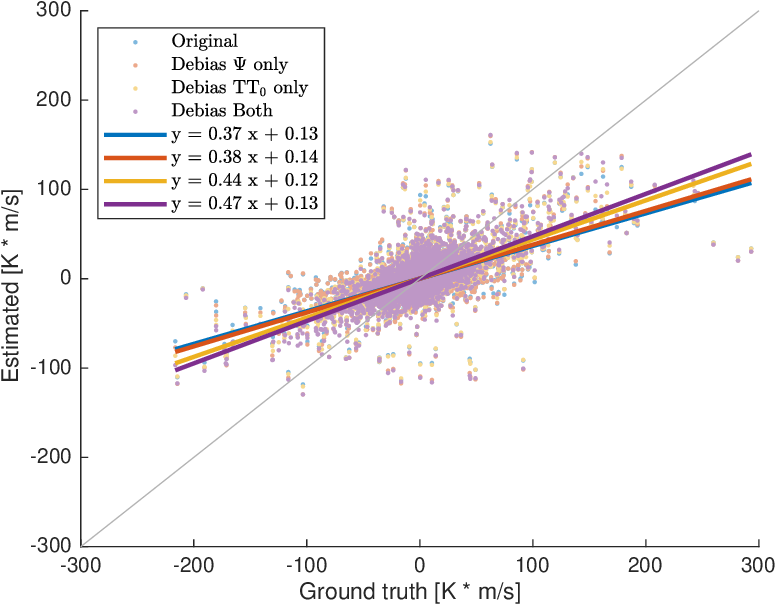}%
    \end{subfigure}
    \caption{Calibration of time-averaged surface temperature transport $\TT$ between ground truth and estimates, depending on which debiasing procedures are applied.}\label{fig: OHT_debias_comparison}
\end{figure}

Figure~\ref{fig: OHT_debias_comparison} and Table~\ref{tab: OHT_debias_comparison} show the calibration and point prediction performance of the debiasing procedure within the two-stage procedure. We separated out the debiasing efficacy based on whether we only debias $\Psi$ but not $\TT$, debias $\TT$ without debiasing $\Psi$, or debias both $\Psi$ and $\TT$. The result indicates that both estimates should be debiased in order to fully leverage the potential that the debiasing procedure can deliver. Notice that debiasing at the second stage is crucial for the performance gain. Only debiasing the first stage leaves an inconsistent result, even though the procedure does improve velocity prediction in the first stage as seen in Figure~\ref{fig: Psi_debias_comparison} and Table~\ref{tab: Psi_debias_comparison}.  Both phenomena can be understood from the fact that the numerical magnitude of the temperature $\theta$ is much greater than that of the velocity $\vel$. The correction accounting for the larger magnitude, i.e., temperature $\theta$, should play the biggest role in accurate prediction of the final $\TT$ field. Meanwhile, this also implies that a slight perturbation in $\vel$, whose magnitude is small, may introduce a large deviation in $\TT$ when multiplied with $\theta$. Without debiasing the $\TT$ in the second stage, predictions made with only the first stage correction are likely to yield subpar performance.

\begin{table}[!htbp] \centering 
  \caption{Point prediction performance of $\text{Av} (\widehat{\TT})$ by the debiasing procedure regarding RMSE, MAD, and correlation~$\rho$. {The best results are highlighted in bold.}}
  \label{tab: OHT_debias_comparison} 
\begin{tabular}{@{\extracolsep{5pt}} llcccc} 
\toprule
  &  & Without Debias & Debias $\Psi$ only & Debias $\TT$ only & Debias both \\ 
\midrule
\multirow{ 3}{*}{Zonal} & RMSE & $16.41$ & $16.38$ & $14.81$ & \bm{$14.30$} \\ 
 & MAD  & $5.653$ & $5.688$ & $5.260$ & \bm{$5.164$}\\ 
 & $\rho$  & $0.811$ & $0.809$ & $0.854$ & \bm{$0.862$}\\ 
\midrule
\multirow{ 3}{*}{Meridional} & RMSE & $12.71$ & $12.71$ & $11.53$ & \bm{$11.22$} \\ 
 & MAD  & $3.281$ & $3.309$ & $3.161$ & \bm{$3.160$}\\ 
 & $\rho$  & $0.588$ & $0.590$ & $0.680$ & \bm{$0.700$}\\ 
\bottomrule
\end{tabular} 
\end{table} 

%% file: sec7_discussion.tex
\section{Discussion and Conclusions}\label{sec: discussion}

In this paper, we introduced a comprehensive spatio-temporal interpolation framework for estimating the global ocean heat transport using in-situ Argo profiles. The framework characterizes a partially observed OHT process as well as the latent velocity process $\vel$ represented by the gradient of an observed quantity, both of which are spatio-temporally correlated in a heterogeneous fashion. Our contributions to OHT estimation are threefold: we formalize the statistical challenges as an end-to-end latent LGPR model accompanied by the two-stage estimation procedure, introduce the approximate EM procedure for jointly estimating both the mean and the covariance parameters, and refine the potentially misspecified mean field model with the debiasing procedure. Our data-driven interpolated fields are on par with state-of-the-art multimission satellite products near the surface, at the spatial resolution that Argo can resolve, and yield sensible new subsurface OHT estimates that can provide useful insights into crucial scientific phenomena. \add{The OHT estimates described in this work are however just one application of the framework, which can be used to map any oceanographic tracer of interest. For example, transport of salt in the ocean is a potential future application that will aid in managing ecosystems and in our overall understanding of changes in regional sea level and ocean stratification, with implications for air-sea interactions.} 

Even though our comprehensive framework was targeted at quantifying ocean heat transport, the techniques involved resonate with broader statistical issues, enabling possible extensions and raising interesting questions that we have not fully addressed in this paper. For example, we handle the latent gradient field by adopting the local Gaussian process. However, nonparametric \emph{derivative estimation} has a long-standing history in statistics with contributions characterizing the optimal derivative estimator based on various criteria as well as data-driven methods to choose the optimal tuning parameters \citep{charnigo_generalized_2012,wang_derivative_2015,dai_optimal_2016,liu_derivative_2018}. Most existing literature on derivative estimation considers a univariate covariate, and extending the established results to the spatio-temporal case is not trivial as spatial local regression itself requires substantial theoretical considerations \citep{hallin_local_2004}.

Another interesting direction is jointly modeling the temperature and velocity fields based on the underlying process for temperature and salinity. Dynamic height anomaly, modeled separately from temperature in this work, is indeed a nonlinear function of temperature and salinity, which are the primitive measurements we obtain from the Argo floats. An exciting extension of our approach could be modeling the temperature and dynamic height anomaly fields instead as a bivariate Gaussian process, similar to the approach taken by \citet{yarger_functional-data_2020} for jointly modeling temperature and salinity. Admittedly, finding an appropriate cross-correlation function between temperature and dynamic height anomaly might not be as straightforward as for temperature and salinity.

\add{Our LGPR model is premised on the assumption that spatio-temporal dependency in the Argo data within a small enough window can be successfully approximated by a stationary process. This assumption shares a common theme with existing methods for modeling non-stationary random fields in spatio-temporal statistics; yet, the premise is challenged if the true underlying field is locally non-stationary. Several refinements are possible, e.g., by adopting a non-stationary covariance function \citep{higdon_process-convolution_1998,paciorek_spatial_2006} directly within each local window, or by better accommodating the spatio-temporal inhomogeneity when local stationarity suffices. The latter might be achieved by allowing the window size to be adaptive, by leveraging a test to determine subregions of stationarity \citep{fuentes_interpolation_2002}, or by adopting a multiresolution wavelet basis \citep{nychka_multiresolution_2002}. These refinements would require a careful treatment as the increased model flexibility would incur a higher variance which may offset the potential bias of the locally stationary model.}

While we relied on the Gaussian process to define the local spatio-temporal process, the complete characterization of the second-order structure (covariance kernel) is not enough to describe the full process when it is non-Gaussian. Even commonly used climate variables, such as temperature, are known to show non-Gaussian properties \citep{kuusela_locally_2018,stein_statistical_2019}. It could therefore be possible to explore non-Gaussian models \citep[e.g.,][]{bolin_multivariate_2020} to improve the prediction of the spatio-temporal process.

Vertical dependence, i.e., correlation across pressure levels, is an important aspect of any profile measurements, including the $T$-$S$ profiles. The dynamic height anomaly $\Psi$ has an additional interesting property in that it is a monotonically increasing function of depth by definition. While our approach partially accommodates the vertical structure, the vertical dependence is not fully modeled. Completely accounting for the vertical structure would yield several improvements, including a truly four-dimensional map of global ocean heat transport, effective confinement of the random fields with respect to depth using the monotonicity condition, and a complete uncertainty quantification of the OHT field. \citet{yarger_functional-data_2020} proposed functional PCA as a way of handling the vertical dependence, even though their focus was on modeling temperature and salinity where the monotone constraint is not required. Fully characterizing the vertical dependence of the global ocean circulation and associated OHT is additionally of great interest to the oceanographic community and thus represents a priority for future work. 

%% file: appendix.tex
 \begin{appendix}

\section{Alternative Decomposition of OHT}\label{sec: Appendix_alternative_decomp}

One may argue that heat transport can be computed by estimating mass transport and temperature grids separately. In particular,
\begin{align}\label{eq: alternative_decomp}
  Q = \int \theta M \d z = \int (\overline{\theta} + \theta') (\overline{M} + M') \d z, \qquad
  M := \rho \vel
\end{align}
Such a decomposition may identify which component---mean or anomaly---in temperature and mass transport drives heat transport. The decomposition does not need a two-step estimation procedure, rather it executes single step estimation for temperature and mass transport, separately. In Section~\ref{sec: Validation}, we empirically demonstrate that this strategy is inferior to the proposed two-stage method when it comes to prediction performance.

 \end{appendix}

%% file: supplement.tex
\begin{center}    
\textbf{\large SUPPLEMENTARY MATERIAL}
\end{center}

\setcounter{equation}{0}
\setcounter{figure}{0}
\setcounter{table}{0}
\setcounter{section}{0}
\setcounter{page}{1}
\makeatletter
\renewcommand{\theequation}{S\arabic{equation}}
\renewcommand{\thefigure}{S\arabic{figure}}
\renewcommand{\thetable}{S\arabic{table}}
\renewcommand{\bibnumfmt}[1]{[S#1]}
\renewcommand{\citenumfont}[1]{S#1}

 \section{Quality Control}\label{sec: QC}
 
 On top of the quality control criteria applied in \citesupp{kuusela_supplementary_2018}, we additionally reject potential duplicates of each profile having the same spatial location with timestamps within 15 minutes since such duplicates are highly unlikely based on the Argo sampling design.
 
In general, salinity measurements are more prone larger biases than the temperature measurements by the nature of the CTD sensor and thus stricter quality control criteria are often required. We inspected the computed dynamic height anomalies and filtered out 34 profiles that were impacted by problematic salinity measurements. These profiles deviate more than 3$\times$interquartile range (IQR) from the median at 10 dbar and more than 10$\times$IQR at deeper depths.

\section{Analytic derivative of Mat{\'e}rn covariance kernel}\label{sec: Matern_derivative_detail}

For our covariance kernel choice~\eqref{eq: Matern_def}, we have the following analytic forms for the gradient and the Hessian of the kernel $k$:
\begin{align}
    \frac{\partial}{\partial x_1} k(\svec_1, \svec_2)
        &= -3 \phi \frac{\Delta_x}{\xi_x^2} \exp (- \sqrt{3}d),\\
    \frac{\partial^2}{\partial x_1 \partial x_2} k(\svec_1, \svec_2) 
        &= \frac{3 \phi }{\xi_x^{2}} \left( 1 - \frac{\sqrt{3}}{d} \left(\frac{\Delta_x}{\xi_x} \right)^2  \right) \exp(- \sqrt{3} d),\\
    \frac{\partial^2}{\partial x_1 \partial y_2} k(\svec_1, \svec_2) 
        &= - \frac{3 \sqrt{3} \phi}{d} \frac{ \Delta_x }{ \xi_y^{2}}\frac{ \Delta_y}{\xi_y^{2}}  \exp(- \sqrt{3} d),
\end{align}
where $\Delta_x = x_1 - x_2$ and  $\Delta_y = y_1 - y_2$.

\section{Alternating Maximization view of the EM procedure}\label{sec: AMview_EM}

One may ask how does the coordinate ascent algorithm proposed in the main text link to the EM algorithm. We follow the maximization-maximization viewpoint of the EM algorithm following \citetsupp{neal_view_1998}. A similar argument was also made in \citetsupp{andresen_convergence_2016}. Under our LGPR model, let $\bm{\Upsilon}$ be a set of observed values of the quantity of interest and let $\bm{Z}$ be a set of unobserved latent data whose joint probability is parameterized using $\bbXi$. The EM algorithm iteratively computes the following two steps, for $l = 1,2,\dots$,
\begin{itemize}
    \item[\bf{E-Step}: ] Compute a distribution $\tilde{P}^{(l)}$ over the range of $\bm{Z}$ such that $\tilde{P}^{(l)} = P(\bm{Z} | \bm{\Upsilon}; \bbXi^{(l-1)} )$.
    \item[\bf{M-Step}: ] Set $\bbXi^{(l)} = \argmax_{\bbXi} \E_{\tilde{P}^{(l)}} \left[ \log P (\bm{\Upsilon}, \bm{Z}; \bbXi) \right]$.
\end{itemize}

When $\bm{Z} = \argmax_{\bbBeta} P( \bm{\Upsilon}; \bbBeta, \bbXi)$ is considered, $\tilde{P}^{(l)}$ is just a point mass at $\bbBeta^{(l)}$, where $\bbBeta^{(l)} = \argmax_{\bbBeta} P( \bm{\Upsilon}; \bbBeta, \bbXi^{(l-1)})$. Thus, the E-step is equivalent to finding the $\bbBeta$ that maximizes the likelihood $\calL$ given the $\bbXi^{(l-1)}$ from the previous iteration. Similarly, $\E_{\tilde{P}^{(l)}} \left[ \log P (\bm{\Upsilon}, \bm{Z}; \bbXi) \right] = \log P (\bm{\Upsilon}, \bbBeta^{(l)}; \bbXi)$ which implies that the M-step is equivalent to finding the maximizer of the likelihood $\calL$ given the $\bbBeta^{(l)}$ found in the previous E-step.

\section{Details of the Approximate EM procedure}\label{sec: AM_detail}

The M-step is performed with the residuals from the previous E-step as described in the paper. In the following, we derive the analytic solution of the E-step assuming that the temporal grid $\calT$ consists of center points of each month. A generalization to more complex grids is straightforward following the LGPR construction. The subscript $i,j$ corresponds to year $i$ and month $j$, i.e., $y_{i,j}$ is a set of observation in year $i$ within month $j$. For every $\xvec^*\in \calX$, denote $\bbbeta := \bbbeta(\xvec^*)$ and $\bbxi_j := \bbxi(\xvec^*, t^*_j)$, where $t^*_j \in \calT$ is the center point of month $j$. Then,
\begin{align*}
    \log \calL (\bbbeta | \bbXi) 
        =& \log(p(y_{1,1:3}); \bbxi_2) + \sum_{j=4}^{12} \log p(y_{1,j} | y_{1:(j-1)}; \bbbeta, \bbxi_2, \dots, \bbxi_{j-1}) \\
         &+ \sum_{i=2}^I \sum_{j=1}^{12} \log p(y_{i,j} | y_{1: i-1,1:12}, y_{i,1:(j-1)}; \bbbeta, \bbXi),\\
    \log \widetilde{\calL} (\bbbeta | \bbXi) 
        =& \log(p(y_{1,1:3}); \bbxi_2) + \sum_{j=4}^{12} \log p(y_{1,j} | y_{(j-2):(j-1)}; \bbbeta, \bbxi_{j-1}) \\
         &+ \sum_{i=2}^I \bigg[ \log p(y_{i,1} | y_{i-1,11:12}; \bbbeta, \bbxi_{12}) + \log p(y_{i,2} | y_{i-1,12}, y_{i,1}; \bbbeta, \bbxi_1)\\
         &+ \sum_{j=3}^{12} \log p(y_{i,j} | y_{i,1:(j-1)}; \bbbeta, \bbxi_{j-1}) \bigg],
\end{align*}
where $\widetilde{\calL}$ is the Vecchia approximated likelihood with 3-month temporal lag.

From hereafter, we show the result for the first year $(i=1)$ and drop the subscript $i$ for conciseness. The summand with all years still maintains the same maximizer form with the only difference in $Q$ and $C$ below.
\begin{align*}
    \log \widetilde{\mathcal{L}}_1 (\bbbeta | \bbXi) 
    \approx& \log p(y_{[1:3]}; \bbxi_2) + \sum_{j=4}^{12} \log p(y_j | y_{(j-2):(j-1)}; \bbxi_{j-1}), \qquad(\text{Vecchia})\\
        =& \log \phi(y_{[1:3]}; \eta_{[1:3]}\bbbeta, K (\bbxi_2))\\ 
         &+ \sum_{j=4}^{12} \log \phi(y_{[j-2:j]}; \eta_{[j-2:j]}\bbbeta + K_{j,-j} K_{-j,-j}^{-1} (y_{-j} - \eta_{-j})\bbbeta, K_{jj} - K_{j,-j} K_{-j,-j}^{-1} K_{-j,j} )\\
        \overset{(*)}{\propto}& \bbbeta^\top \underbrace{\left[ \eta_{[3]}^\top K (\bbxi_2)^{-1} \eta_{[3]} + \sum_{j=4}^{12} (\eta_j - K_{j,-j} K_{-j,-j}^{-1} \eta_{-j})^\top [K^{-1}]_{jj} (\eta_j - K_{j,-j} K_{-j,-j}^{-1} \eta_{-j}) \right]}_{:=Q} \bbbeta \\
        &- 2 \underbrace{\left[ y_{[3]}^\top K (\bbxi_2)^{-1} \eta_{[3]}  +  \sum_{j=4}^{12}  (y_j - K_{j,-j} K_{-j,-j}^{-1} y_{-j})^\top [K^{-1}]_{jj}  (\eta_j - K_{j,-j} K_{-j,-j}^{-1} \eta_{-j})  \right]}_{:=C} \bbbeta,
\end{align*}
where $\phi(y; \mu, \Sigma)$ is the pdf of multivariate Normal distribution with mean $\mu$ and variance $\Sigma$; $K = K(\bbxi_{j-1})$ is a covariance matrix constructed from the parameter $\bbxi_{j-1}$ and subscript $-j$ corresponds to all indices except month $j$. $(*)$ follows from Woodbery matrix identity,  
\begin{align*}
 (K_{jj} - K_{j,-j} K_{-j,-j}^{-1} K_{-j,j} )^{-1} 
    &= K_{jj}^{-1}-K_{jj}^{-1}K_{j,-j} \left(K_{-j,-j} + K_{-j,j} K_{jj}^{-1}K_{j,-j} \right)^{-1}K_{-j,j} K_{jj}^{-1}\\ 
    &= [K^{-1}]_{jj}
\end{align*}
Therefore, the maximizer is
$$\widehat{\bbbeta} = Q^{-1} C.$$

\section{Predictive distribution for a single depth OHT}\label{sec: predictive_simple}

From Equation~\eqref{eq: joint_GP}, the predictive distribution of $\hat{\vel} (\svec_i) := \vel (\svec_i) \;|\; \Data$ for coordinates $\svec_i \in \svec_\Data \cap \widetilde{\Window} (\svec^*)$ is $n_i$-dimensional Gaussian with mean $\overline{\mu}_{\nabla_{\xvec}} (\svec_i)$ and variance $\overline{\Sigma}_{\nabla_{\xvec}} (\svec_i)$ as follows:
\begin{align*}
  \overline{\mu}_{\nabla_{\xvec}} (\svec_i) &= \mu_{\vel_{\rm ref}} (\svec_i) + R \left[ \nabla_{\xvec} m_{\Psi} (\svec_i) + \nabla_{\xvec} k_{\Psi} (\svec_i, \svec_i) \widetilde{\bm{K}}_{\Psi,i}^{-1} \left[\Psi_i - m_{\Psi} (\svec_i) \right] \right] \\
  \overline{\Sigma}_{\nabla_{\xvec}} (\svec_i) &= \Sigma_{\vel_{\rm ref}} (\svec_i) +  R\left[ H_{\Psi} (\svec_i, \svec_i) - \nabla_{\xvec^*} k_{\Psi} (\svec_i, \svec_i)^\top \widetilde{\bm{K}}_{\Psi,i}^{-1} \nabla_{\xvec^*} k_\Psi (\svec_i, \svec_i) \right] R^\top
\end{align*}
where \add{$\mu_{\vel_{\rm ref}}$, $\Sigma_{\vel_{\rm ref}}$ are the estimates and the mapping error from reference velocity estimates}, $H (\svec, \svec')$ is the Hessian of the kernel $k(\svec, \svec')$ and $\widetilde{\bm{K}}_{\Psi,i} = \bm{K}_{\Psi,i} + \widehat{\sigma}_\epsilon^2 I_{n_i}$.

Similarly, the predictive distribution of $\OHT (\svec^{**}) | {\vel} (\svec_\Data), \Data$ for any $\svec^{**} \in \widetilde{\Window} (\svec^*)$ is a Gaussian with mean $\overline{\mu}_{\OHT| {\vel} }$ and variance $\overline{\sigma^2}_{\OHT |  {\vel} }$ where
\begin{align*}
  \overline{\mu}_{\OHT| {\vel} } (\svec^{**}) &= m_{\OHT|{\vel}} (\svec^{**}) +  k_{\OHT|{\vel}} (\svec^{**}, \svec_i) \widetilde{\bm{K}}_{\OHT| {\vel},i}^{-1} \left[\theta_i \odot \vel_i - m_{\OHT|{\vel}} (\svec_i) \right] \\
  \overline{\sigma^2}_{\OHT| {\vel} } (\svec^{**}) &= k_{\OHT|{\vel}} (\svec^{**}, \svec^{**}) - k_{\OHT|{\vel}} (\svec^{**}, \svec_i)^\top \widetilde{\bm{K}}_{\OHT|{\vel},i}^{-1} k_{\OHT|{\vel}}  (\svec_i, \svec^{**})
\end{align*}
where $\theta_i := \theta (\svec_i)$ and $\odot$ is a Hadamard product.

Since $\OHT | \vel, \Data $ and ${\vel} | \Data$ forms a linear Gaussian system, the predictive distribution $\OHT_i (\svec^{**}) \;|\; \Data$ at spatio-temporal coordinate $\svec^{**}$ within $\widetilde{\Window} (\svec^*)$ is a Gaussian with mean $\overline{\mu}_{\OHT}$ and variance $\overline{\sigma^2}_{\OHT}$.
\begin{align*}
   \OHT_i (\svec^{**}) \;|\; \Data 
    \propto& 
      \Normal (\overline{\mu}_{\OHT| {\vel} } (\svec^{**}), 
      \overline{\sigma^2}_{\OHT| {\vel} } (\svec^{**}))
      \cdot \Normal \left( \theta_i \odot \overline{\mu}_{\nabla_{\xvec}} (\svec_i), [\theta_i \theta_i^\top] \odot \overline{\Sigma}_{\nabla_{\xvec}} (\svec_i) \right)\\
    =& \Normal (\overline{\mu}_{\OHT} (\svec^{**}), \overline{\sigma^2}_{\OHT} (\svec^{**}))\\
  \overline{\mu}_{\OHT} (\svec^{**})
    =& m_{\OHT|{\vel}} (\svec^{**}) +  k_{\OHT| {\vel}} (\svec^{**}, \svec_i) \widetilde{\bm{K}}_{\OHT| {\vel},i}^{-1} \left[\theta_i \odot \overline{\mu}_{\nabla_{\xvec}} (\svec_i) - m_{\OHT|{\vel}} (\svec_i) \right]\\
  \overline{\sigma^2}_{\OHT} (\svec^{**})
    =& k_{\OHT|{\vel}} (\svec^{**}, \svec_i) \widetilde{\bm{K}}_{\OHT| {\vel},i}^{-1}  [\theta_i \theta_i^\top] \odot \overline{\Sigma}_{\nabla_{\xvec}} (\svec_i) \widetilde{\bm{K}}_{\OHT| {\vel},i}^{-1}  k_{\OHT|{\vel}} (\svec_i, \svec^{**}) \\
        &+  \overline{\sigma^2}_{\OHT| {\vel} } (\svec^{**}).
\end{align*}

\section{Upscaled surface satellite estimates}\label{sec: Appendix_more_fig}
\
\begin{figure}[!htb]
    \centering
    \begin{subfigure}[b]{.49\textwidth}
        \centering
        \includegraphics[trim={0 0 0 0},clip,width=\textwidth, origin=c]{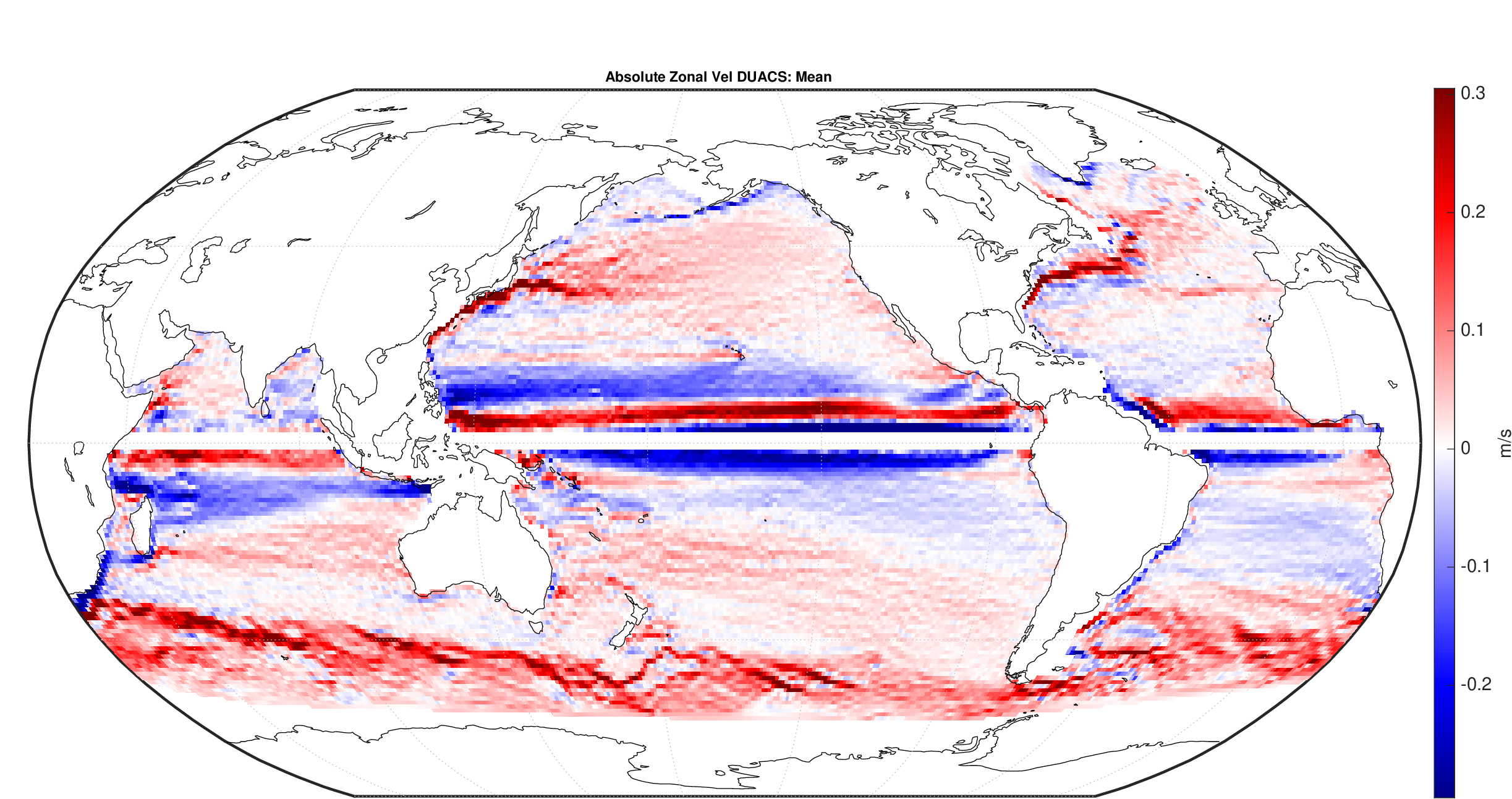}
        \caption{Zonal (East-West)}\label{fig: ground_truth_zon_vel}%
    \end{subfigure}
    \begin{subfigure}[b]{.49\textwidth}
        \centering
        \includegraphics[trim={0 0 0 0},clip,width=\textwidth, origin=c]{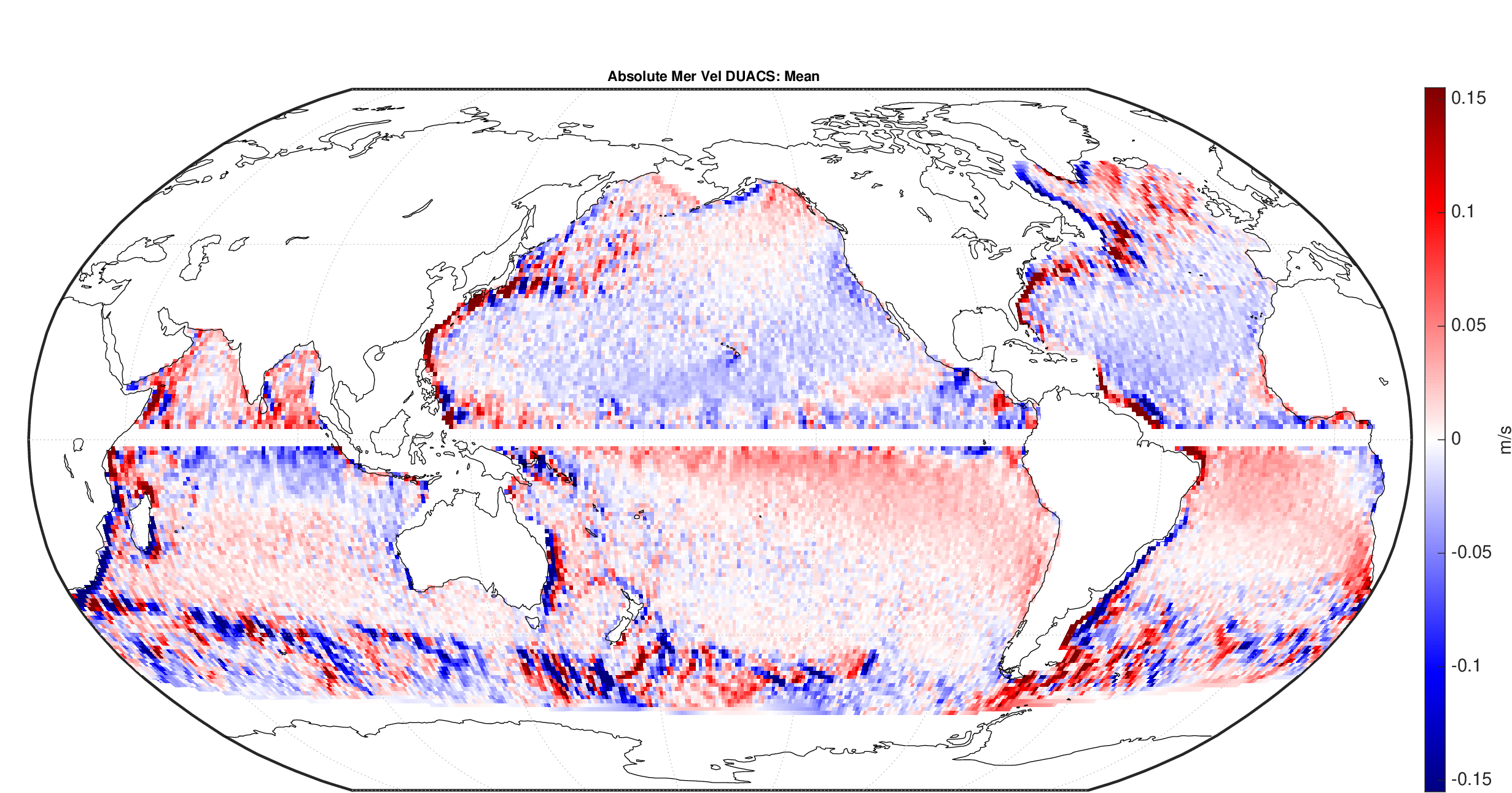}
        \caption{Meridional (North-South)}\label{fig: ground_truth_mer_vel}%
    \end{subfigure}
    \caption{Upscaled time averaged surface velocity from satellite product (DUACS)}\label{fig: ground_truth_vel}
\end{figure}
\begin{figure}[!htb]
    \centering
    \begin{subfigure}[b]{.49\textwidth}
        \centering
        \includegraphics[trim={0 0 0 0},clip,width=\textwidth, origin=c]{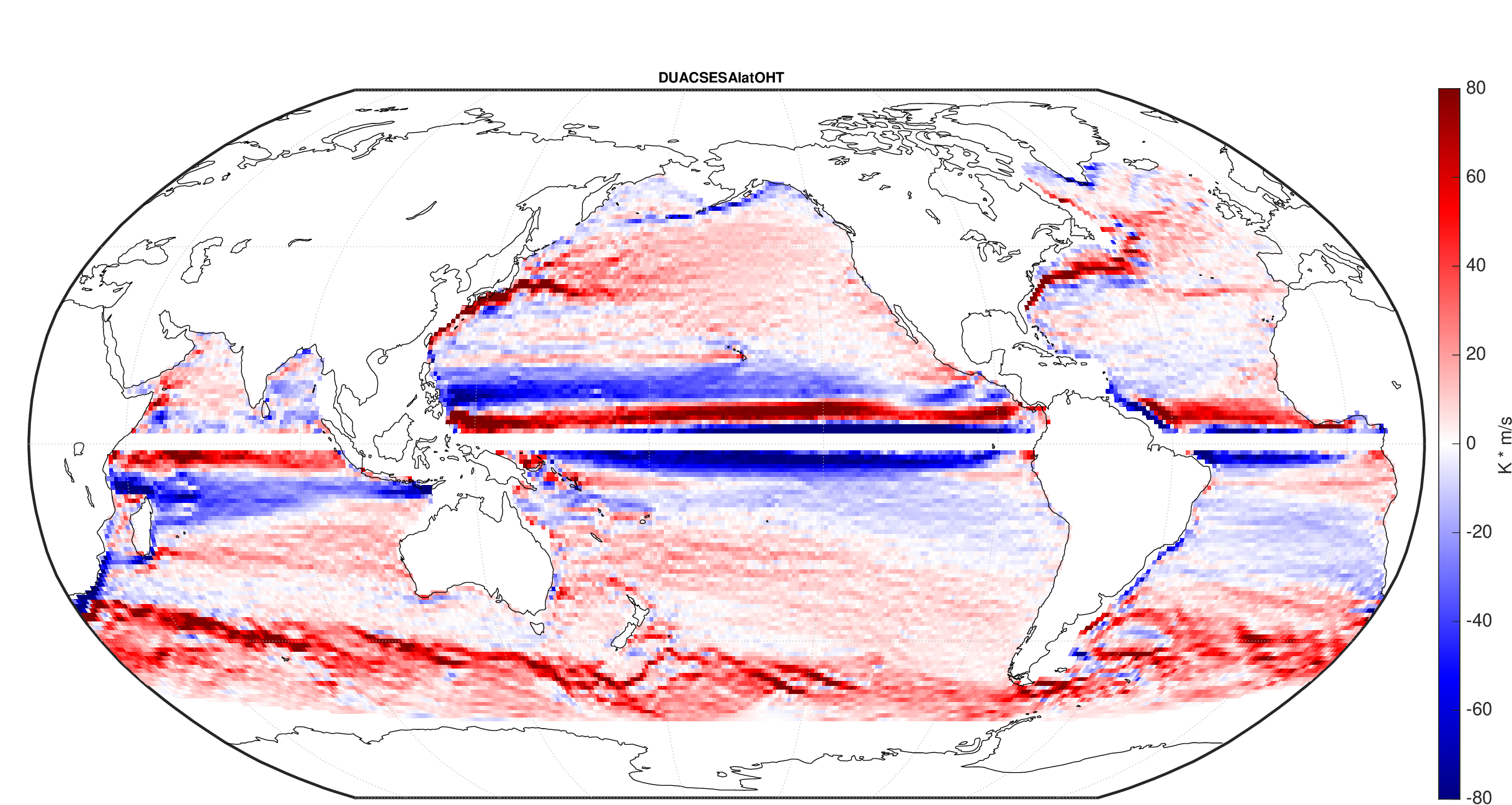}%
        \caption{Zonal (East-West)}\label{fig: ground_truth_zon}%
    \end{subfigure}
    \begin{subfigure}[b]{.49\textwidth}
        \centering
        \includegraphics[trim={0 0 0 0},clip,width=\textwidth, origin=c]{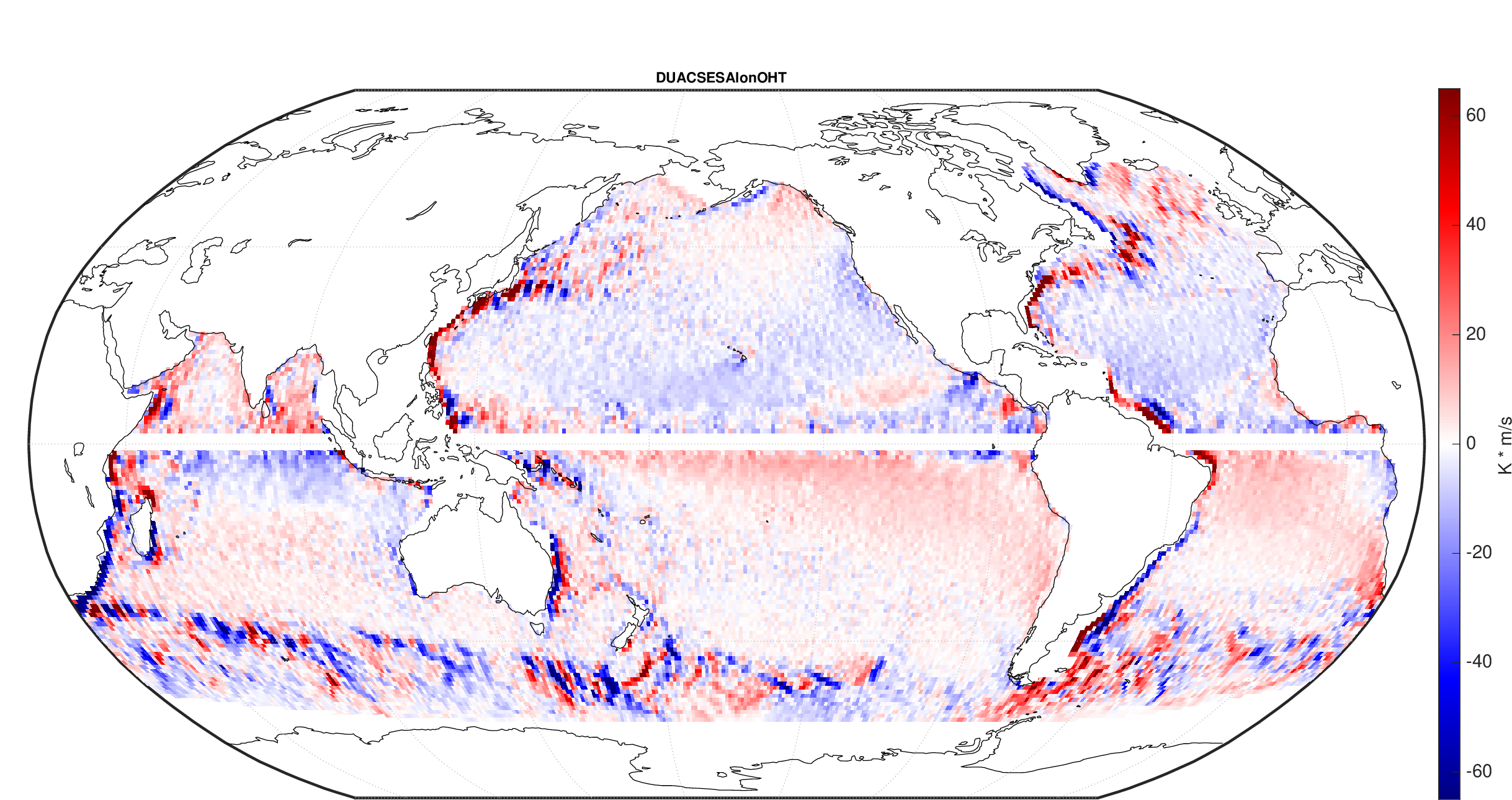}%
        \caption{Meridional (North-South)}\label{fig: ground_truth_mer}%
    \end{subfigure}
    \caption{Upscaled time averaged surface temperature transport from satellite products (DUACS, CCI-C3S)}\label{fig: ground_truth_TT0}
\end{figure}


\section{\add{Additional comparative numerical studies}}

Given the pseudo-observations generated from the multimission satellite observations used in the validation study in Section~5, we compare the predicted monthly full field at every grid point in the global open ocean $(\calX)$ during 2007--2018 $(\calT)$ to the ground truth full fields, i.e., the upscaled absolute dynamic topography $\Psi$, velocity $\vel$, and surface heat transport $\TT$ gridded fields (Figs.~\ref{fig: ground_truth_vel} and \ref{fig: ground_truth_TT0}). We evaluate the predictive performance with two metrics: ignorance score \citepsupp[IGN,][]{roulston_evaluating_2002} and root mean squared error (RMSE). Let $\Upsilon$ be either $\Psi$, $\vel$, or $\TT$. Then,
\begin{align*}
    \operatorname{IGN}  = - \sum_{\svec \in \calS } \log p( \Upsilon_{\sf T} (\svec) \;|\; \Data \, ; \widehat{\bbBeta}, \widehat{\bbXi}), \qquad 
    \operatorname{RMSE} = \sqrt{ \frac{1}{|\calS| } \sum_{\svec \in \calS } \left( \Upsilon_{\sf T} (\svec) - \widehat{\Upsilon} (\svec \, ; \widehat{\bbBeta}, \widehat{\bbXi}) \right)^2},
\end{align*}
where $\Upsilon_{\sf T} (\svec)$ is the ground truth $\Upsilon$ at $\svec$, and $p(\cdot \;|\; \Data)$ is the predictive Gaussian density function. While RMSE measures the deterministic accuracy, IGN measures the probabilistic accuracy by assessing the goodness of fit of the predictive distribution learned from the data to the ground truth.

\subsection{Effect of window size}

\begin{figure}[!htb]
    \begin{subfigure}{.48\textwidth}
        \centering
        \includegraphics[width=0.9\textwidth]{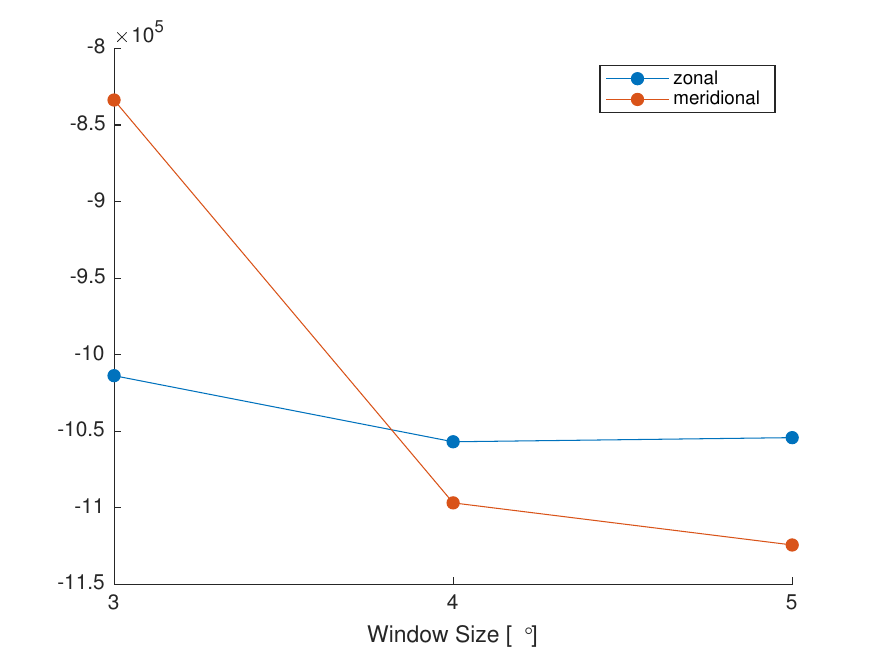}
        \caption{IGN}\label{fig: nll_vel}
    \end{subfigure}
    \centering
    \begin{subfigure}{.48\textwidth}
        \centering
        \includegraphics[width=0.9\textwidth]{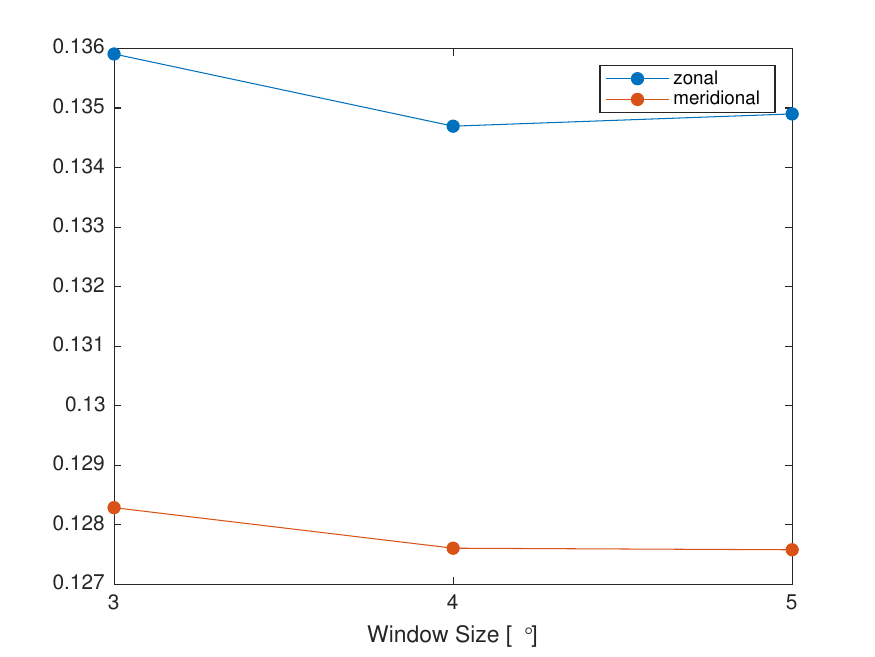}
        \caption{RMSE}\label{fig: rmse_vel}
    \end{subfigure}
    \caption{Performance metrics for predicting the velocity $(\vel)$ field by varying window size.}\label{fig: vel_metric_window}
\end{figure}

We investigate how the spatial bandwidth choice $\lambda_G$ affects the prediction of the velocity and $\TT$ fields. Figure~\ref{fig: vel_metric_window} shows the predictive performance metrics of velocities for $\lambda_G \in \{3^\circ, 4^\circ, 5^\circ\}$. This result highlights the impact of $\lambda_G$ to the predictive performance in the first stage of our procedure. A $4^\circ$ spatial window is the optimal choice in minimizing RMSE of both zonal and meridional velocities and in minimizing IGN of zonal velocity. One could consider $5^\circ$ if minimizing IGN of meridional velocity is the prime concern, but at the expense of worse zonal prediction performance and larger computational cost.

\begin{figure}[!htb]
    \begin{subfigure}{.48\textwidth}
        \centering
        \includegraphics[width=0.9\textwidth]{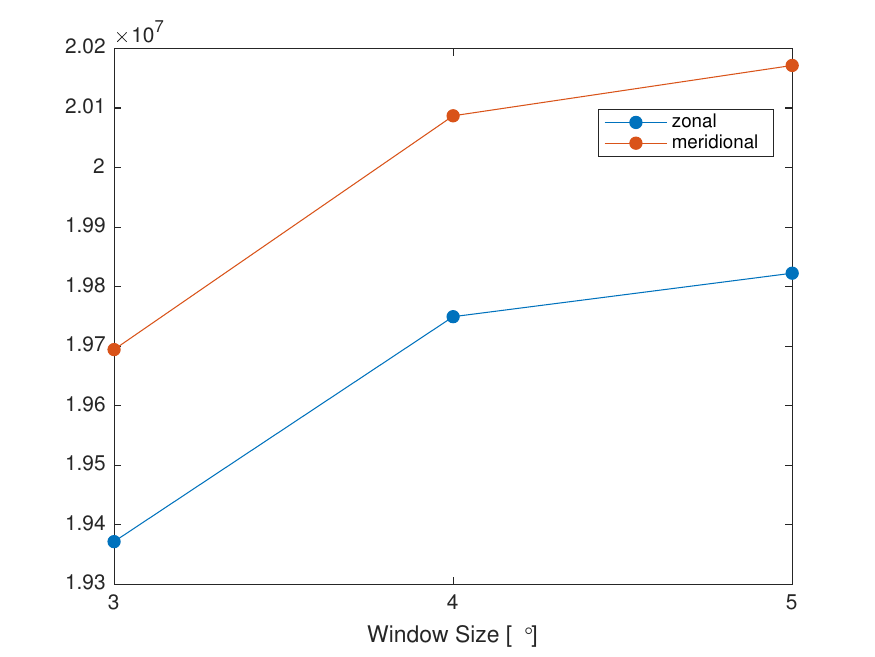}
        \caption{IGN}\label{fig: nll_OHT}
    \end{subfigure}
    \centering
    \begin{subfigure}{.48\textwidth}
        \centering
        \includegraphics[width=0.9\textwidth]{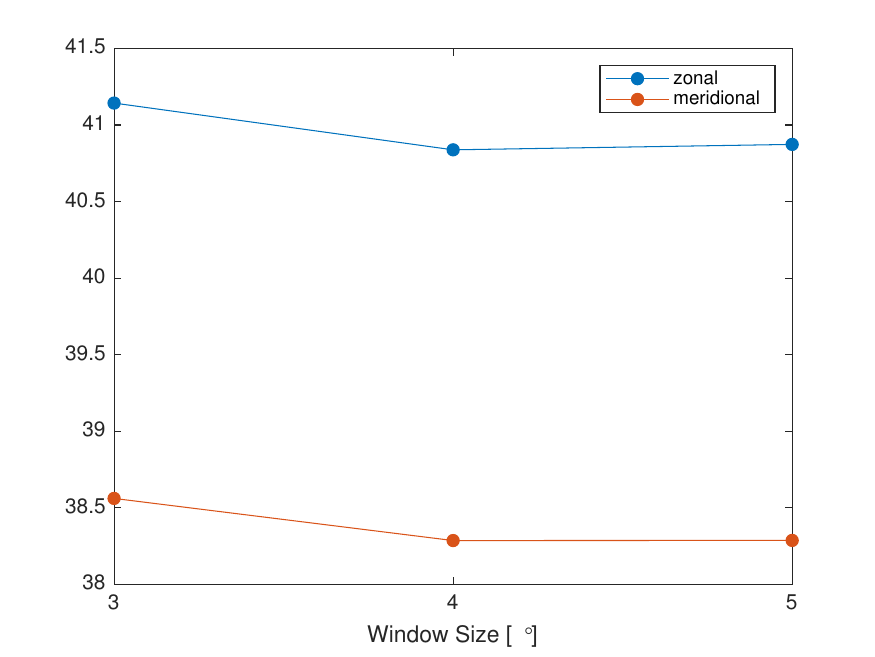}
        \caption{RMSE}\label{fig: rmse_OHT}
    \end{subfigure}
    \caption{Performance metrics for predicting the $\TT$ field by varying window size.}\label{fig: oht_metric_window}
\end{figure}

Figure~\ref{fig: oht_metric_window} shows the predictive metrics of $\TT$ by $\lambda_G$. In computing the metrics, the same $\lambda_G$ is used at both stages of the two-stage procedure. As it was the case for the velocities, a $4^\circ$ spatial window is optimal for minimizing RMSE of both zonal and meridional $\TT$. While IGN decreases as $\lambda_G$ decreases, we do not choose the $3^\circ$ window size since that choice leads to losing essential boundary dynamics due to too scarce data within the windows near the coastal boundaries. In conclusion, we adopt $\lambda_G = 4^\circ$ for our main results in Sections~\ref{sec:Results} and~\ref{sec: Validation}.

\subsection{Approximate EM algorithm} \label{sec: EM_iteration}

The goal of this analysis is to numerically investigate the claim that the proposed EM procedure improves over \citetsupp{kuusela_locally_2018} in both predictive performance and uncertainty quantification.
Since the approximate EM procedure is agnostic to the quantity of interest, we focus on $\Psi$ and $\vel$ in this comparison. This way, we can compare the interpolation performance of the procedure to the ground truth $\Psi$ field, as well as the prediction performance to the latent ground truth $\vel$. Recall that \citetsupp{kuusela_locally_2018} use the \citetsupp{roemmich_20042008_2009} mean field which is estimated by OLS and then estimate the covariance parameters from the residuals within the temporal window of interest. Even though the approximate EM procedure can aggregate the spatio-temporal covariance structure across different temporal windows in estimating the mean field, we limit the comparison to gridded full fields predicted at November 15th of every year, which is the center point of the temporal window for the local Gaussian process ranging from October 1st to December 31st, to make a fair comparison between the EM procedure and \citetsupp{kuusela_locally_2018}.

\begin{figure}[!htb]
    \begin{subfigure}{.48\textwidth}
        \centering
        \includegraphics[width=0.9\textwidth]{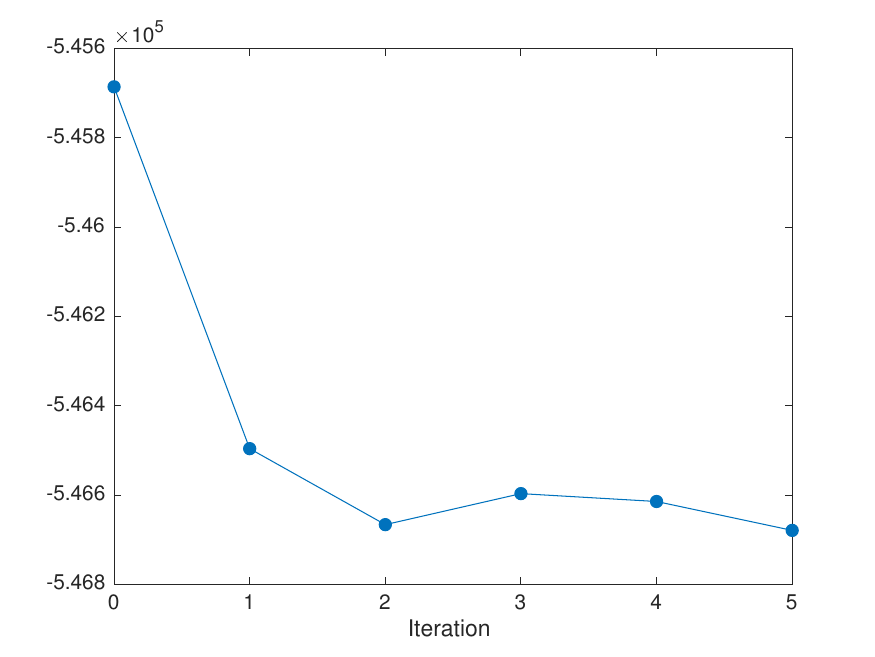}
        \caption{IGN}\label{fig: nll_adt}
    \end{subfigure}
    \centering
    \begin{subfigure}{.48\textwidth}
        \centering
        \includegraphics[width=0.9\textwidth]{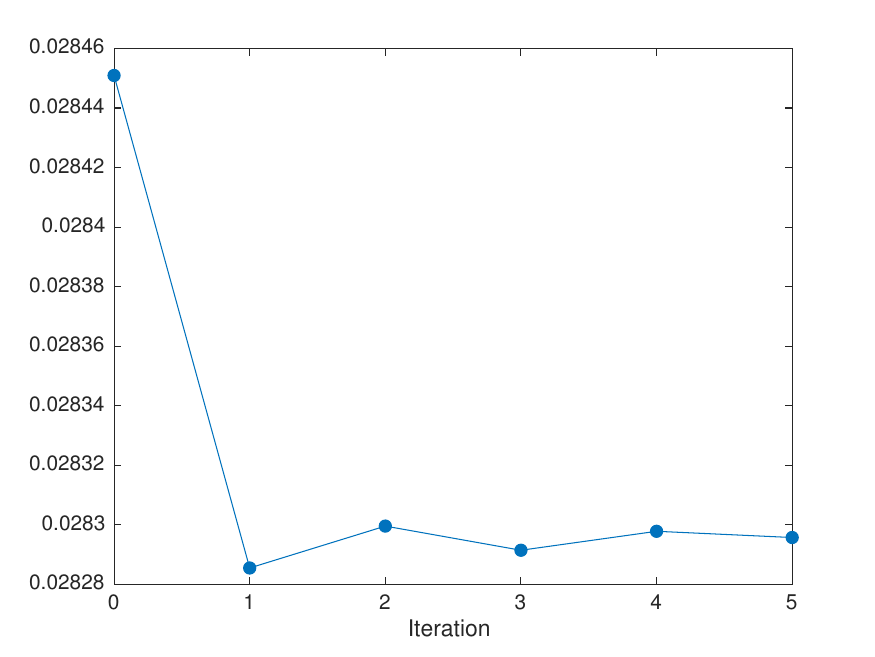}
        \caption{RMSE}\label{fig: rmse_adt}
    \end{subfigure}
    \caption{Performance metrics for predicting the absolute dynamic topography $(\Psi)$ field by EM iteration.}\label{fig: adt_metric_EM}
\end{figure}

\begin{figure}[!htb]
    \begin{subfigure}{.48\textwidth}
        \centering
        \includegraphics[width=0.9\textwidth]{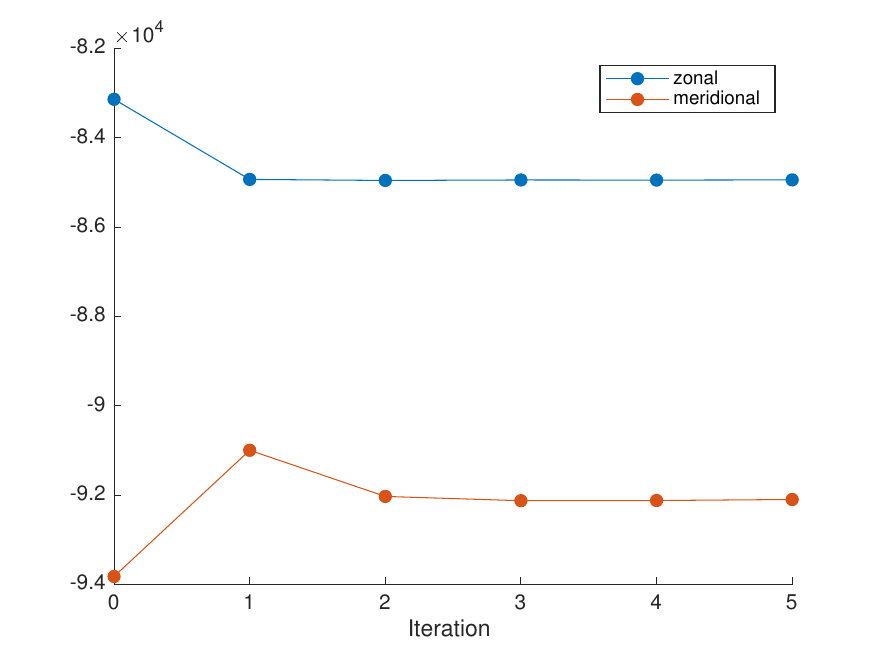}
        \caption{IGN}\label{fig: nll_vel_EM}
    \end{subfigure}
    \centering
    \begin{subfigure}{.48\textwidth}
        \centering
        \includegraphics[width=0.9\textwidth]{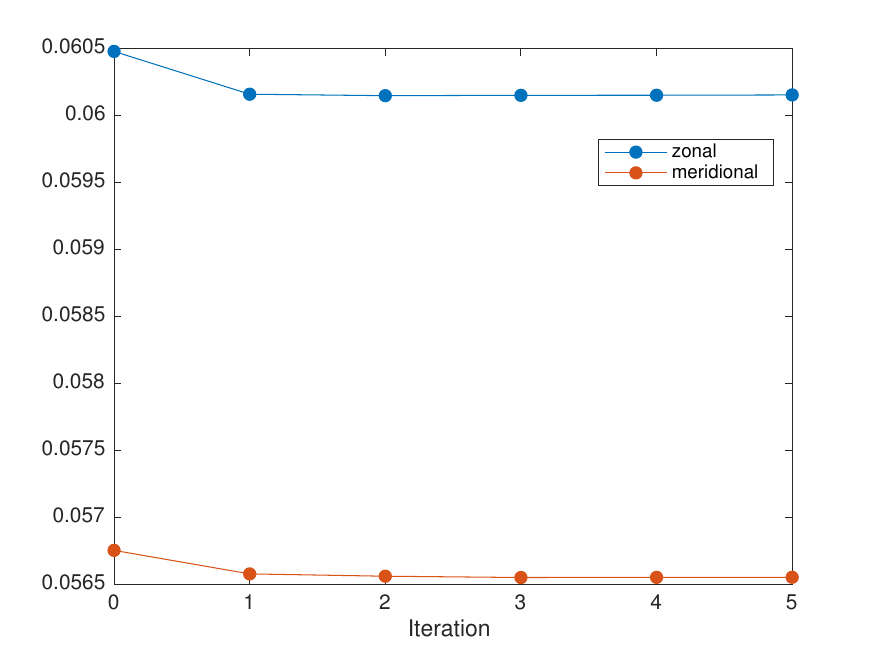}
        \caption{RMSE}\label{fig: rmse_vel_EM}
    \end{subfigure}
    \caption{Performance metrics for predicting the velocity $(\vel)$ field by EM iteration.}\label{fig: vel_metric_EM}
\end{figure}

Figures~\ref{fig: adt_metric_EM} and \ref{fig: vel_metric_EM} show the predictive performance metrics---IGN and RMSE---of absolute dynamic topography and zonal and meridional velocities by EM iteration, respectively. Since iteration 0 corresponds to the performance of \citetsupp{kuusela_locally_2018}, we conclude that jointly estimating the mean and the covariance parameters using the proposed EM procedure improves the predictive performance in both chosen metrics.

\vspace{10em}

\section{Joint Analysis of Argo and Spray Gliders}\label{sec: Supp_Spray}

Spray underwater gliders \citepsupp{sherman_autonomous_2001,rudnick_spray_2016} are buoyancy driven autonomous profiling vehicles that drive along strong fronts in a sawtooth path. With float density decreasing dramatically within the Gulf Stream on its shoreward side, the Argo program cannot (and is not intended to) thoroughly sample the Gulf Stream along the continental shelf, but gliders can ably fill this role \citepsupp{todd_underwater_2017}.

\begin{figure}[!htb]
    \begin{subfigure}{.48\textwidth}
        \centering
        \includegraphics[width=0.9\textwidth]{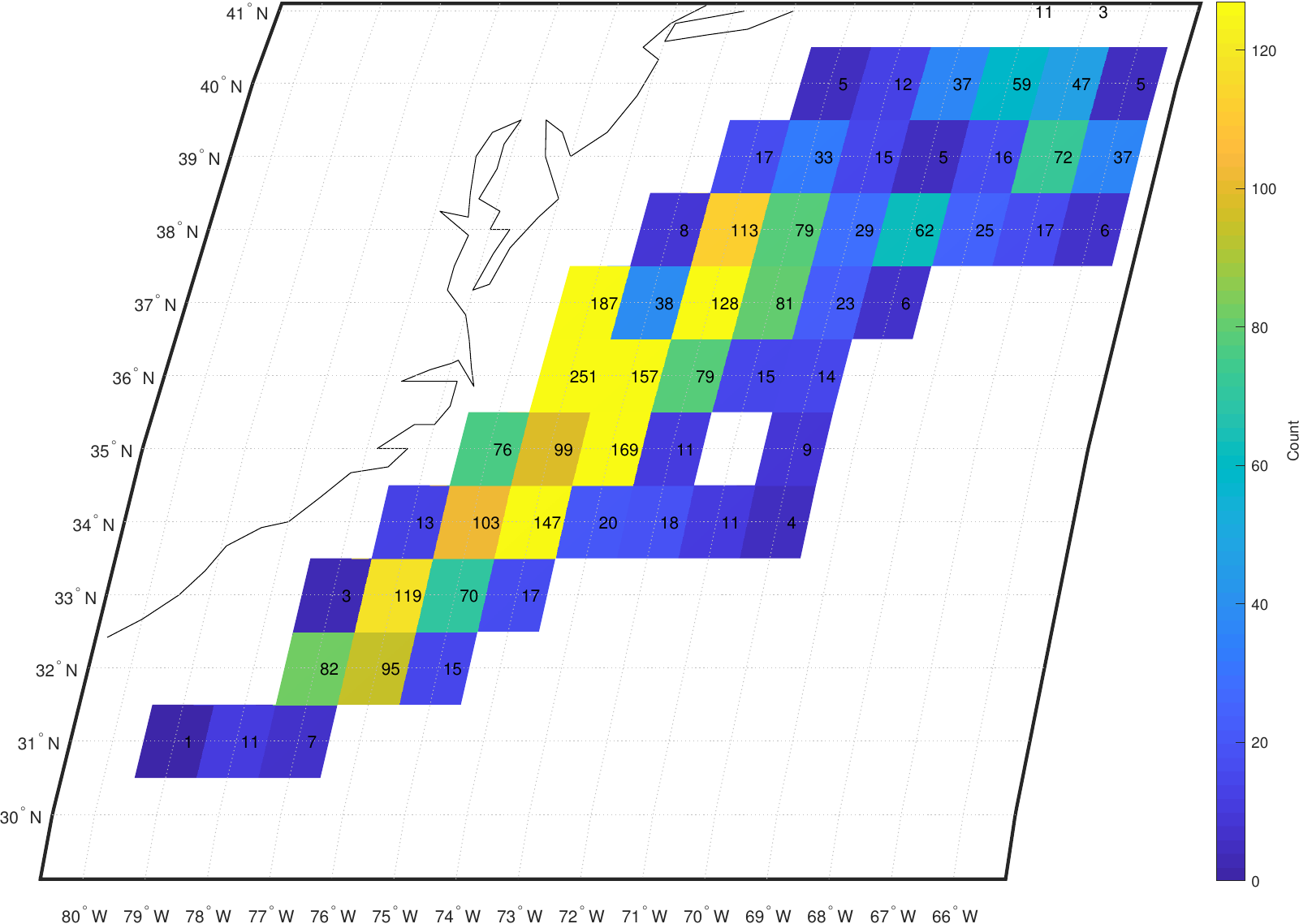}
        \caption{Spray}\label{fig: nobs_Spray}
    \end{subfigure}
    \centering
    \begin{subfigure}{.48\textwidth}
        \centering
        \includegraphics[width=0.9\textwidth]{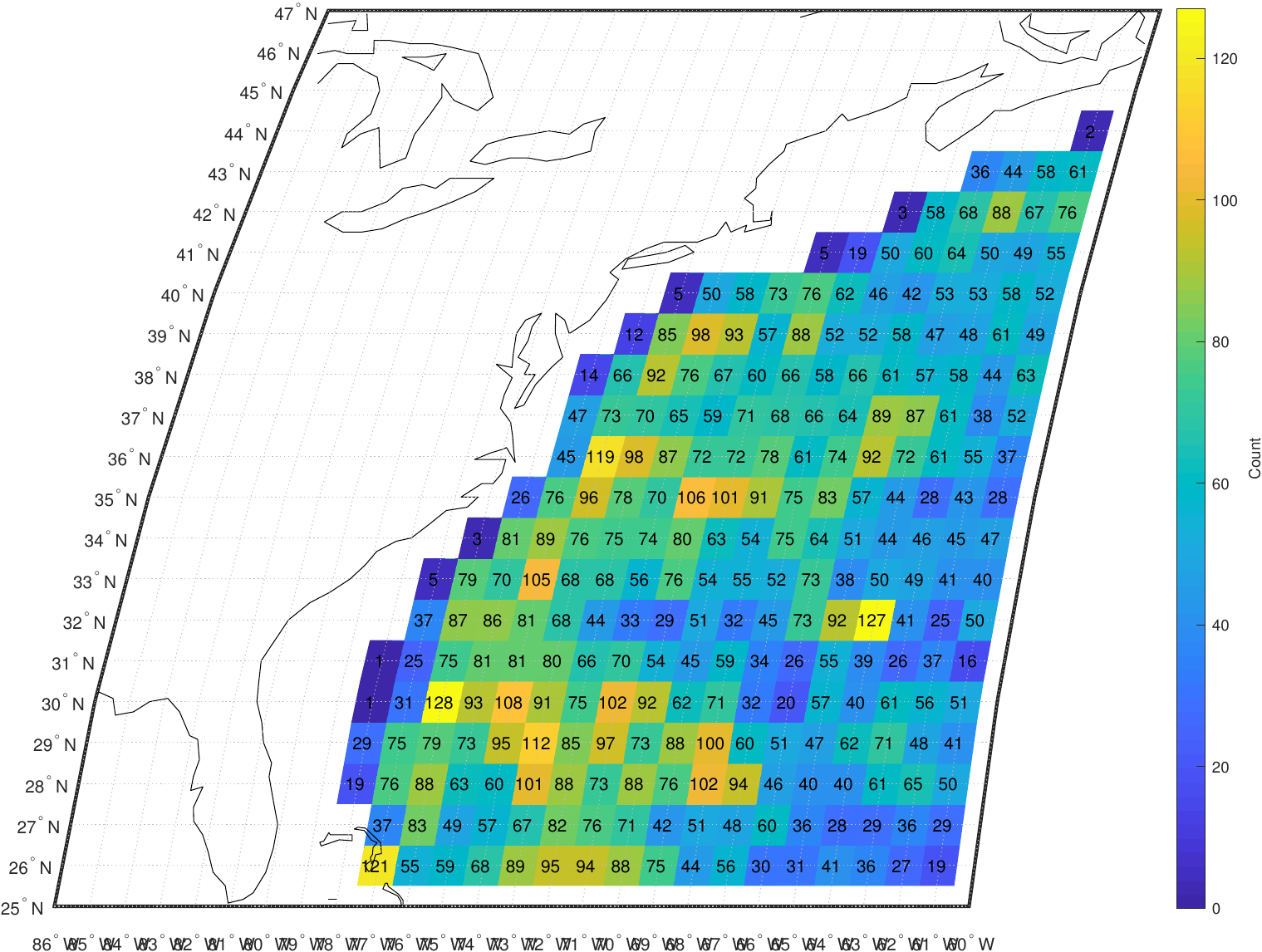}
        \caption{Argo}\label{fig: nobs_Argo}
    \end{subfigure}
    \caption{Number of profiles in $1^\circ \times 1^\circ$ grid at $15$dbar.}\label{fig: nobs}
\end{figure}

 Similarly to the Argo program, Spray gliders measure (in-situ) temperature, salinity, and pressure. In addition, they also measure absolute velocity. Out of 10,577 available profiles  recorded between January 2007 and December 2018 \citepsupp{rudnick_spray_2016}, 2,791 ($26.4\%$) profiles have measurements down to $900$ dbar. These profiles can be used along with the Argo profiles to improve the geostrophic velocity estimate. Figure~\ref{fig: nobs} shows the number of profiles binned in $1^\circ \times 1^\circ$ grid at $15$ dbar from Spray and Argo, respectively.

Argo-based estimates are in a surprisingly good agreement with satellite estimates as we may see from the validation section in the main paper. Most regions with relatively larger residuals, however, are close to the continental shelves where Argo float density decreases dramatically within the shallow coastal region since the floats cannot dive to 2000 dbar. This decreased sampling density degrades the quality of the estimates. We empirically show that this is indeed the case by coalescing the Spray glider observations with the Argo array. Under our data-driven statistical framework, we can simply form the union of the two data sets and re-run the mapping algorithms with the combined set of data. We re-estimate all the relevant coefficients, covariance parameters, and apply the debiasing procedure to the aggregated data.

\subsection{Geostrophic Velocity \texorpdfstring{$\hat{\vel}$}{\textbf{v}}}\label{sec: Spray_Vel}

\begin{figure}[!htb]
    \begin{subfigure}{.48\textwidth}
        \centering
        \includegraphics[width=0.75\textwidth]{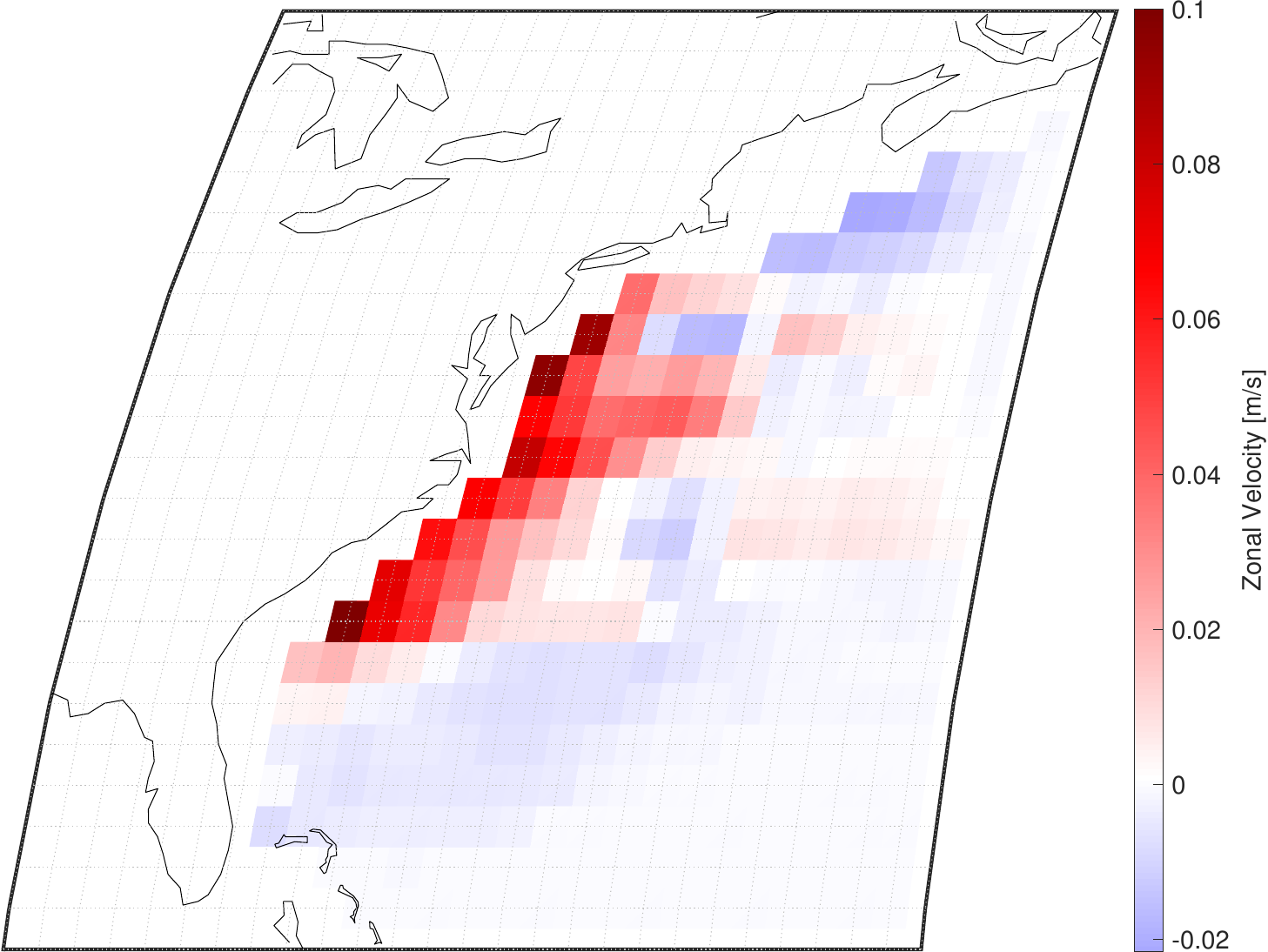}
        \caption{Zonal (Debiased)}\label{fig: mean_zonal15_Diff}
    \end{subfigure}
    \centering
    \begin{subfigure}{.48\textwidth}
        \centering
        \includegraphics[width=0.75\textwidth]{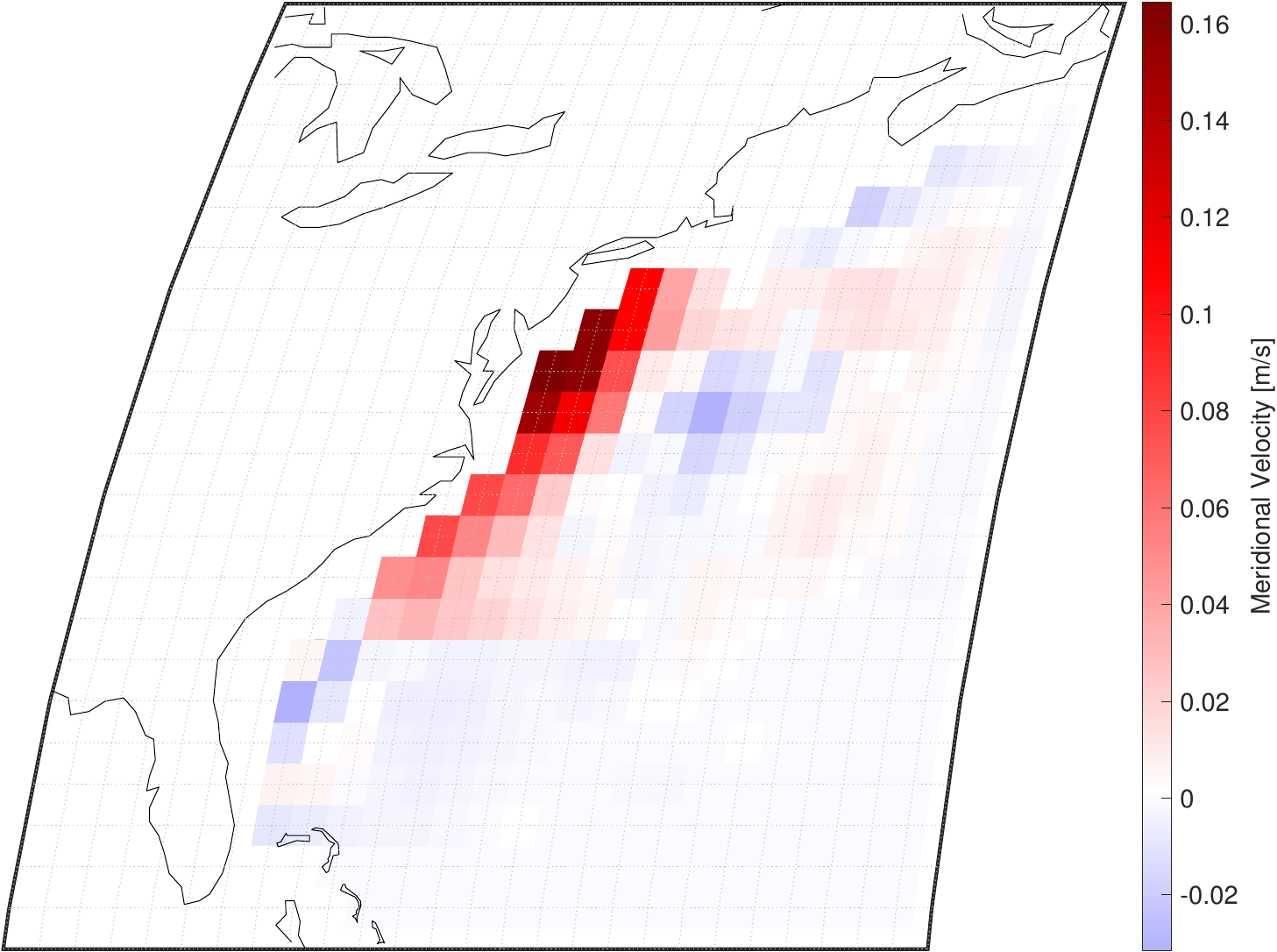}
        \caption{Meridional (Debiased)}\label{fig: mean_meridional15_Diff}
    \end{subfigure}
    \caption{Time averaged Mean velocity $\text{Av} (\hat{\vel})$ differences at 15 dbar.}\label{fig: mean_geovel15_Diff}
\end{figure}

Figure~\ref{fig: mean_geovel15_Diff} illustrates the difference in the time averaged mean geostrophic velocity $\text{Av} (\hat{\vel})$ at 15 dbar between estimates from aggregated Argo and Spray profiles and from Argo profiles only. Positive value in red means that the estimate from the aggregated data has a larger value than that from Argo by itself. While the magnitude difference is smaller in deeper depth, the qualitative differences in deeper depths were consistent with Figure~\ref{fig: mean_geovel15_Diff}.

Clearly, including the Spray gliders helps resolve underestimated signals close to coastal shelves. The improvement is not only substantial in magnitude but also has a critical impact for quantifying ocean heat transport, in that WBCs are the key driving component in the large-scale ocean circulation. This also implies that the miscalibration near the coastal shelves is due to an innate limitation in the spatial coverage of the Argo array rather than due to insufficient statistical modeling.

\subsection{Heat Transport \texorpdfstring{$\hat{\OHT}$}{\textbf{OHT}} and MHT}

\begin{figure}[!htb]
    \centering
    \begin{subfigure}{.48\textwidth}
        \centering
        \includegraphics[width=0.75\textwidth]{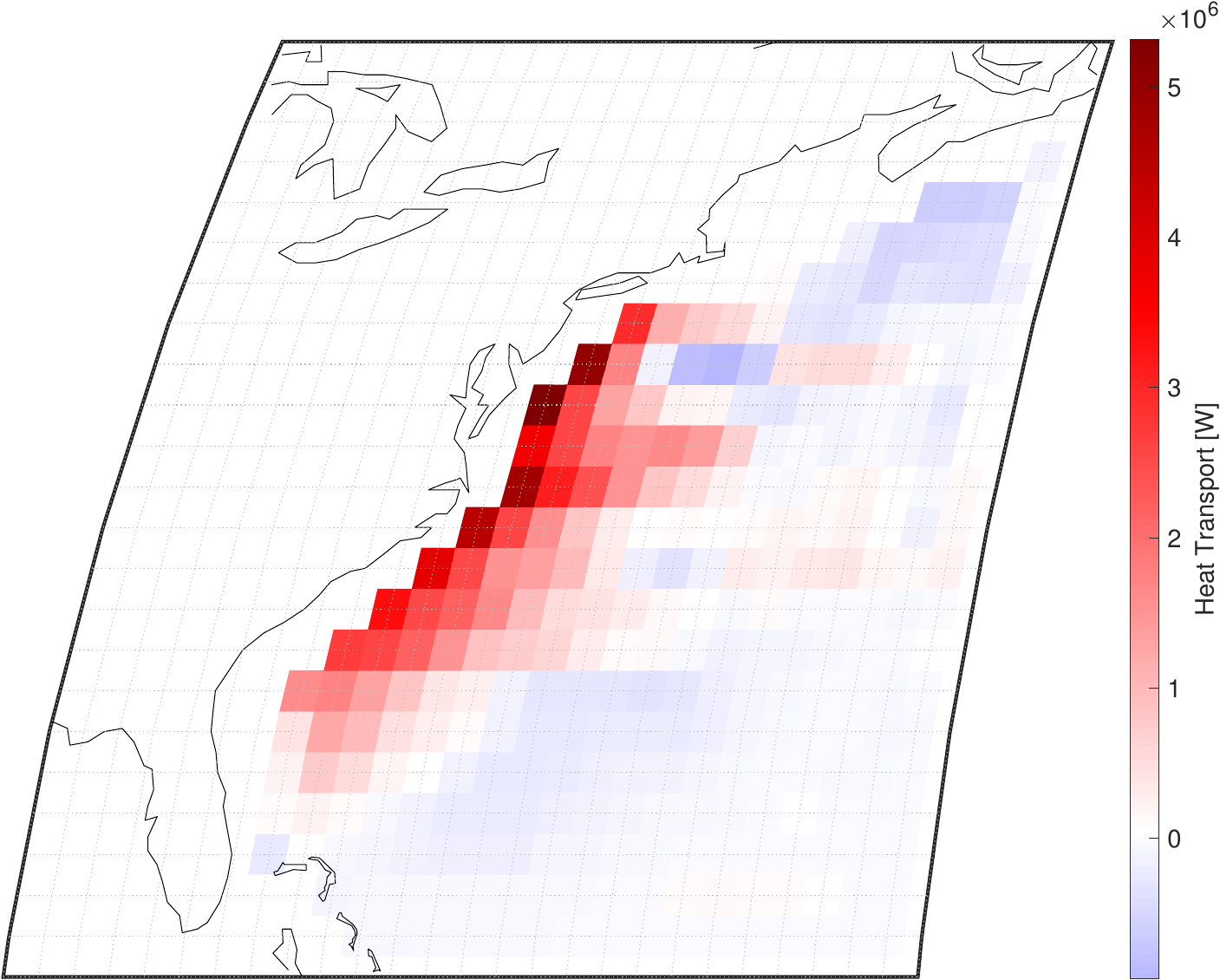}
        \caption{Zonal (Debiased)}\label{fig: meanht_zonal10_Diff}
    \end{subfigure}
    \begin{subfigure}{.48\textwidth}
        \centering
        \includegraphics[width=0.75\textwidth]{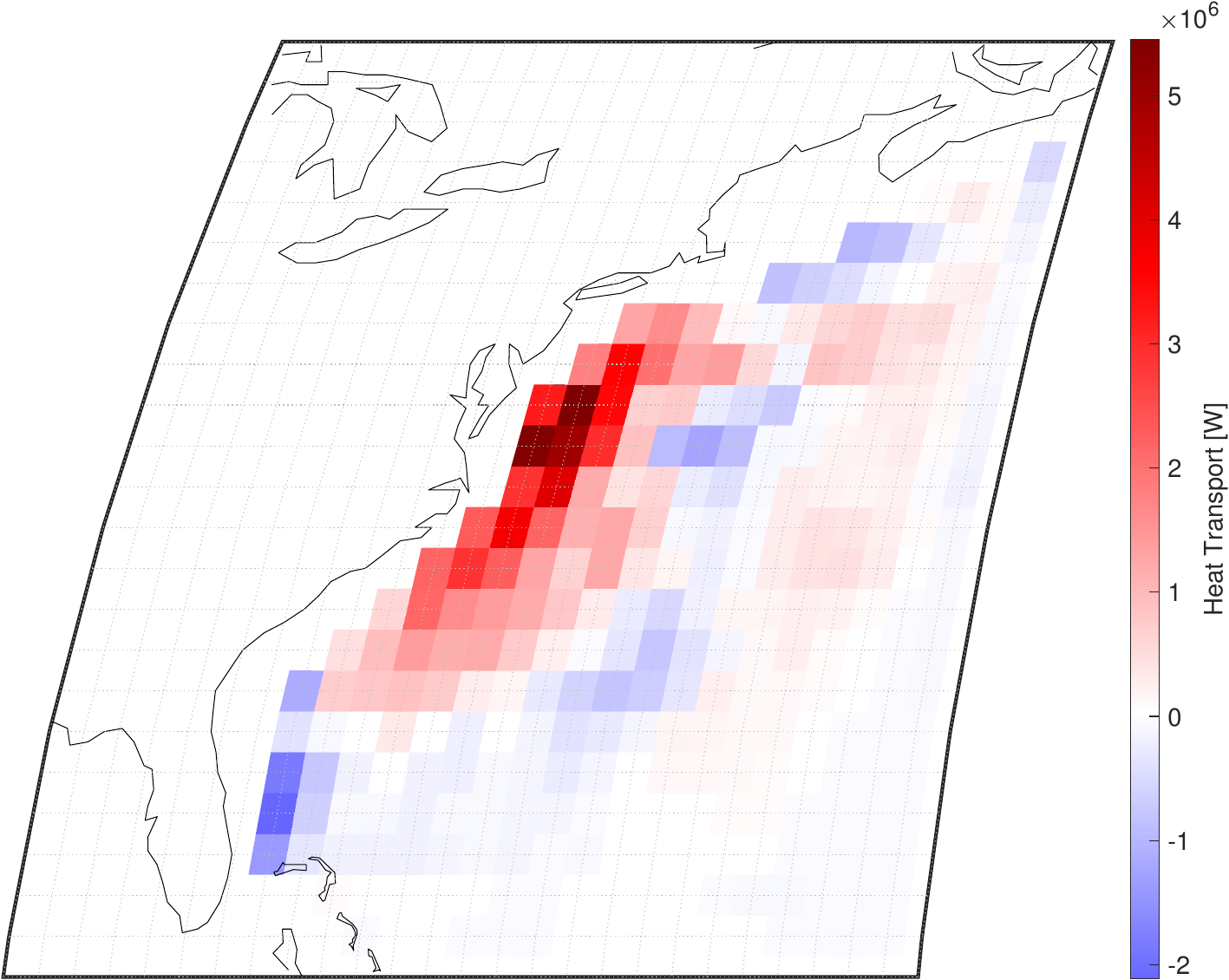}
        \caption{Meridional (Debiased)}\label{fig: meanht_meridional10_Diff}
    \end{subfigure}
    \caption{Mean absolute heat transport $\text{Av} (\widehat{\OHT})$ differences over $10$ to $900$ dbar.}\label{fig: meanht_10_SprayArgo_Diff}
\end{figure}

The Argo-only underestimate in the geostrophic velocity $\hat{\vel}$ has a direct consequence on the heat transport. Figure~\ref{fig: meanht_10_SprayArgo_Diff} illustrates the difference in the upper ocean time averaged mean heat transport $\text{Av} (\widehat{\OHT})$ between estimates from the aggregated profiles and those from Argo profiles only. Positive values in red mean that there is more transport when Spray profiles were included in the analysis. The result shown is based on the seasonally averaged mean transport suppressing the seasonal cycle explicitly modeled for $\OHT$. Similar spatial underestimation near the coastal shelves for both $\text{Av} (\widehat{\OHT})$ and $\text{Av} (\hat{\vel})$ confirms that the velocity underestimation was consistent throughout varying depths.

\begin{figure}[!htb]
        \centering
        \includegraphics[trim={0 0 0 0},clip,width=0.5\textwidth, origin=c]  {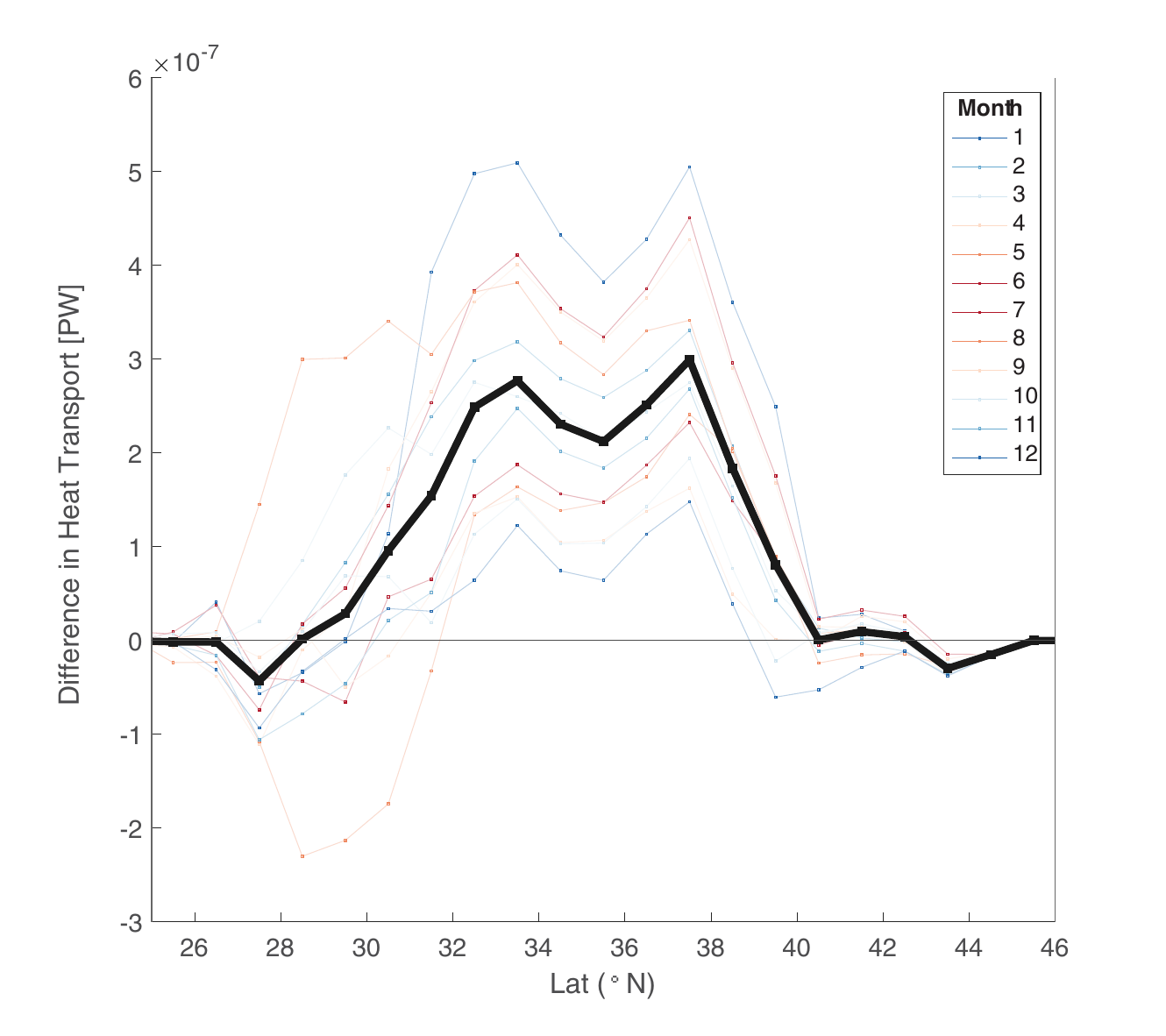}
    \caption{Estimated difference in mean Meridional Heat Transport from $10$ dbar to $900$ dbar.}\label{fig: mean_MHT_Spray}
\end{figure}

Mean meridional heat transport (MHT) differences in Figure~\ref{fig: mean_MHT_Spray} summarize the aggregated underestimated signals over the relevant latitudes. Positive difference implies that there is more transport when Spray profiles were jointly analyzed with Argo. The black bold line corresponds to the seasonally averaged transport as shown in Figure~\ref{fig: meanht_10_SprayArgo_Diff} and separate monthly transports accounting for the seasonal cycle are overlaid. There is a large fluctuation between $28^\circ$N to $32^\circ$N in early and late summer months. Nevertheless, the Spray gliders effectively capture the underestimated heat transport in general. 

